\documentclass[%
 aip,
 amsmath,amssymb,
 reprint,%
]{revtex4-1}

\usepackage{graphicx}
\usepackage{dcolumn}
\usepackage{bm}

\usepackage[utf8]{inputenc}
\usepackage[T1]{fontenc}
\usepackage{mathptmx}
\usepackage{etoolbox}
\usepackage{float}
\usepackage{caption}
\usepackage{subcaption}
\usepackage{xcolor}
\usepackage{url}
\usepackage{hyperref}

\newcommand{\bM}{\bm M}

\newcommand{\bD}{\bm D}

\newcommand{\bx}{\bm x}

\newcommand{\bz}{\bm z}

\newcommand{\bU}{\bm U}
\newcommand{\bv}{\bm v}
\newcommand{\bV}{\bm V}

\newcommand{\bSigma}{\bm{\Sigma}}

\makeatletter
\def\@email#1#2{%
 \endgroup
 \patchcmd{\titleblock@produce}
  {\frontmatter@RRAPformat}
  {\frontmatter@RRAPformat{\produce@RRAP{*#1\href{mailto:#2}{#2}}}\frontmatter@RRAPformat}
  {}{}
}%
\makeatother
\begin{document}

\preprint{AIP/123-QED}

\title[]{Generalization capabilities and robustness of hybrid models grounded in physics compared to purely deep learning models}
\author{R. Abadía-Heredia}%
\email{sr.abadia@upm.es}%
\affiliation{%
ETSI Aeronáutica y del Espacio, Universidad Politécnica de Madrid, Plaza Cardenal Cisneros, 3, Madrid, 28040, Madrid, Spain.
}
\author{A. Corrochano}%
\affiliation{%
ETSI Aeronáutica y del Espacio, Universidad Politécnica de Madrid, Plaza Cardenal Cisneros, 3, Madrid, 28040, Madrid, Spain.
}
\author{M. Lopez-Martin}%
\affiliation{%
ETSI Aeronáutica y del Espacio, Universidad Politécnica de Madrid, Plaza Cardenal Cisneros, 3, Madrid, 28040, Madrid, Spain.
}
\author{S. Le Clainche}%
\affiliation{%
ETSI Aeronáutica y del Espacio, Universidad Politécnica de Madrid, Plaza Cardenal Cisneros, 3, Madrid, 28040, Madrid, Spain.
}%

\date{\today}

\begin{abstract}
This study \footnote{This article has been submitted to Physics of Fluid. After it is published, it will be found at \url{https://pubs.aip.org/aip/pof}.} investigates the generalization capabilities and robustness of purely deep learning (DL) models and hybrid models based on physical principles in fluid dynamics applications, specifically focusing on iteratively forecasting the temporal evolution of flow dynamics. Three autoregressive models were compared: a hybrid model (POD-DL) that combines proper orthogonal decomposition (POD) with a long-short term memory (LSTM) layer, a convolutional autoencoder combined with a convolutional LSTM (ConvLSTM) layer and a variational autoencoder (VAE) combined with a ConvLSTM layer. These models were tested on two high-dimensional, nonlinear datasets representing the velocity field of flow past a circular cylinder in both laminar and turbulent regimes. The study used latent dimension methods, enabling a bijective reduction of high-dimensional dynamics into a lower-order space to facilitate future predictions. While the VAE and ConvLSTM models accurately predicted laminar flow, the hybrid POD-DL model outperformed the others across both laminar and turbulent flow regimes. This success is attributed to the model's ability to incorporate modal decomposition, reducing the dimensionality of the data, by a non-parametric method, and simplifying the forecasting component. By leveraging POD, the model not only gained insight into the underlying physics, improving prediction accuracy with less training data, but also reduce the number of trainable parameters as POD is non-parametric. The findings emphasize the potential of hybrid models, particularly those integrating modal decomposition and deep learning, in predicting complex flow dynamics. 
\end{abstract}

\maketitle

\section{Introduction}


Fluid dynamics is the study of fluid motion and is fundamental to various natural and industrial processes involving flow, including combustion systems \cite{sole_fluid_mixing}, power generation, aerodynamics \cite{Marusicetal2003}, marine propulsion \cite{DeMontG88}, blood flow in tumors \cite{AbadiaetalRIENG2021}, and the wakes of flying insects \cite{Insect}, among others. This broad range of applications has driven significant interest in the characterization of fluid flows. However, they are governed by nonlinear, high-dimensional system of equations, posing substantial challenges for their resolution. In most cases, analytical solutions are intractable, and numerical approaches require discretization schemes with a large number of degrees of freedom, leading to high computational costs.

This complexity is exacerbated in complex flows, namely flows in transitional and turbulent regime, which describe the majority of flows encountered in both industry and nature. These complex flows are characterized by the existence of multiple spatio-temporal scales within their dynamics. The number of scales increases nearly quadratic with the Reynolds number (Re), a non-dimensional number defined as $\hbox{Re}=VD/\nu$, where $V$ and $D$ are the velocity and the characteristic length of the problem, respectively, and $\nu$ denotes the kinematic viscosity. However, resolving all flow scales numerically can become computationally prohibitive for many problems.

Therefore, there is a high interest in reducing the computational cost associated with obtaining high-fidelity predictions of flow dynamics \cite{brunton_etal_2020_ML_fluid}. To address this issue, reduced order models (ROMs) have been proposed as an effective solution \cite{roms}. They are designed to capture the most essential dynamics of a flow within a given context. There are two primary types of ROMs: intrusive projection-based ROMs and non-intrusive data-driven ROMs. Intrusive projection-based techniques, such as the Galerkin projection, aim to reduce the degrees of freedom in the governing equations by expanding them within a transformed space. This space is typically constructed using an orthogonal basis that retains only the most significant or energetic modes. The orthogonal basis is commonly derived through proper orthogonal decomposition (POD) \cite{aubry_1991_pod}, a method first applied to analyze a turbulent flow by Lumley \cite{lumley_1967_pod}. By projecting the governing equations onto this energetic basis, the ROM effectively captures the primary flow dynamics. The basis must be carefully computed depending on the specific problem \cite{vinuesa_forecasting}. 
Conversely, a non-intrusive data-driven ROM approach can infer the underlying physics from the data itself without requiring explicit knowledge of the governing equations. In this context, techniques such as POD \cite{aubry_1991_pod}, spectral POD \cite{spectral_pod_2016}, dynamic mode decomposition (DMD) \cite{schmid2010} and higher order dynamic mode decomposition (HODMD) \cite{LeClaincheAndVega2017} are commonly utilized. A complete review of non-intrusive data-driven ROM techniques applied to fluid dynamics can be found in \cite{modal_decomposition_review}.

Furthermore, the advent of deep learning (DL) algorithms has significantly expanded the potential for developing non-intrusive ROMs to predict flow dynamics \cite{Kutz_2017, brunton_etal_2020_ML_fluid}. For instance, a DL model has been employed to predict small-scale structures in turbulent flows at high Reynolds numbers, by previously training the model on data from turbulent flows at low and medium Reynolds numbers \cite{rom_small_scales_from_big}. Another study utilized a generative adversarial network (GAN) to predict the water depth field for various coastlines, using only high-fidelity simulation data for training \cite{gan_coastal_line}. Note in these cases, there was no need to modify the governing equations to predict the new dynamics; instead, the dynamics were extrapolated from data-driven models that were able to capture the physics behind the data. These DL-ROM techniques are considered surrogate models, offering significant reductions in computational time while maintaining the ability to predict complex flow dynamics accurately.

In other scenarios these DL-ROM methodologies can be used to forecast the flow dynamics based on a historical record, allowing a significant reduction in the computational cost associated with obtaining the solutions required. This is the scope of this work.

Time series forecasting, applied to fluid dynamics, has been extensively studied in the literature, and different deep learning architectures and methodologies have been proposed. For example, Pathak {\it et al.} \cite{forecasting_chaos_echo_state_2018} used an echo-state-network based on recurrent neural networks to forecast the state evolution of a chaotic system. In another work, Vlachas {\it et al.} \cite{forecasting_chaos_echo_state_2018_b} proposed the use of long short-term memory (LSTM) networks \cite{hochreiter_etal_1997_lstm} to forecast chaotic systems. In works proposed by R. Han {\it et al.} \cite{forecasting_convAutoEnc_2019} and K. Hasegawa {\it et al.} \cite{forecasting_convAutoEnc_2D_2020}, a DL architecture combining a convolutional autoencoder with LSTM networks was introduced to capture the spatial and temporal features of a spatio-temporal flow dynamics. Similarly, T. Nakamura 
{\it et al.} \cite{forecasting_convAutoEnc_3D_2021} extended this approach to three-dimensional flow fields, enhancing the model's ability to handle more complex spatial patterns.

A similar methodology can be applied to probabilistic forecasting, where the goal is to determine the probability distribution of future states rather than a single deterministic prediction \cite{gneiting_2014_probForecast}. This has been achieved combining variational autoencoders (VAEs) with recurrent neural networks (RNNs), which provide a probabilistic approach to encoding and decoding the latent space \cite{prob_forecasting_vae_2020, prob_forecasting_vae_2022}.

The use of DL autoencoders can be substitute by non-intrusive ROMs. For example, Ali {\it et al.} \cite{forecasting_dmdc_dl_2021} proposed a hybrid model combining dynamic mode decomposition with control (DMDc) and deep learning architectures. POD can also be employed to reduce the dimensionality, with forecasting performed on the POD coefficients to predict the temporal evolution of the POD modes. Several authors have studied these types of hybrid models, combining POD with deep learning architectures \cite{forecasting_pod_dl_2018, forecasting_pod_dl_2019, parish_etal_2020_LSTM_traditional, abadiaherediaetal_2022, forecasting_pod_dl_2022, corrochano_etal2023_NNComb}. Moreover, a probabilistic forecasting version of these hybrid POD-DL models has been proposed by \cite{prob_forecasting_pod_dl_2020}.

The previously mentioned methods have in common to project the high-dimensional dynamics in a low-dimensional latent space, either using autoencoders or modal decomposition techniques, and then forecast in this reduced space. Hasegawa {\it et al.} \cite{forecasting_convAutoEnc_2D_2020} studied the role of latent spaces in forecasting high-dimensional dynamics.

This study evaluates three autoregressive methodologies for forecasting spatiotemporal flow dynamics using a small dataset, with the aim of reducing computational costs compared to numerically solving the governing equations. The selected models are grounded in the contemporary literature about spatiotemporal forecasting and include the following: (1) A residual autoencoder, chosen for its proven efficacy in handling spatiotemporal data by integrating a convolutional autoencoder for spatial feature extraction with a convolutional long short-term memory (ConvLSTM) network for temporal forecasting; (2) a variational autoencoder (VAE), which adopts a similar architecture—combining a convolutional autoencoder with an LSTM network—while leveraging a probabilistic framework; and (3) a POD-DL model, which integrates modal decomposition techniques with deep learning architectures to incorporate physical insights and enhance predictive accuracy.

Another objective of this study is to compare various methods of dimensionality reduction for the creation of latent dimensions. Specifically, the performance of deep learning autoencoders—encompassing both the residual and the variational autoencoders—is evaluated against POD, which is employed by the POD-DL model.

Deep learning autoencoders rely entirely on neural networks to compress high-dimensional data into a lower-dimensional latent space, enabling the capture of nonlinear relationships within the spatial domain. In contrast, POD achieves dimensionality reduction by projecting data onto an orthogonal basis constructed from a set of POD modes, which effectively represent the most energetically significant structures in the system's dynamics.

While nonlinear DL autoencoders generally exhibit superior compression capabilities compared to POD, they often lack interpretability regarding the underlying physical phenomena \cite{evazi_etal_2022, pod_autoencoders_ConvAE_Fukami_2020, pod_autoencoders_ConvAE_Fukami_2020_ordered, pod_autoencoders_ConvAE_Murata_2020, munoz_etal_2023_sort_latent_variables}. An exception to this is the linear autoencoder, whose solutions have been observed to closely resemble those of POD modes \cite{pod_autoencoders_petros_2002}. To maintain clarity, the term "autoencoder" will hereafter refer exclusively to DL autoencoders, unless explicitly stated otherwise.

To the best of the authors' knowledge, no comprehensive comparison of these existing methods has been conducted to date. This paper presents, for the first time, an in-depth evaluation of the capabilities of this class of models and demonstrates the advantages of generating hybrid models based on physics-informed approaches, like the POD-DL. The results indicate that the predictive power, generalization ability, and robustness of these hybrid models surpass those of purely deep-learning-based methods. This work represents a significant breakthrough in the field, achieving an optimal balance between computational efficiency—enabled by dimensionality reduction—and the generation of new, high-fidelity data.

This work is organized as follows: Section \ref{sec: methodology} describes the three models studied in this work. Section \ref{sec: results_discussion} explains how the models were trained and shows the results from each model and dataset. Section \ref{sec: discussion} discusses the results obtained. Finally, Section \ref{sec: conclusion} presents the conclusions drawn from this work.

\section{Methodology} \label{sec: methodology}

In time series forecasting, the goal is to produce accurate predictions of future events $\{\bv_{t+1}, \bv_{t+2}, \dots\}$ based on historical data $\{\bv_{1}, \dots, \bv_{t}\}$. In this study, we investigate and compare three distinct approaches: two that rely entirely on deep learning (DL) models, and a third that combines proper orthogonal decomposition (POD) with DL, which we refer to as the POD-DL model. The latter was previously developed and tested in \cite{abadiaherediaetal_2022,corrochano_etal2023_NNComb}.

These approaches fall into two main categories: point forecasting and probabilistic forecasting \cite{gneiting_2014_probForecast}. The former aims to approximate a function $f$ as follows,
\begin{equation}
    f:[J_{1} \times J_{2} \times J_{3} \times J_{4} \times q] \rightarrow [J_{1} \times J_{2} \times J_{3} \times J_{4} \times 1].
    \label{eq: deterministic_func}
\end{equation}
Here, $J_{1}$ represents the components dimension of the dataset (i.e., different velocity components, different Reynolds numbers, different initial conditions, etc.). The indices $J_{2}$, $J_{3}$, $J_{4}$ correspond to the spatial discretization along the $x$-, $y$- and $z$-axis, respectively. The variable $q$ denotes the length of the historical data used for prediction, while the value $1$ indicates that the objective is to forecast a single time-step ahead. The computation of more than one prediction, as shown in Sec. \ref{sec: results_discussion}, is obtained by an autoregressive procedure. A more detailed explanation of the dataset structure and preprocessing steps is provided in Appendix \ref{sec: datasets_and_preprocessing}.

More precisely this function $f$ represents the forecasting process, as follows,
\begin{equation}
    f(\{\bv_{k}\}_{k=t-q+1}^{t}) = \hat{\bv}_{t+1} \simeq \bv_{t+1}.
\end{equation}
In this work, a residual autoencoder and POD-DL models are utilized to approximate this function. The loss function used to fit these models is the mean squared error (MSE), defined as follows,
\begin{equation}
    \hbox{MSE} = \frac{1}{n}\sum_{i=1}^{n}(\bv_i - \hat{\bv}_i)^{2},
    \label{eq: mse}
\end{equation}
where $\hat{\bv}_i$ is the prediction obtained from the forecasting model.
 
The probabilistic forecasting framework is similar to point forecasting, with the key difference being that instead of approximating a function, we aim to approximate the probability distribution $\mathbb{F}_{\bv}$ underlying the forecasting process. To achieve this, a variational autoencoder (VAE) \citep{kingma2022autoencoding} is employed. In the following section we explain the proposed models in more detail, starting with the two models within the point forecast framework.

\subsection{Point forecasting: Residual autoencoder and POD-DL}
\label{sec: deterministic_framework}
This subsection presents the models that belong to the point forecasting methodology, where the objective is to do a deterministic prediction by approximating a function as described in eq. \eqref{eq: deterministic_func}. The models presented reduce data dimensionality by means of an encoder (residual autoencoder) or POD (POD-DL). In the reduced-dimensional space, deep learning architectures are used to forecast the evolution of the flow. Finally, to recover the original (high-dimensional) space, the residual autoencoder uses a decoder and the POD-DL reverses the POD.

More precisely, in the POD-DL model, singular value decomposition (SVD) \cite{Sirovich87} is applied to decompose a high-dimensional dataset $\bD$ in a product of matrices, as follows,

\begin{equation}
    \bD = \bU \bSigma \bV^{T}.
\end{equation}

Here, the matrix $\bU$ represents the POD modes, the diagonal matrix $\bSigma$ contains the singular values, and $\bV^{T}$ corresponds to the POD coefficients.

In most applications is enough to retain only the most energetic POD modes (i.e., columns of matrix $\bU$) representing important structures (i.e., coherent structures) within the flow dynamics. This truncation of modes is usually done by looking at the singular values (i.e., diagonal elements of matrix $\bSigma$) that contain the relevance of each POD mode within the flow dynamics. The temporal evolution of these POD modes is carried out by the POD coefficients (i.e., columns of matrix $\bV^{T}$).

The forecasting is then performed in the weighted POD coefficients, $\bSigma \bV^{T}$, where their temporal evolution is predicted using deep learning architectures, drawing parallels to Galerkin methods.

Here is where resides the principal advantages of the POD-DL model in contrast to the other models: (1) truncating the POD modes enables forecasting to focus specifically on the most important flow scales for a required problem, as in most cases there is no need to solve the smaller flow scales, generally related to uncorrelated flow dynamics; (2) SVD is a computationally efficient approach to compute the POD modes; and (3) by going from a high-dimensional tensor to a matrix (i.e., $\bSigma \bV^{T}$) where columns represent the temporal dimension, POD simplifies the model architecture required for forecasting, reducing the risk of overfitting and improving generalization.

In this work the deep learning architecture used to predict the evolution of the POD temporal coefficients is a long-short term memory (LSTM), where in a previous study \cite{parish_etal_2020_LSTM_traditional} was observed to outperform traditional forecasting methods in both accuracy and computational efficiency. This advantage arises primarily because traditional methods often require significant computational resources and struggle to generalize to new or unseen conditions. This was also observed in the M4 forecasting competition where machine learning models outperformed the classical probabilistic models \cite{MAKRIDAKIS_etal_2018_M4}.

\begin{figure}[h]
	\centering
	\includegraphics[height=0.3\columnwidth]{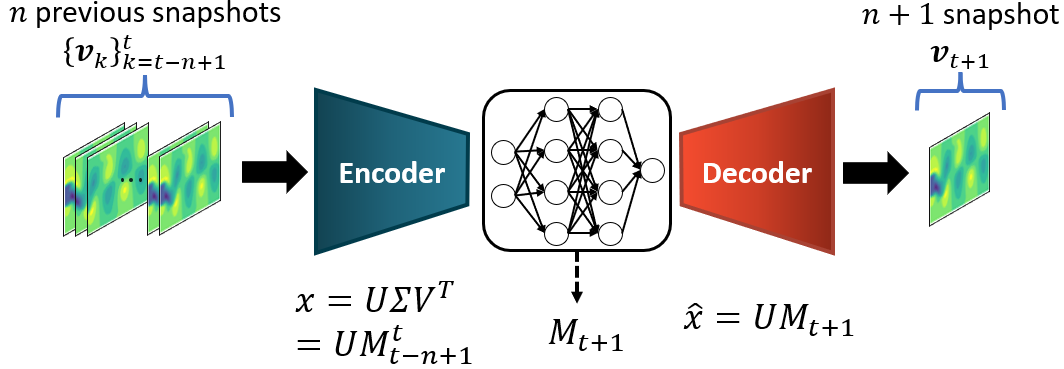}
	\caption{Overview of the architecture of the POD-DL model. Note, it can be represented as an encoder-decoder structure. However, in this case the encoder is just the matrix decomposition ($\bU, \bSigma, \bV^{T}$) from SVD. The temporal information is formed by a matrix multiplication $\bM_{t-n+1}^{t} = \bSigma * (\bV_{t-n+1}^{t})^{T}$, which is send to a DL model, usually a long-short term memory (LSTM), to predict the time-ahead sample, i.e., $\bM_{t+1}$. Finally, the decoder reconstructs the snapshot by multiplying matrices $\bx_{t+1} = \bU * \bM_{t+1}$.}
	\label{fig: svd_nn_arch}
\end{figure}

The detailed architecture of the POD-DL model is provided in Table \ref{tab: pod_dl}. This architecture was already used in previous works \cite{parish_etal_2020_LSTM_traditional,abadiaherediaetal_2022,corrochano_etal2023_NNComb} in different applications to the ones presented in this article, illustrating the generalization capabilities of this model and the robustness of the architecture presented. Figure \ref{fig: svd_nn_arch} provides a visual depiction of the POD-DL methodology.

\begin{figure}[H]
	\centering
	\includegraphics[height=0.22\columnwidth]{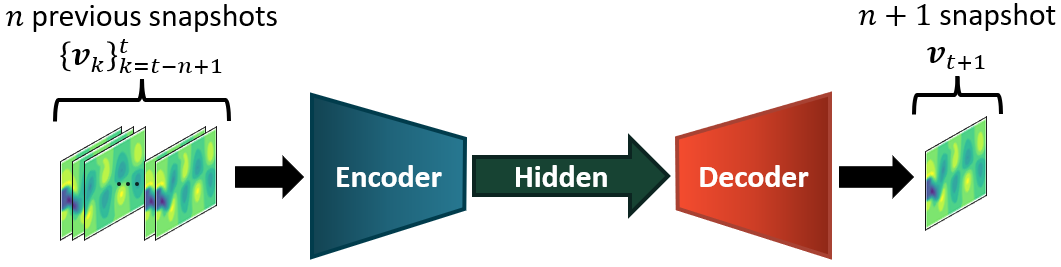}
	\caption{Overview of the architecture of the residual autoencoder. Unlike the POD-DL, the encoder and decoder of this model are composed of convolutional neural networks.}
	\label{fig: autoencoder_arch}
\end{figure}

The second model, which is based on the works presented by \cite{forecasting_convAutoEnc_2019, forecasting_convAutoEnc_2D_2020, forecasting_convAutoEnc_3D_2021}, uses a residual convolutional autoencoder, as illustrated in Fig. \ref{fig: autoencoder_arch}.  The detailed architecture of the model, when applied to the laminar flow, is provided in Tables \ref{tab: encoder}, \ref{tab: hidden}, and \ref{tab: decoder} in Appendix \ref{appen: model_arch_tab}. 

\begin{figure*}
	\centering
	\begin{subfigure}[b]{0.7\textwidth}
		\centering
		\includegraphics[width=\textwidth]{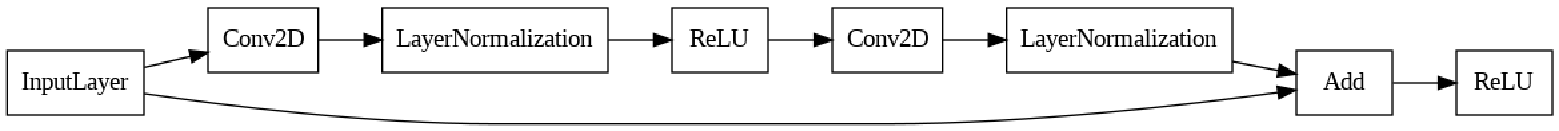}
        \caption{}
	\end{subfigure}
    \hfill
	\begin{subfigure}[b]{0.7\textwidth}
		\centering
		\includegraphics[width=\textwidth]{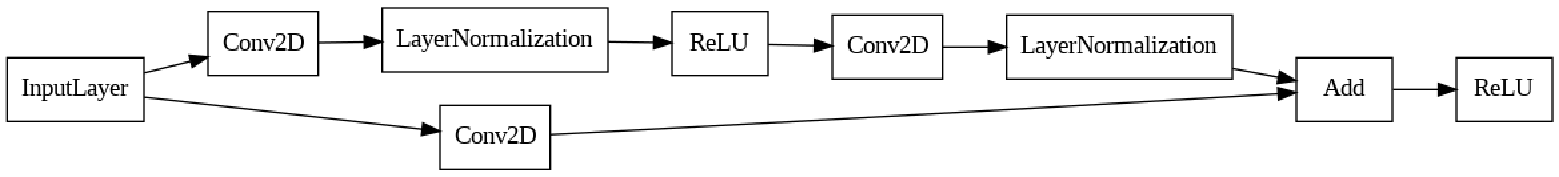}
        \caption{}
	\end{subfigure}
	\caption{Visual representation of both types of blocks, identity (top) and convolutional (bottom). Both of them add up the input and output values of the block. However the identity block (top) does not reduce the spatial dimension of the input snapshot while the convolutional block (bottom) does it.}
	\label{fig: conv_ident_blocks}
\end{figure*}

The encoder, on the one hand, consists of residual convolutional networks (ResCNN) \citep{kaiming2016_ResCNN}, where the input and output of a CNN are added together. These ResCNNs are primarily defined by the identity and convolutional blocks, depicted in Fig. \ref{fig: conv_ident_blocks}. Both blocks add the input and output values, but with a key difference: the identity block neither reduces the size nor changes the number of channels of the input snapshot, while the convolutional block does. This architecture accelerates training and enhances robustness. The underlying principle is that it is easier for a block to represent the identity function (requiring only the convolutional kernels to be zero) than for the CNN to do so. On the other hand, the decoder is composed of transpose convolutional networks, which perform the inverse operation of convolutional kernels, allowing to go back from the low-dimensional latent space to the high-dimensional dynamics, as shown in Figure \ref{fig: autoencoder_arch}.

Note that both the encoder and decoder process only the spatial dimensions of the snapshots and do not alter the temporal dimension. In Appendix \ref{appen: model_arch_tab}, Fig. \ref{fig: enc_out_kth} displays some outputs from the encoder when incoming snapshots are from the laminar flow. Unlike the latent variables obtained from SVD, these variables are not hierarchically ordered based on the energy they contain. However, several methods have been proposed to sort the latent variables produced by an autoencoder \citep{pod_autoencoders_ConvAE_Fukami_2020_ordered, munoz_etal_2023_sort_latent_variables}, providing information on which latent variable contributes most to the overall dynamics.

A hidden layer composed of a convolutional LSTM layer (ConvLSTM)  \citep{shi_etal_2015_convlstm} is used to capture the temporal correlation in the input sequence of snapshots $\{\bv_{i-n+1}, \bv_{i-n+2}, \dots, \bv_{i}\}$. For detailed architectural information, refer to Table \ref{tab: hidden}. Note in Fig. \ref{fig: autoencoder_arch}, this hidden layer is situated between the encoder and decoder to forecast on the latent variables.

For both the residual and variational autoencoders a ConvLSTM is chosen instead of the regular LSTM \citep{hochreiter_etal_1997_lstm}, which is used in the POD-DL model. This is because the inputs to these models are snapshots, and the LSTM layer only accepts vector-formatted entries. Using regular LSTM would have required flattening the latent variables returned by the encoder, which would result in losing spatial information and increasing the number of trainable parameters. In contrast, ConvLSTM accepts matrix-formatted entries, allowing for a more natural flow of information through the model.

This problem is not encountered in the POD-DL model because the entries to this model are vectors (the POD coefficients), allowing to use the LSTM architecture.

\subsection{Probabilistic forecasting: Variational autoencoder}

\begin{figure}[H]
	\centering
	\includegraphics[trim = {0cm, 0cm, 0cm, 0cm}, height=0.26\columnwidth]{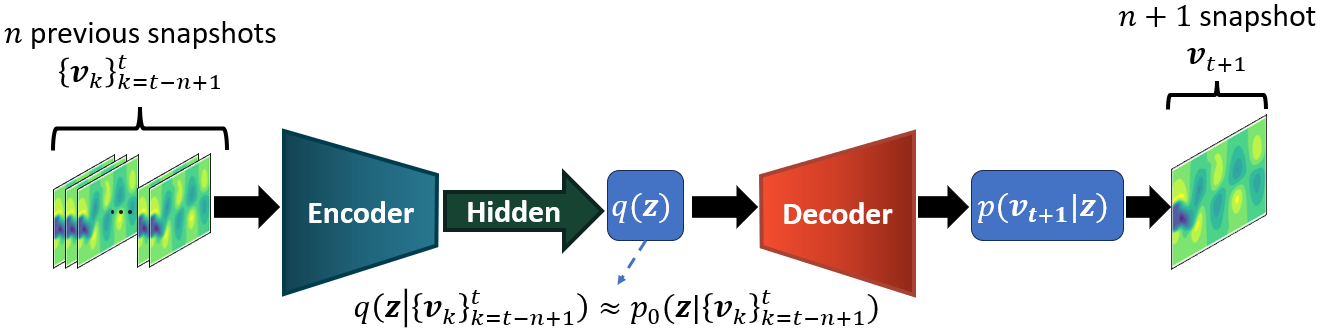}
	\caption{Overview of the architecture of the VAE. Similar to the residual autoencoder, the encoder and decoder of this model are composed of convolutional neural networks.}
	\label{fig: vae_arch}
\end{figure}

In contrast to point forecasting, our goal here is to approximate the probability distribution $\mathbb{F}_{\bv}$ that governs the forecasting process from a sequence of previous snapshots $\{\bv_{i-n+1}, \bv_{i-n+2}, \dots, \bv_{i}\}$ to the next snapshot $\{\bv_{i+1}\}$. To achieve this, we define a third model which adopts the Bayesian perspective of statistical inference and develop a variational inference model \citep{VarInf_Blei_2017} based on variational autoencoders (VAEs) \citep{kingma2022autoencoding}. This model is presented in this article for the first time, to the authors' knowledge. For a visual depiction of the proposed model, see Fig. \ref{fig: vae_arch}.

As with the residual autoencoder, VAE adopts an encoder-decoder structure that operate on the spatial dimension of the snapshots, and there is a hidden block that operates on the temporal dimension to capture the temporal correlations within the latent space. The specific architecture used for the laminar flow is detailed in Tables \ref{tab: vae_encoder} and \ref{tab: vae_decoder}.

In contrast to both the residual convolutional autoencoder and the POD-DL model, VAE aims to approximate the conditional posterior probability function $p_{0}(\bz|\{\bv_i\}_{i=t-n+1}^{t})$ of the latent variables $\bz$ based on the input sequence of snapshots $\{\bv_{t-n+1}, \dots, \bv_{t}\}$. 

This conditional posterior $p_{0}(\cdot)$ is parametric, and the information from the encoder and hidden layer is utilized to select the optimal parameters. Additionally, the likelihood $p(\bv_{t+1}|\bz)$ of the future snapshot $\{\bv_{t+1}\}$ is approximated based on the latent variables $\bz$.

Note from Fig. \ref{fig: vae_arch} that the decoder input comes from the posterior distribution $p_{0}(\bz|\{\bv_i\}_{i=t-n+1}^{t})$. Therefore, it is necessary to generate a sample from this posterior and use it as input to the decoder. The same is done with the likelihood $p(\bv_{t+1}|\bz)$ to obtain the time-ahead predicted snapshot.

The original variational autoencoder presented in \citep{kingma2022autoencoding} was applied to approximate the probability distribution underlying a set of images. After training, it is possible to randomly generate a sample from the approximated posterior, pass that sample into the decoder, and ultimately generate a new image from the likelihood. In this way, the VAE can be understood as a random image generator. The loss function used in the study is the Evidence Lower BOund (ELBO), which is described as follows:
\begin{equation}
    \hbox{ELBO} = \mathbb{E}[\hbox{log} \ p(\bx|\bz)] - \hbox{KL}[q(\bz|\bx)||p_0(\bz|\bx)].
    \label{eq: elbo_orig}
\end{equation}
Where $\bx$ is the input image, $p_{0}(\bz|\bx)$ is the exact posterior of the latent variables $\bz$, $q(\bz|\bx)$ is the approximate posterior, $p(\bx|\bz)$ is the likelihood of the image $\bx$ given the latent variables $\bz$ and KL$[\cdot||\cdot]$ is the Kullback-Leibler divergence. Note that in the original formulation \eqref{eq: elbo_orig}, the ELBO is evaluated for a single image $\bx$. 

In this work, we explore a modification of this loss function \eqref{eq: elbo_orig}, where the exact and approximate posterior are conditioned on the input sequence of snapshots $\{\bv_i\}_{i=t-n+1}^{t}$ instead of a single image $\bx$. Additionally, we aim to approximate the likelihood of the time-ahead snapshot $p(\bv_{t+1}|\bz)$. Therefore the loss function used to train the VAE is as follows,
\begin{eqnarray}
    \hbox{ELBOf} &=& \mathbb{E}[\hbox{log} \ p(\bv_{t+1}|\bz)] -\nonumber \\
    & & \hbox{KL}[q(\bz|\{\bv_i\}_{i=t-n+1}^{t})||p_0(\bz|\{\bv_i\}_{i=t-n+1}^{t})].
    \label{eq: elbo_new}
\end{eqnarray}
This newly defined loss function allows to approximate the likelihood of the time-ahead snapshot based on the latent variables $\bz$ that are obtained from the previous sequence of snapshots $\{\bv_i\}_{i=t-n+1}^{t}$. Since this study focuses on a forecasting problem, the loss function used to train the VAE is the ELBOf \eqref{eq: elbo_new}, rather than the standard ELBO \eqref{eq: elbo_orig}.

\section{Results} \label{sec: results_discussion}

The three models presented in Section \ref{sec: methodology} are tested on two distinct datasets, each representing a laminar and turbulent flow problem, respectively. The models were evaluated on their ability to predict the future evolution of the flow dynamics for each dataset, using only a limited amount of data for training. A brief description of the two datasets, as well as the organization and preprocessing process carried out can be found in Appendix \ref{sec: datasets_and_preprocessing}.

\begin{table}
\caption{Training length measured in number of epochs, training time measured in minutes and prediction time of the time-ahead snapshots measured in seconds for each one of the three forecasting models and each dataset, respectively. Note the laminar flow is three-dimensional, and the turbulent flow is two-dimensional.\label{tab: training_length}}
\begin{ruledtabular}
\begin{tabular}{cccccl}

\multicolumn{1}{l}{\textbf{Dataset}} & \multicolumn{1}{c}{\textbf{Model}} & \multicolumn{1}{c}{\begin{tabular}[c]{@{}c@{}}\textbf{Training length}\\ (\textbf{epochs})\end{tabular}} & \multicolumn{1}{c}{\begin{tabular}[c]{@{}c@{}c@{}}\textbf{Training} \\ \textbf{time} \\ \textbf{(min.)}\end{tabular}} & \multicolumn{1}{c}{\begin{tabular}[c]{@{}c@{}c@{}}\textbf{Prediction} \\ \textbf{time} \\ \textbf{(sec.)}\end{tabular}} & \\ \hline

\multicolumn{1}{l}{Laminar} & \multicolumn{1}{c}{} & \multicolumn{1}{c}{} & \multicolumn{1}{c}{} & \multicolumn{1}{c}{}  & \\ \hline

\multicolumn{1}{l}{} & \multicolumn{1}{c}{POD-DL} & \multicolumn{1}{c}{420} & \multicolumn{1}{c}{$4.677$} & \multicolumn{1}{c}{$2.415$} &  \\

\multicolumn{1}{l}{} & \multicolumn{1}{c}{Residual AE} & \multicolumn{1}{c}{30} & \multicolumn{1}{c}{$280.623$} & \multicolumn{1}{c}{$165.071$} &  \\

\multicolumn{1}{l}{} & \multicolumn{1}{c}{VAE} & \multicolumn{1}{c}{30} & \multicolumn{1}{c}{$86.269$} & \multicolumn{1}{c}{$45.631$} &  \\ \hline

\multicolumn{1}{l}{Turbulent} & \multicolumn{1}{c}{} & \multicolumn{1}{c}{} & \multicolumn{1}{c}{} & \multicolumn{1}{c}{} &  \\ \hline

\multicolumn{1}{l}{} & \multicolumn{1}{c}{POD-DL} & \multicolumn{1}{c}{$420$} & \multicolumn{1}{c}{$30.418$} & \multicolumn{1}{c}{$1.683$} &  \\

\multicolumn{1}{l}{} & \multicolumn{1}{c}{Residual AE} & \multicolumn{1}{c}{$120$} & \multicolumn{1}{c}{$329.665$} & \multicolumn{1}{c}{$29.094$} &  \\

\multicolumn{1}{l}{} & \multicolumn{1}{c}{VAE} & \multicolumn{1}{c}{$14$} & \multicolumn{1}{c}{$13.966$} & \multicolumn{1}{c}{$14.746$} &  \\

\end{tabular}
\end{ruledtabular}
\end{table}

The following sections present the temporal predictions obtained from the three models when applied to the laminar (Section \ref{sec: results_synthetic}) and turbulent flows (Section \ref{sec: results_experimental}). Table \ref{tab: training_length} shows the number of epochs used to train each model, the training time and the inference time to generate time-ahead predictions. These times were computed by repeating the training and prediction processes multiple times and taking the mean values. The training of each model was halted once the loss function ceased to decrease, indicating that the training had reached at least a local minimum.

Note that the models studied in this work are autoregressive, i.e., these models use the time-ahead prediction $\{\bv_{t+1}\}$ from previous time steps $\{\bv_{t-n+1}, \dots, \bv_{t}\}$ as a new input (i.e., $\{\bv_{t-n+2}, \dots, \bv_{t}, \bv_{t+1}\}$) to generate a new prediction $\{\bv_{t+2}\}$, and so on. In both datasets, we ask the models to generate $200$ time-ahead snapshots.

\begin{table}
\caption{Learning rate used to train each model. Note the length of the input sequence of snapshots varies depending on the dataset for the residual autoencoder and the variational autoencoder, but not for the POD-DL model.\label{tab: learning_rate}}
\begin{ruledtabular}
\begin{tabular}{ccccl}

\multicolumn{1}{l}{\textbf{Model}} & \multicolumn{1}{c}{\begin{tabular}[c]{@{}c@{}}\textbf{Learning rate} \\ ($\alpha$)\end{tabular}} & \multicolumn{1}{c}{\begin{tabular}[c]{@{}c@{}}\textbf{Length}\\ \textbf{input sequence}\end{tabular}} & \multicolumn{1}{c}{\begin{tabular}[c]{@{}c@{}}\textbf{$\#$ predictions} \\ \textbf{generated}\end{tabular}} &  \\ \hline

\multicolumn{1}{l}{POD-DL} & \multicolumn{1}{c}{$10^{-3}$} & \multicolumn{1}{c}{$10$} & \multicolumn{1}{c}{$6$} &  \\

\multicolumn{1}{l}{Residual AE} & \multicolumn{1}{c}{$5 \times 10^{-4}$} & \multicolumn{1}{c}{$10$ (laminar) / $5$ (turbulent)} & \multicolumn{1}{c}{$1$} &  \\

\multicolumn{1}{l}{VAE} & \multicolumn{1}{c}{$5 \times 10^{-5}$} & \multicolumn{1}{c}{$10$ (laminar) / $5$ (turbulent)} & \multicolumn{1}{c}{$1$} &  \\

\end{tabular}
\end{ruledtabular}
\end{table}

Independently of the datasets, the Adam method \cite{kingma2017adam} is used the optimize the training of the models, with the default values for the parameters $\beta_{1} = 0.9$, $\beta_{2} = 0.999$ and $\epsilon = 10^{-8}$ (see details in \cite{kingma2017adam}) to train the models. 

The following parameters were identified as optimal for each model after conducting multiple experiments. These parameters varied between models, as they were fine-tuned to achieve the best performance for each specific case. The learning rate $\alpha$ differs between models, as shown in Table \ref{tab: learning_rate}. Additionally, both the length of the input sequence of snapshots and the number of time-ahead predictions generated by the models differ between models and datasets, as also detailed in Table \ref{tab: learning_rate}. Predicting multiple future samples simultaneously is a complex task that increases the overall training complexity, necessitating a larger training dataset.

After extensive experimentation, it was observed that the residual autoencoder and VAE models perform better when tasked with predicting a single time-ahead sample, in contrast to the POD-DL model. The architectural simplicity of the POD-DL model enables it to leverage a smaller training set while still predicting multiple time-ahead predictions effectively, i.e., a longer horizon.

The number of training and testing samples for each dataset is summarized in Table \ref{tab: num_training_testing}. These values represent the optimal training sample sizes determined for the residual autoencoder and VAE models. However, Appendix \ref{appen: different_training_size} presents an analysis of the POD-DL model's performance when trained with varying numbers of samples, highlighting its behavior under different training set sizes.

The selection of hyperparameters for all models was informed by extensive experimentations, aiming to balance the model complexity with training efficiency. For the residual autoencoder, hyperparameter tuning prioritized optimizing the capture of both spatial and temporal dependencies, as a smaller latent dimension may simplify forecasting but risks insufficient representation of spatial dependencies, potentially degrading the accuracy of dynamic predictions. Effective forecasting hinges on the encoder's ability to accurately capture the spatial structures within the flow, necessitating a careful trade-off in the latent dimension size.

For the VAE, hyperparameter tuning followed a similar approach to the residual autoencoder, with an additional focus on balancing the reconstruction loss and the Kullback-Leibler divergence.

Finally, in the case of the POD-DL model, hyperparameter tuning was centered on selecting an appropriate number of POD modes. This process aimed to capture the most energetically significant structures within the flow dynamics while ensuring computational feasibility, since the greater the number of modes retained, the longer and more complex the training will be. This is because the less energetic modes has a noisier dynamics. Truncating to the optimal number of modes was critical to preserving essential dynamics and maintaining a balance between model accuracy and efficiency.

\begin{table}
\caption{Number of snapshots used for both training and testing the models. Note the laminar flow is three-dimensional, while the turbulent flow is two-dimensional.\label{tab: num_training_testing}}
\begin{ruledtabular}
\begin{tabular}{cccccl}

\multicolumn{1}{l}{\textbf{Dataset}} & \multicolumn{1}{c}{\textbf{Snapshot dimension}} & \multicolumn{1}{c}{\textbf{Training}} & \multicolumn{1}{c}{\textbf{Testing}} & \multicolumn{1}{c}{\textbf{Total}} &  \\ \hline

\multicolumn{1}{l}{Laminar} & \multicolumn{1}{c}{3D} & \multicolumn{1}{c}{$299$} & \multicolumn{1}{c}{$200$} & \multicolumn{1}{c}{$499$} &  \\

\multicolumn{1}{l}{Turbulent} & \multicolumn{1}{c}{2D} & \multicolumn{1}{c}{$2800$} & \multicolumn{1}{c}{$1200$} & \multicolumn{1}{c}{$4000$} &  \\

\end{tabular}
\end{ruledtabular}
\end{table}

\begin{figure*}
     \centering
     \begin{subfigure}[b]{0.33\textwidth}
         \centering
         \includegraphics[width=\textwidth]{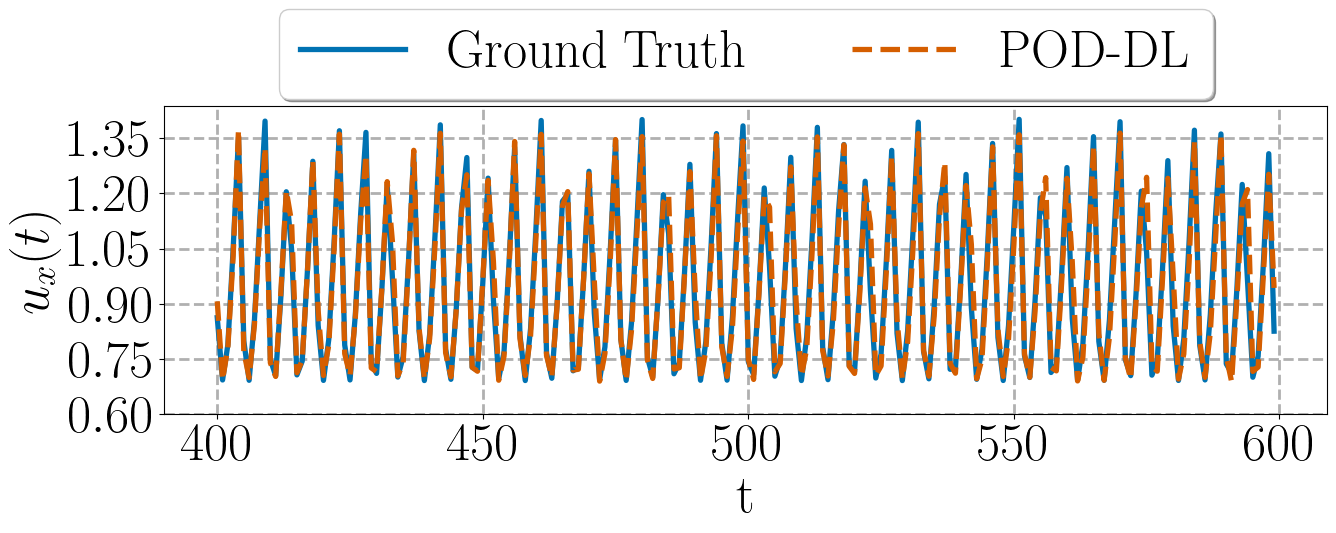}
         \caption{}
     \end{subfigure}
     \hfill
     \begin{subfigure}[b]{0.33\textwidth}
         \centering
         \includegraphics[width=\textwidth]{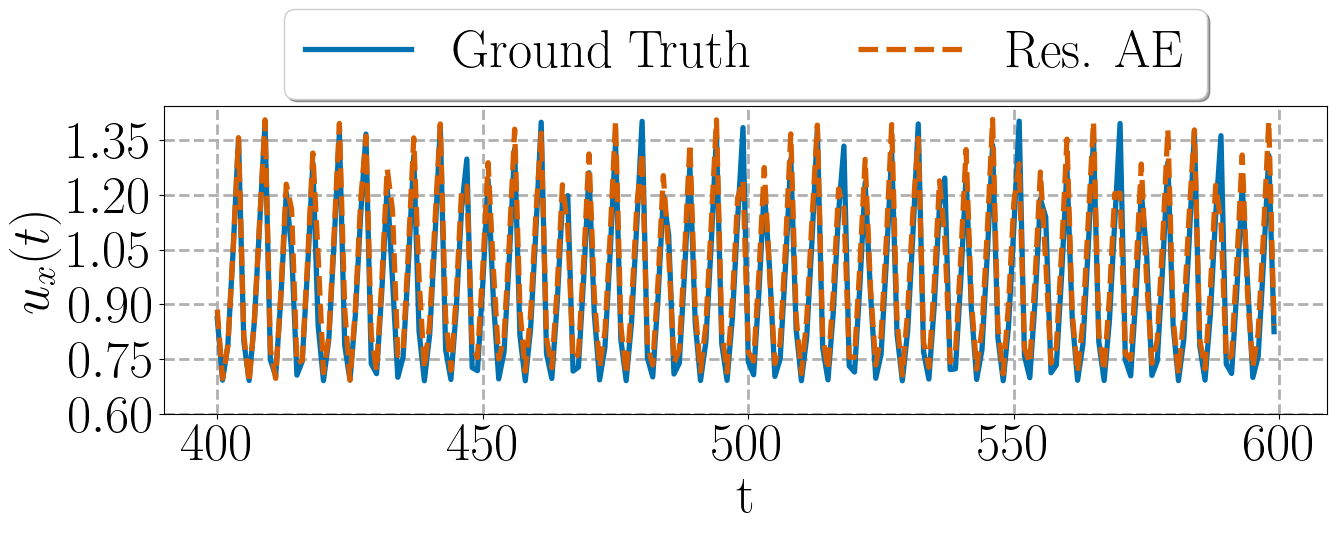}
         \caption{}
     \end{subfigure}
     \hfill
     \begin{subfigure}[b]{0.33\textwidth}
         \centering
         \includegraphics[width=\textwidth]{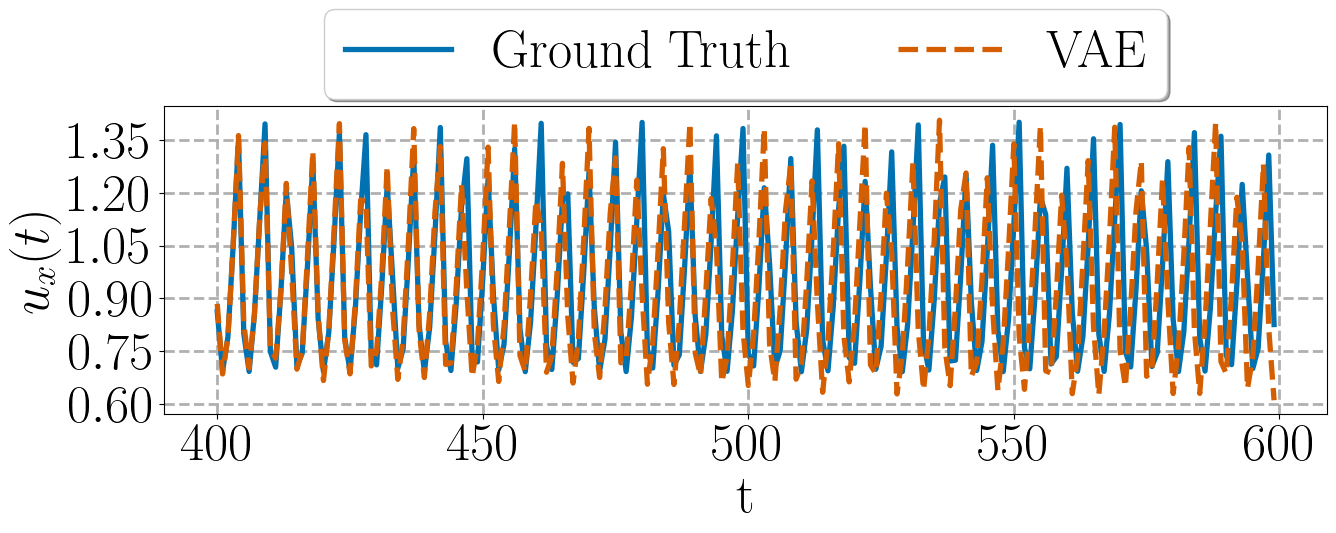}
         \caption{}
     \end{subfigure}
     \hfill
     \begin{subfigure}[b]{0.33\textwidth}
         \centering
         \includegraphics[width=\textwidth]{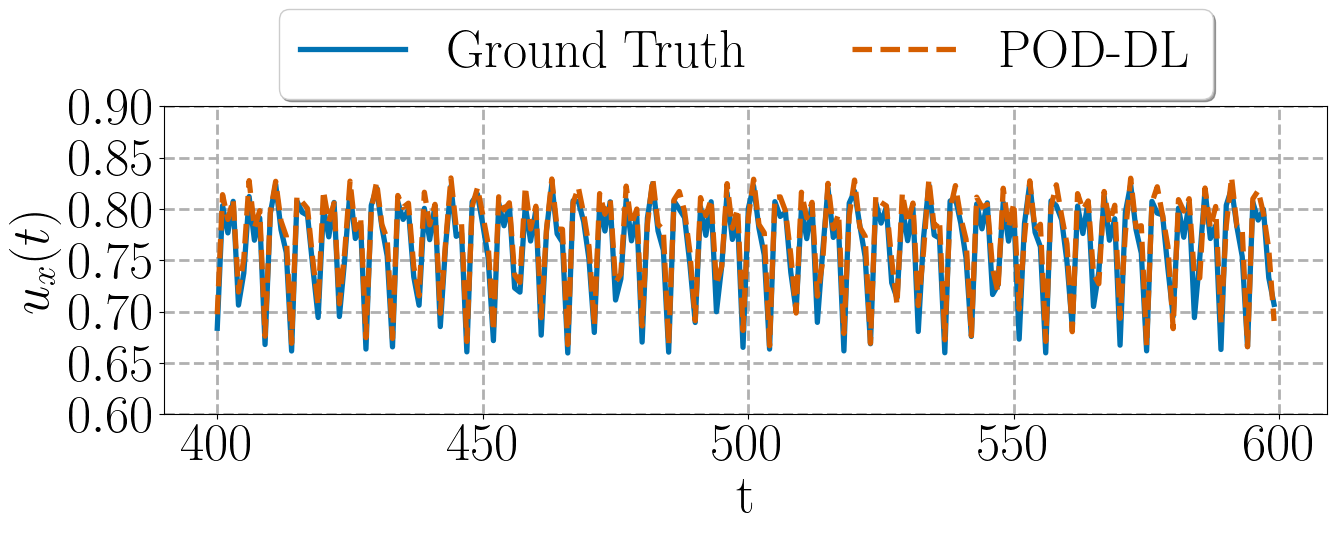}
         \caption{}
     \end{subfigure}
     \hfill
     \begin{subfigure}[b]{0.33\textwidth}
         \centering
         \includegraphics[width=\textwidth]{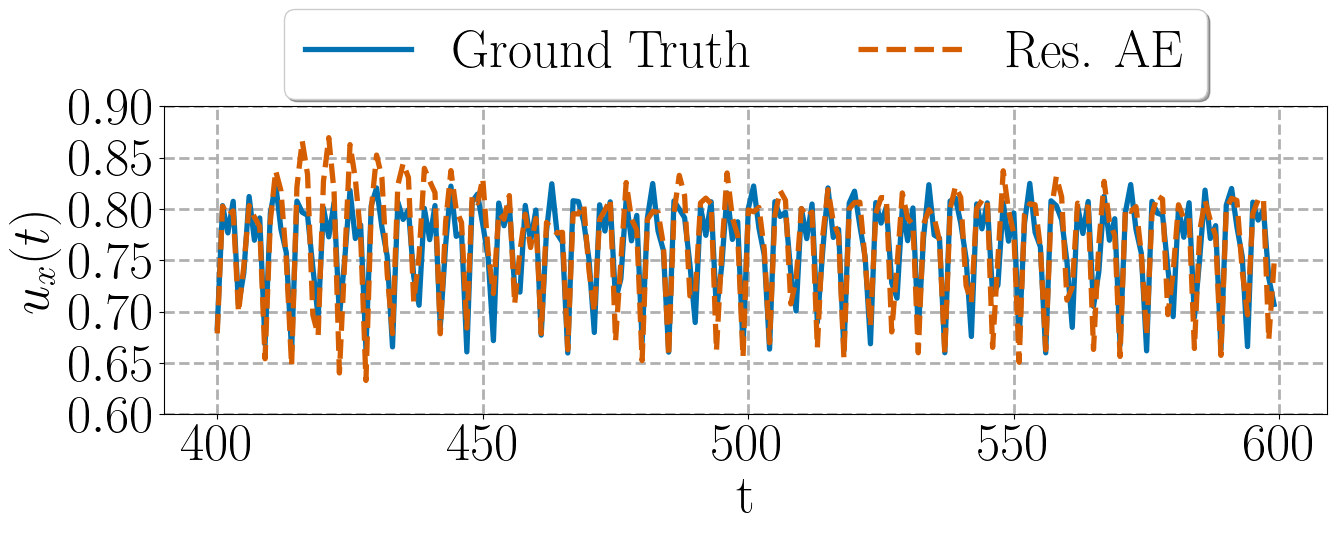}
         \caption{}
     \end{subfigure}
     \hfill
     \begin{subfigure}[b]{0.33\textwidth}
         \centering
         \includegraphics[width=\textwidth]{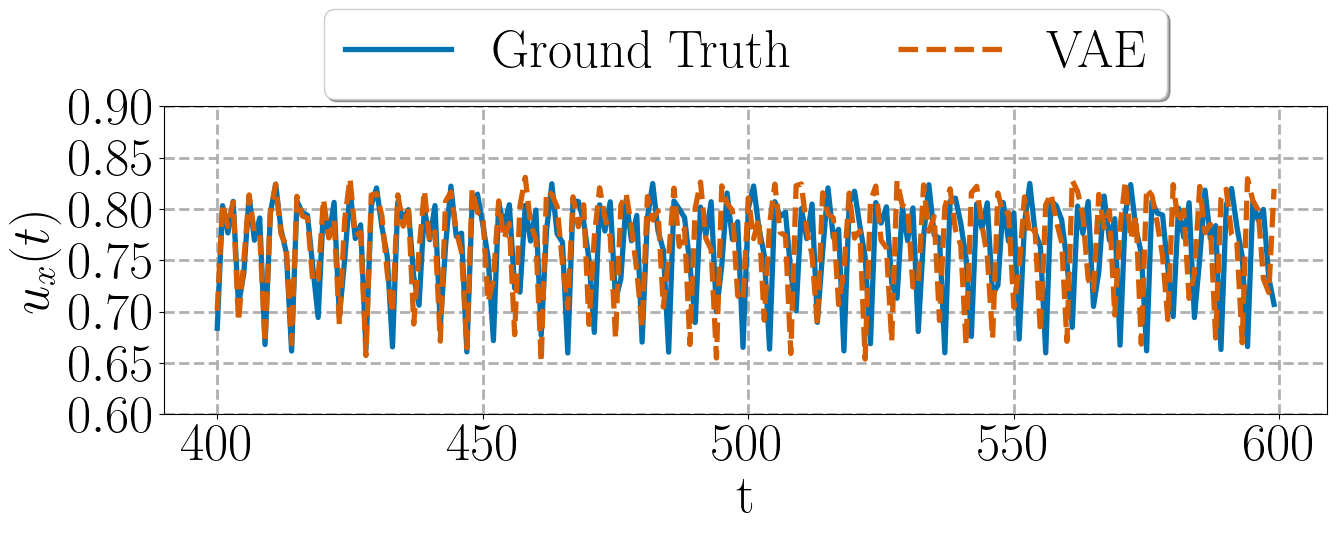}
         \caption{}
     \end{subfigure}
        \caption{Comparison of the ground truth streamwise velocity evolution $u_{x}(t)$, at spatial coordinate $P_{1}$ shown in Fig. \ref{fig: synthetic_ref_points}, against the predictions obtained from the (a) POD-DL, (b) residual autoencoder and (c) variational autoencoder models, respectively. Counter part for the spatial coordinate $P_{2}$ in figures (d), (e) and (f). In all figures the ground truth is represented by a solid line and the prediction by a dashed line.}
        \label{fig: synthetic_strmw_velocity_compare}
\end{figure*}

Two metrics are used to quantitatively measure how close are the predictions to the ground truth data: the relative root mean squared error (RRMSE) and the structural similarity index measure (SSIM) \cite{ssim_2004, ssim_2009}. The former is defined as follows,
\begin{equation}
    \hbox{RRMSE} = \sqrt{\frac{\sum_{k = 1}^K || \bv_k - \hat{\bv}_k ||^2}{\sum_{k=1}^K || \bv_k ||^2}}.
    \label{eq: rrmse}
\end{equation}
Where $\bv_{k}$ is the $k$-th ground truth snapshot and $\hat{\bv}_{k}$ is the $k$-th prediction obtained from the forecasting model.
\begin{figure}
     \centering
     \begin{subfigure}[b]{0.4\textwidth}
         \centering
         \includegraphics[width=\textwidth]{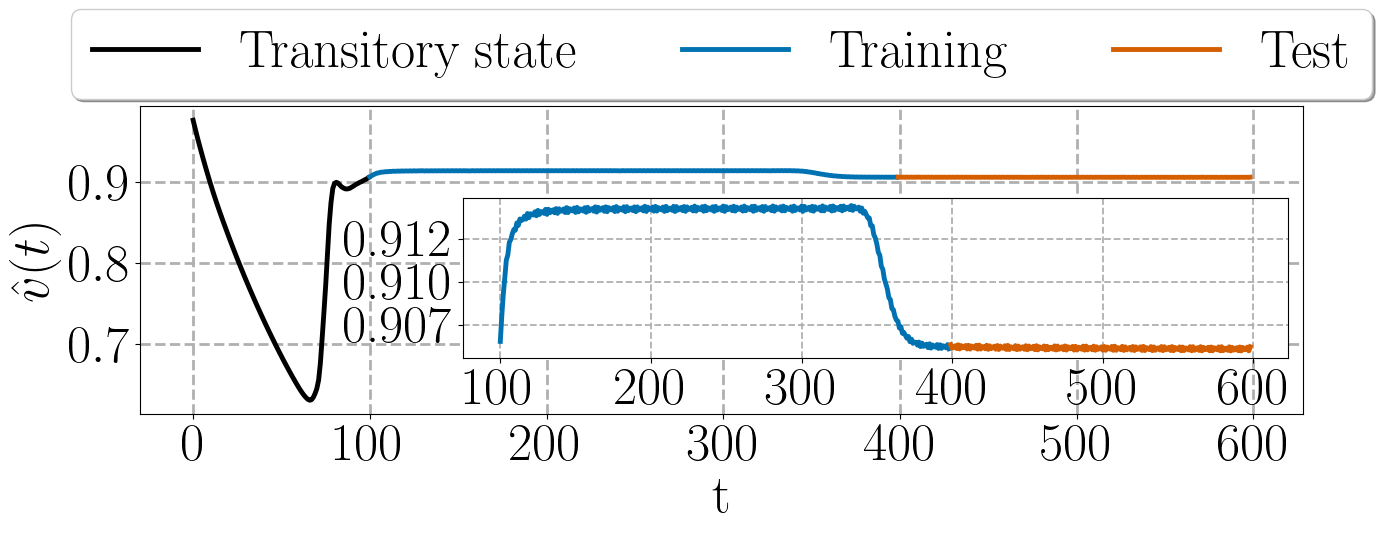}
         \caption{}
     \end{subfigure}
     \hfill
     \begin{subfigure}[b]{0.4\textwidth}
         \centering
         \includegraphics[width=\textwidth]{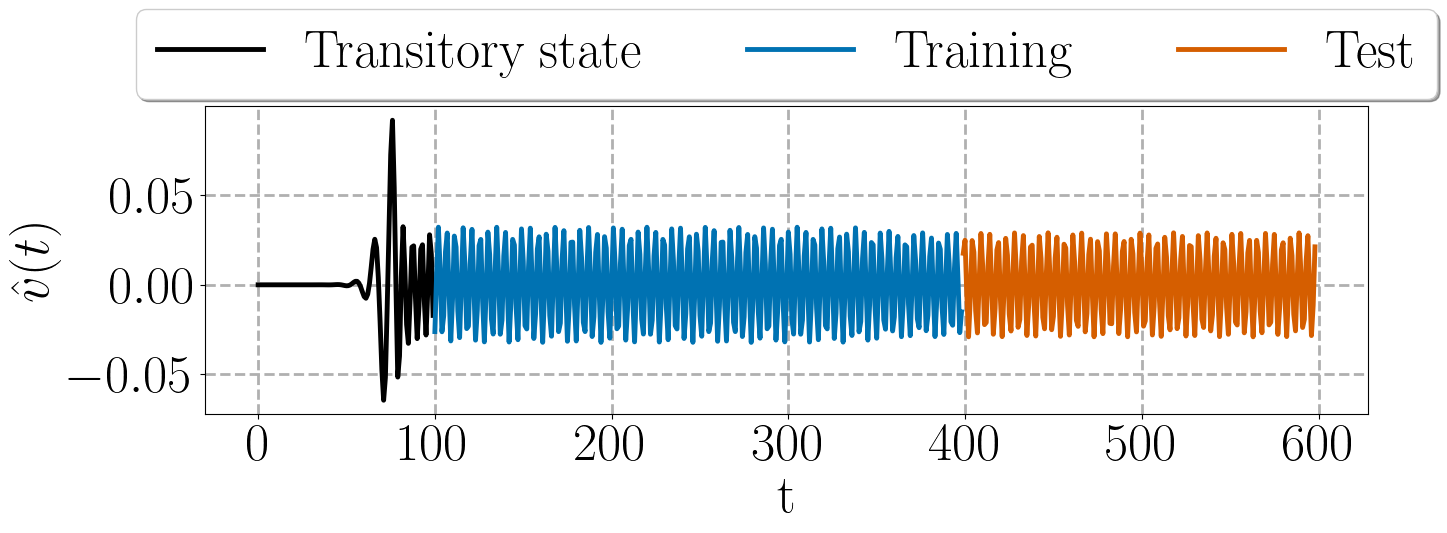}
         \caption{}
     \end{subfigure}
        \caption{(a) Streamwise velocity $\hat{u}(t)$ and (b) wall-normal velocity $\hat{v}(t)$ evolution in the laminar flow. Black, blue and orange lines represent the transitory state, of the numerical simulation, and the data used for training and testing the models, respectively.}
        \label{fig: synthetic_train_test_split}
\end{figure}
As both generated predictions and ground truth data are snapshots, the SSIM is used to measure the degree of similarity between them based on their structural information, as follows,
\begin{equation}
    \hbox{SSIM}(\bv_{k}, \hat{\bv}_{k}) = \frac{(2 \mu_{\bv_{k}} \mu_{\hat{\bv}_{k}} + c_{1})(2\sigma_{\bv_{k}\hat{\bv}_{k}} + c_{2})}{(\mu_{\bv_{k}}^{2} + \mu_{\hat{\bv}_{k}}^{2} + c_{1})(\sigma_{\bv_{k}}^{2} + \sigma_{\hat{\bv}_{k}}^{2} + c_{2})},
    \label{eq: ssim}
\end{equation}
where $\mu_{\bv_{k}}$ and $\mu_{\hat{\bv}_{k}}$ represents the mean of the ground truth and predicted snapshots, respectively. Consequently, $\sigma_{\bv_{k}}$ and $\sigma_{\hat{\bv}_{k}}$ denotes the standard deviation of the ground truth and predicted snapshots. Moreover, $\sigma_{\bv_{k}\hat{\bv}_{k}}$ is the covariance between both snapshots, while $c_{1}$ and $c_2$ are fixed constants to avoid instability.

The RRMSE is an unbounded metric, whose values are non-negative. These indicate better performance (less error) when they are closer to $0$. However, the SSIM is bounded in the interval $[0, 1]$, where $1$ indicates the best performance (high similarity) and $0$ indicates the worst performance (low similarity).
\begin{figure*}
     \centering
     \begin{subfigure}[b]{0.325\textwidth}
         \centering
         \includegraphics[width=\textwidth]{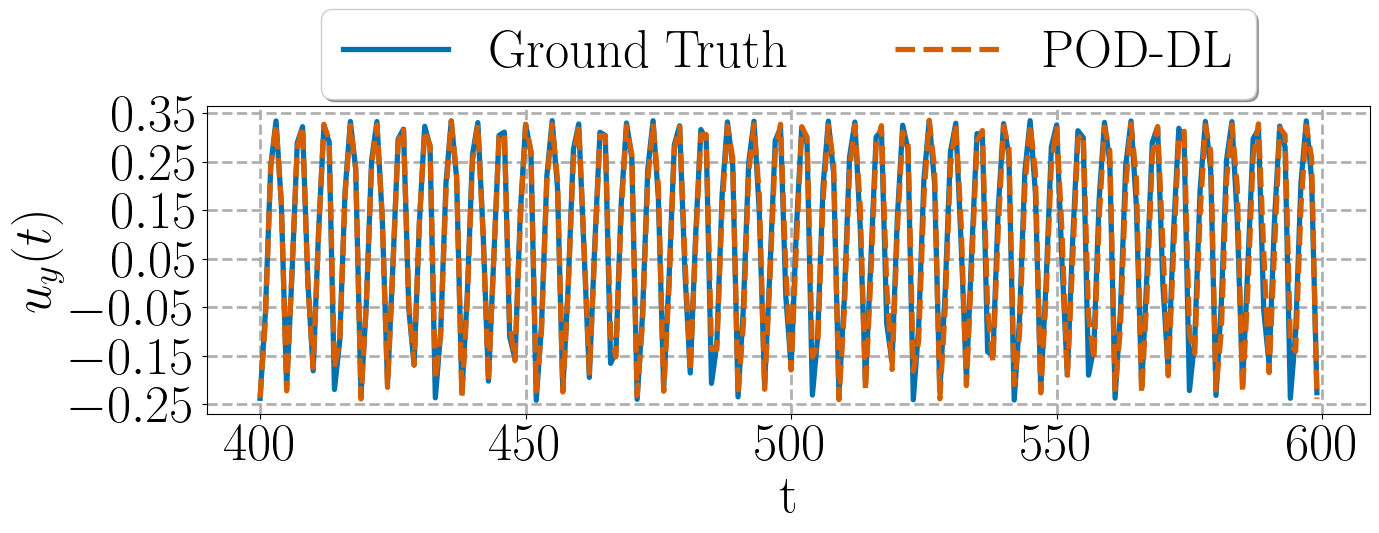}
         \caption{}
     \end{subfigure}
     \hfill
     \begin{subfigure}[b]{0.325\textwidth}
         \centering
         \includegraphics[width=\textwidth]{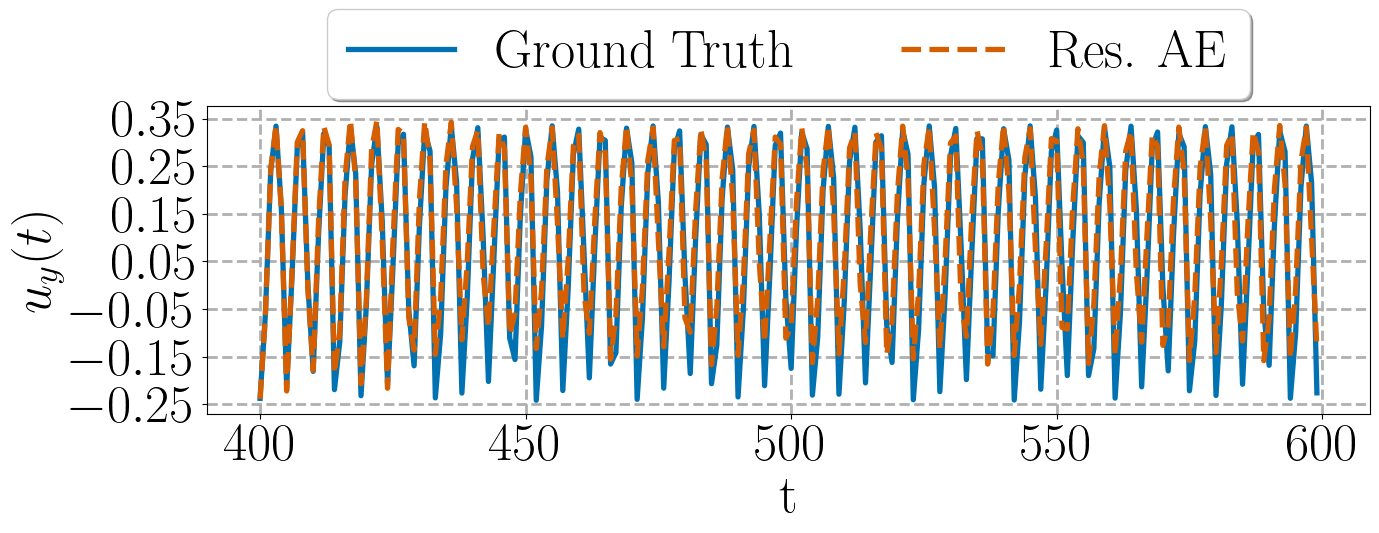}
         \caption{}
     \end{subfigure}
     \hfill
     \begin{subfigure}[b]{0.325\textwidth}
         \centering
         \includegraphics[width=\textwidth]{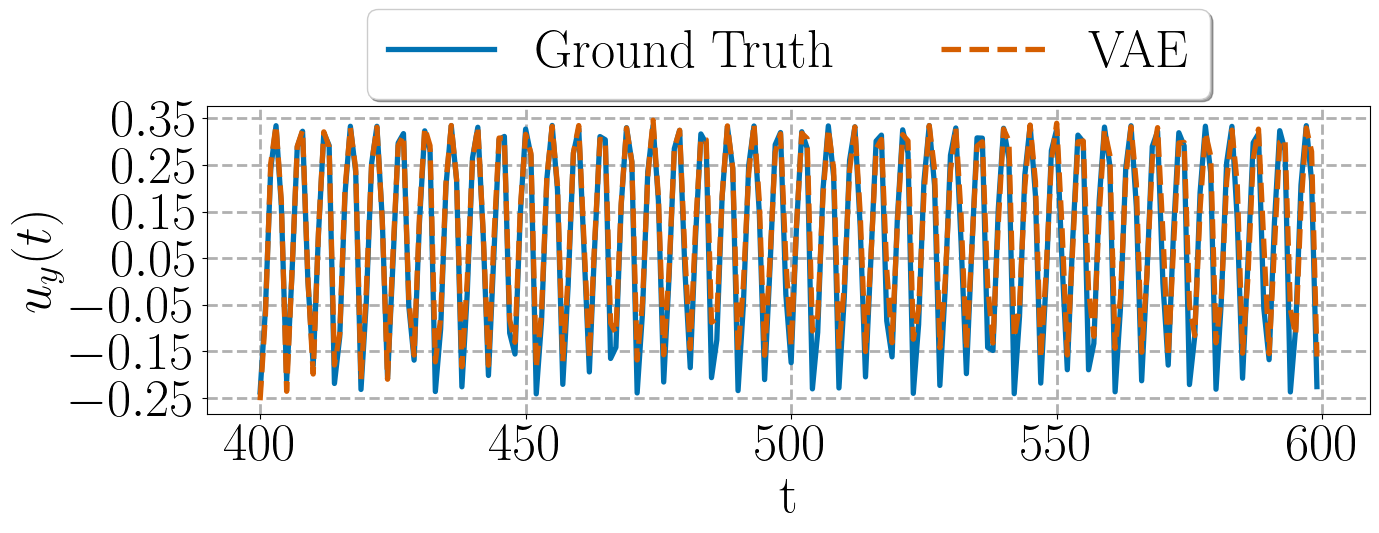}
         \caption{}
     \end{subfigure}
     \hfill
     \begin{subfigure}[b]{0.325\textwidth}
         \centering
         \includegraphics[width=\textwidth]{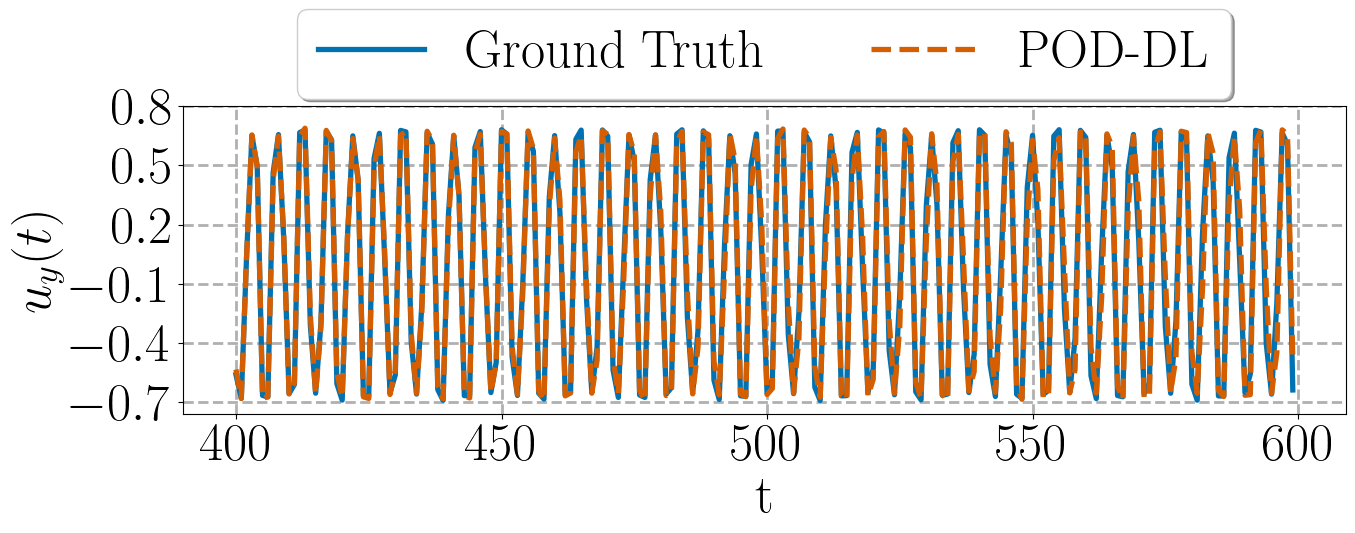}
         \caption{}
     \end{subfigure}
     \hfill
     \begin{subfigure}[b]{0.325\textwidth}
         \centering
         \includegraphics[width=\textwidth]{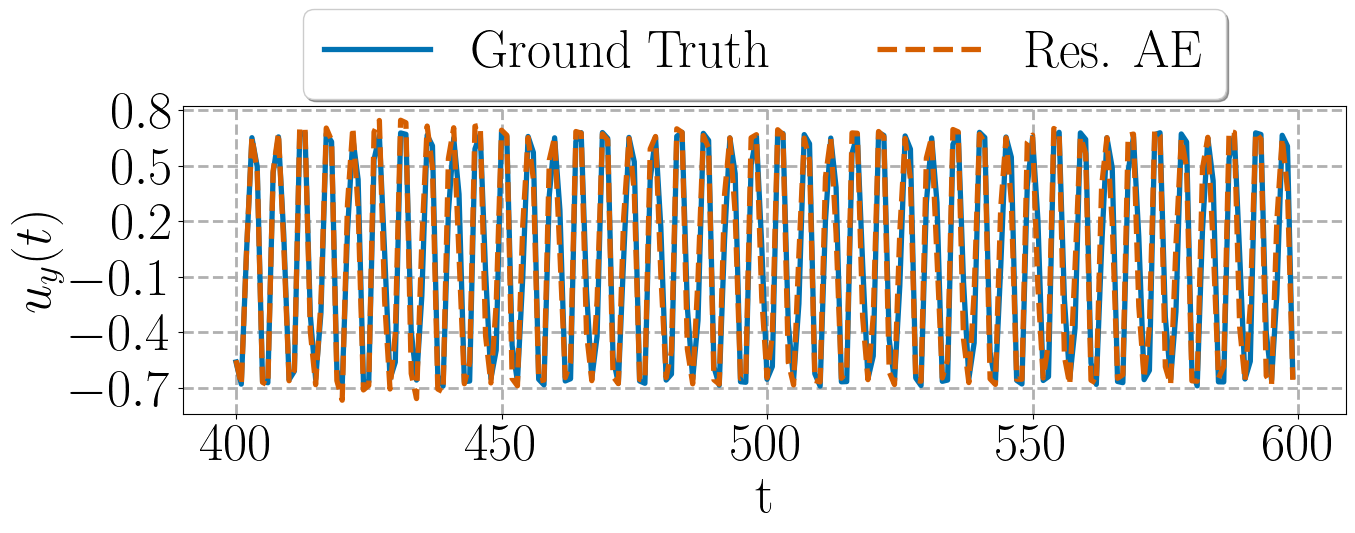}
         \caption{}
     \end{subfigure}
     \hfill
     \begin{subfigure}[b]{0.325\textwidth}
         \centering
         \includegraphics[width=\textwidth]{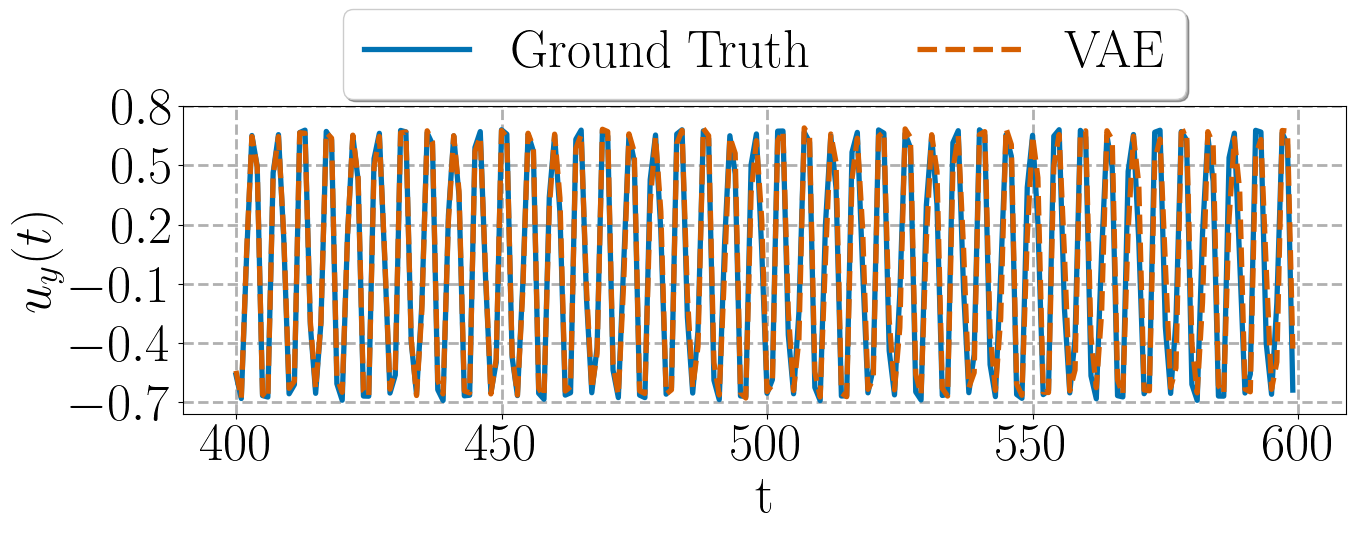}
         \caption{}
     \end{subfigure}
        \caption{Same as Figure \ref{fig: synthetic_strmw_velocity_compare} for the wall-normal velocity $u_{y}(t)$.}
        \label{fig: synthetic_normal_velocity_compare}
\end{figure*}
\subsection{Three-dimensional wake behind a circular cylinder: laminar flow} \label{sec: results_synthetic}
This dataset represents a three-dimensional wake of a circular cylinder, from a viscous, incompressible, and Newtonian flow, which is described by the Navier–Stokes equations, written in non-conservative form, as follows,
\begin{align}
    \nabla \cdot \bv & = 0, \\
    \frac{\partial \bv}{\partial t} + (\bv \cdot \nabla)\bv & = -\nabla p + \frac{1}{\hbox{Re}}\Delta \bv.
\end{align}
Where $p$ and $\bv$ are the pressure and velocity, respectively, and Re is the Reynolds number, defined as $\hbox{Re} = VD/\nu$. In this case, $V$ is the free stream velocity, $D$ the diameter of the circular cylinder and $\nu$ the kinematic viscosity of the fluid . In this case, Re = $280$. The database was generated and validated in Ref. \cite{LeClainche_2018_cilind}.

The first $100$ snapshots of the simulation correspond to the transitory state of the flow, as is shown in Fig. \ref{fig: synthetic_train_test_split}. It is  known that in this regime the flow statistics are constantly varying, thus these first snapshots are not considered for training and testing the models. Figure \ref{fig: synthetic_train_test_split} shows how snapshots from $100$ to $399$ are used to train the models, while the rest ($400$ to $599$) are used to testing. Note how these sets are disjoint allowing to test the models on unseen data during training, in this way the models are evaluated on their extrapolation ability to predict long-term temporal dynamics accurately.

Since this dataset is three-dimensional in space, the number of training samples can be increased using the data augmentation technique described in Appendix \ref{sec: datasets_and_preprocessing}, which allows us to go from $299$ samples to $299 * J_{4} = 299 * 64 = 19136$ samples, where 64 represents the number of discretization points along the $z$-axis. Note, this data augmentation is only used for the residual and variational autoencoders. In the POD-DL model, the POD is a non-parametric technique that already projects the high-dimensional snapshots into a low-dimensional space. 

\begin{figure}
     \centering
     \begin{subfigure}[b]{0.22\textwidth}
         \centering
         \includegraphics[width=\textwidth]{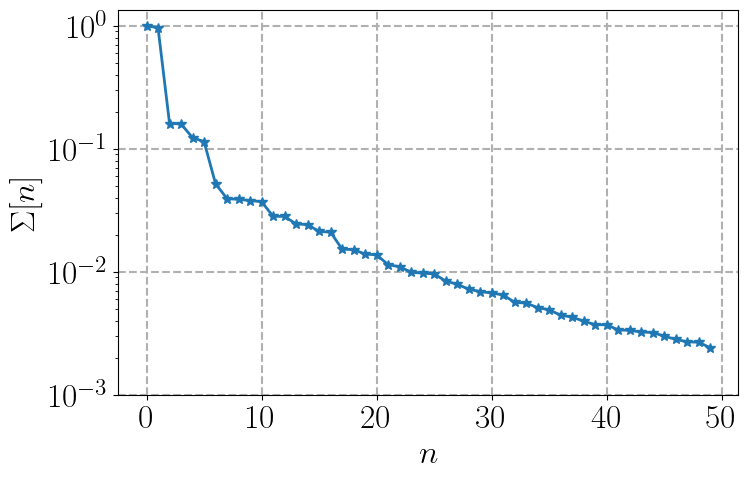}
         \caption{}
     \end{subfigure}
     \hfill
     \begin{subfigure}[b]{0.22\textwidth}
         \centering
         \includegraphics[width=\textwidth]{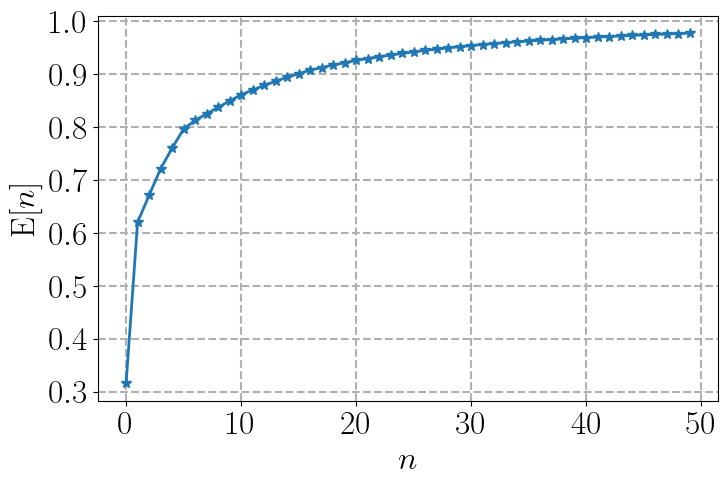}
         \caption{}
     \end{subfigure}
        \caption{(a) Mode decay and (b) cummulative energy for the $50$ most energetic modes in the laminar flow.}
        \label{fig: synth_singular_values}
\end{figure}

\begin{figure}
     \centering
     \begin{subfigure}[b]{0.236\textwidth}
         \centering
         \includegraphics[width=\textwidth]{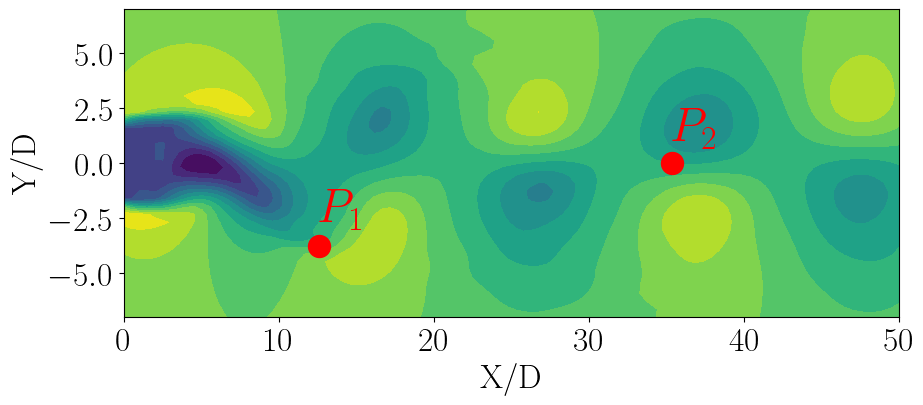}
         \caption{}
     \end{subfigure}
     \hfill
     \begin{subfigure}[b]{0.236\textwidth}
         \centering
         \includegraphics[width=\textwidth]{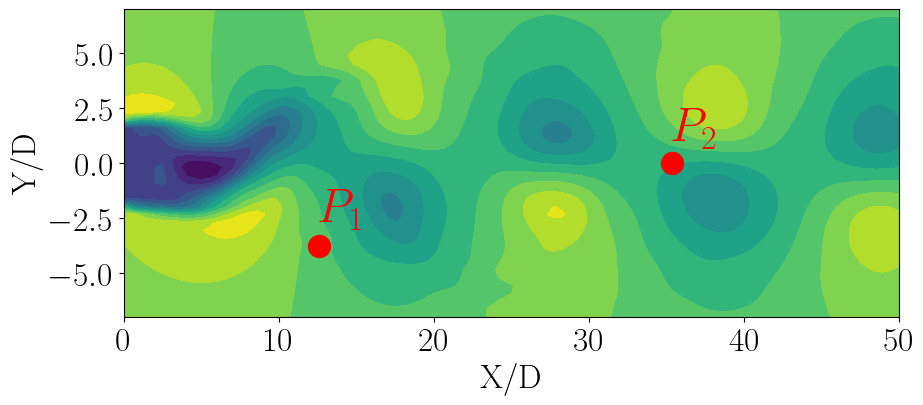}
         \caption{}
     \end{subfigure}
    \caption{Snapshots from the laminar flow streamwise velocity, at time instants (a) $400$ and (b) $440$, representing by red points, $P_{1}$ (left) and $P_{2}$ (right), the spatial coordinates used to compare the streamwise, $u_{x}$, and wall-normal, $u_{y}$, velocity evolution shown in Fig. \ref{fig: synthetic_strmw_velocity_compare} and Fig. \ref{fig: synthetic_normal_velocity_compare}.}
    \label{fig: synthetic_ref_points}
\end{figure}

The decay of the $50$ most energetic modes and cummulative energy $E$ are shown in Fig. \ref{fig: synth_singular_values}. The cummulative energy is defined as follows,
\begin{equation}
    E = \frac{\sum^{\kappa}_{i=1}\lambda_{i}}{\sum^{K}_{i = 1} \lambda_{i}}, \label{eq: cummEnerg}
\end{equation}
where $\lambda_{i}$ is the $i$-th singular value, $K$ is the number of samples and $\kappa$ the number of retained modes. Following \cite{abadiaherediaetal_2022}, the $20$ most energetic modes are retained to perform the forecasting in the laminar flow. Note in Fig. \ref{fig: synth_singular_values} these number of modes contained more than $90\%$ of the overall energy.
Once the models have been trained, they are tasked to iteratively predict $200$ snapshots ahead in time (from $400$ to $599$) and compare the predicted snapshots with the ground truth. This comparison is done at two representative spatial coordinates shown in Fig. \ref{fig: synthetic_ref_points}, where Fig. \ref{fig: synthetic_strmw_velocity_compare} presents the predictions obtained for the streamwise velocity evolution from the POD-DL, residual autoencoder, and variational autoencoder models, respectively. Figure \ref{fig: synthetic_normal_velocity_compare} shows the same comparison for the wall-normal velocity evolution.

\begin{figure}
     \centering
     \begin{subfigure}[b]{0.4\textwidth}
         \centering
         \includegraphics[width=\textwidth]{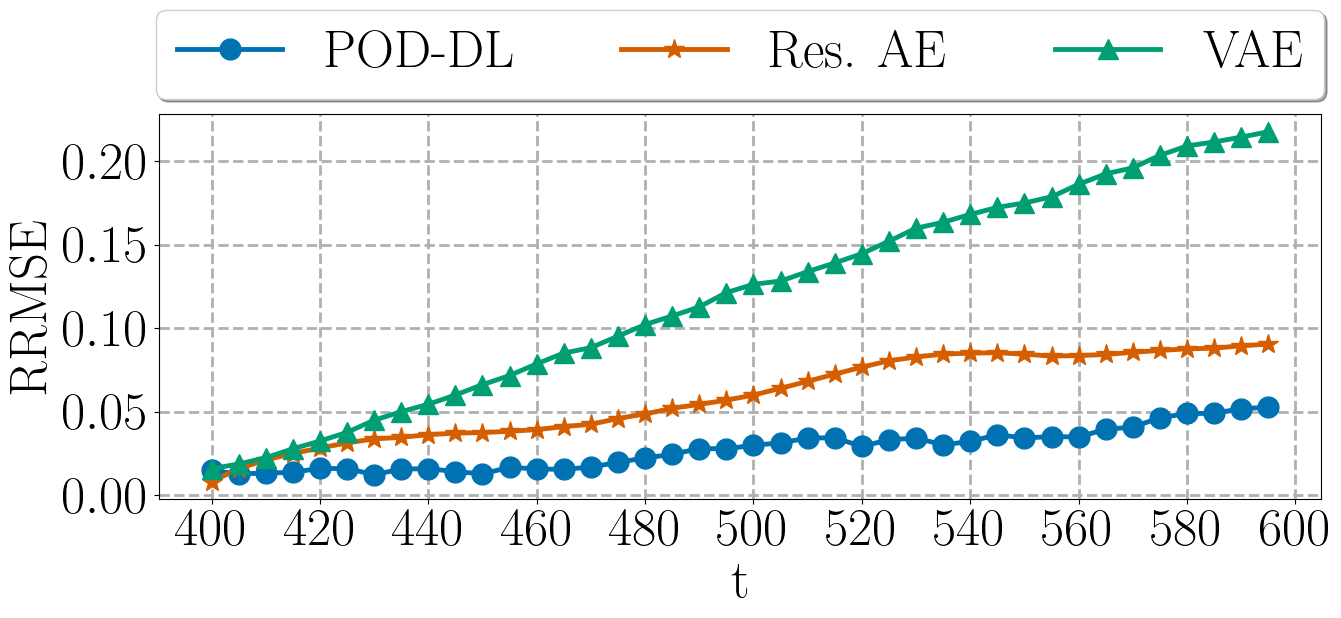}
         \caption{}
     \end{subfigure}
     \hfill
     \begin{subfigure}[b]{0.4\textwidth}
         \centering
         \includegraphics[width=\textwidth]{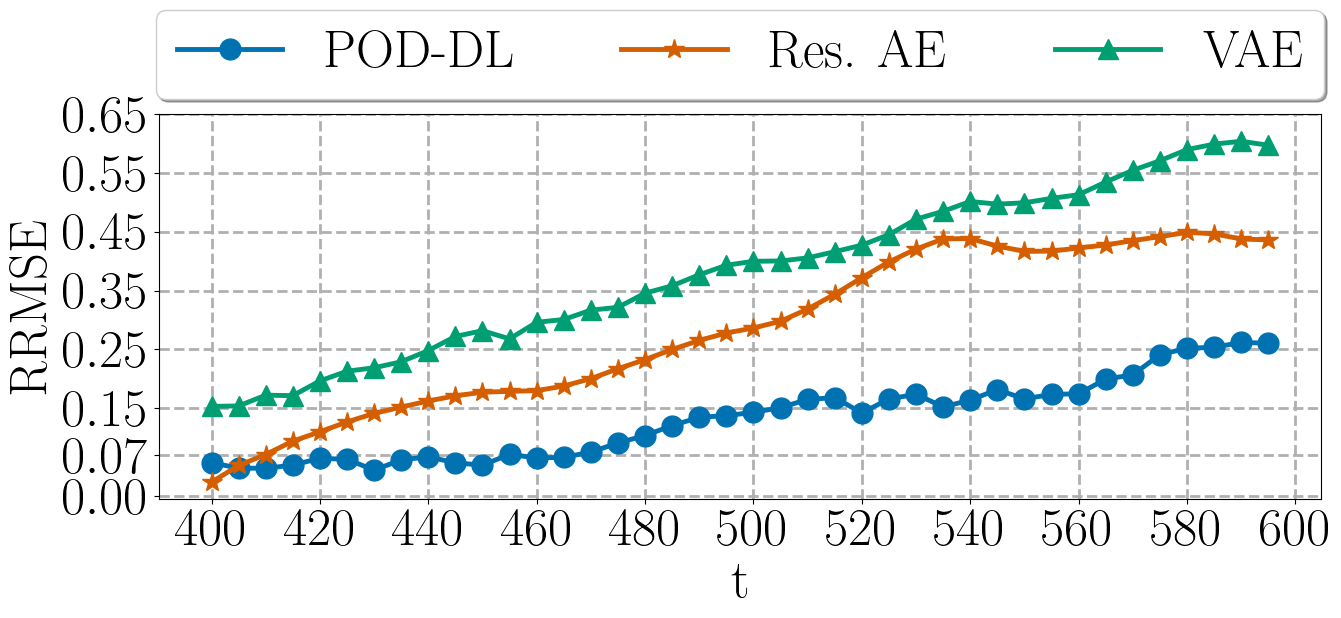}
         \caption{}
     \end{subfigure}
     \hfill
     \begin{subfigure}[b]{0.4\textwidth}
         \centering
         \includegraphics[width=\textwidth]{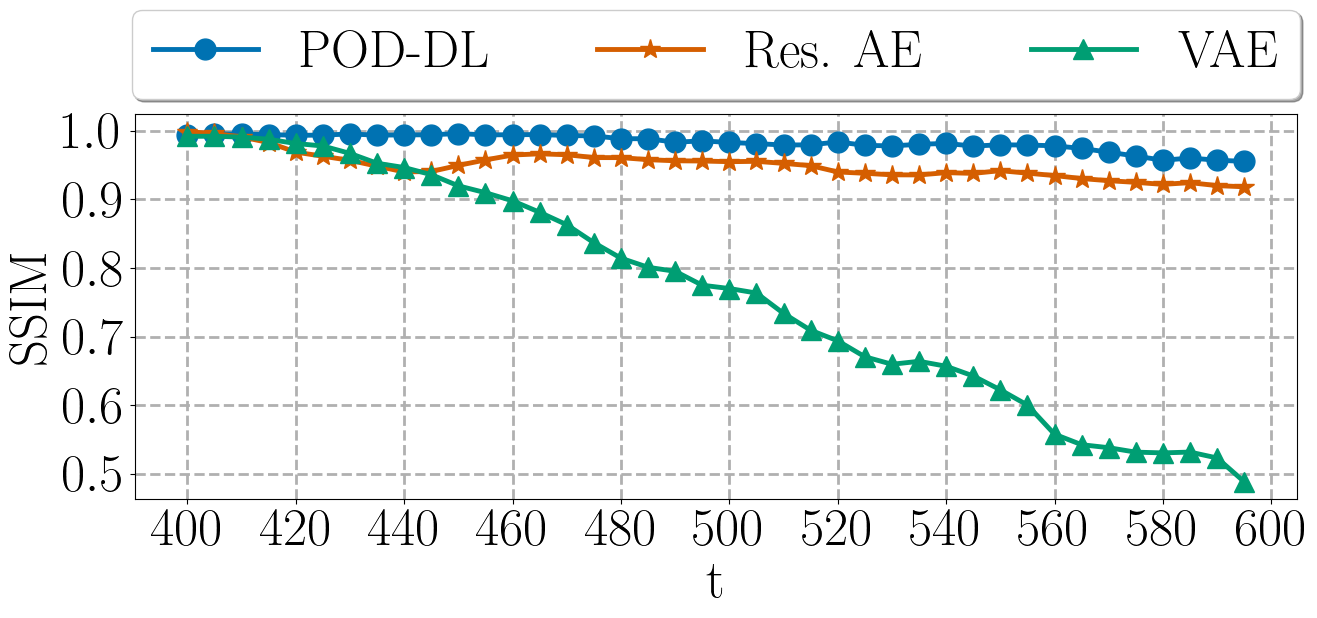}
         \caption{}
     \end{subfigure}
     \hfill
     \begin{subfigure}[b]{0.4\textwidth}
         \centering
         \includegraphics[width=\textwidth]{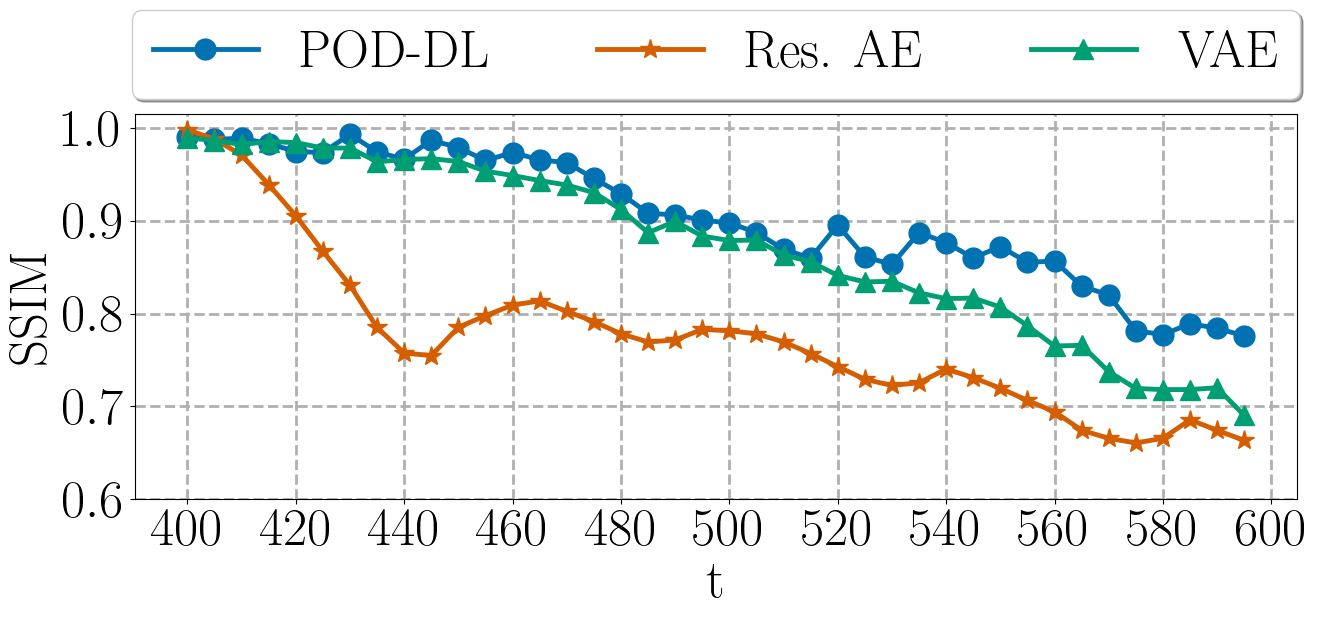}
         \caption{}
     \end{subfigure}
        \caption{Prediction error committed by the models, in the laminar flow, measured by the RRMSE in the (a) streamwise and (b) wall-normal velocity component, respectively, and the SSIM in the (c) streamwise and (d) wall-normal velocity component, respectively.}
        \label{fig: synthetic_rrmse_ssim}
\end{figure}

Note how in the streamwise velocity (Fig. \ref{fig: synthetic_strmw_velocity_compare}) the predictions from the POD-DL are the closest to the ground truth, in comparison to the other two models. A similar scenario can be observed for the evolution of the wall-normal velocity (Fig. \ref{fig: synthetic_normal_velocity_compare}), where predictions from the POD-DL model are also the most accurate, with respect to the reference. Nonetheless, predictions from the residual autoencoder are more accurate than the VAE model.
To evaluate the accuracy of the predictions, both the RRMSE, defined in eq. \ref{eq: rrmse}, and the SSIM, defined in eq. \ref{eq: ssim}, are used. The SSIM measures the similarity between two figures based on their structural information. Figure \ref{fig: synthetic_rrmse_ssim} shows the measures obtained by these two metrics for the laminar flow. Following the results in Fig. \ref{fig: synthetic_strmw_velocity_compare} and Fig. \ref{fig: synthetic_normal_velocity_compare}, the most accurate predictions, according to these metrics, are obtained from the POD-DL model. As the predictions are generated iteratively, it is intuitive to think that accumulated error will make predictions further out in time worse. However, the POD-DL model is found to be more stable over time.

\begin{figure*}
     \centering
     \begin{subfigure}[b]{0.62\textwidth}
         \centering
         \includegraphics[width=\textwidth]{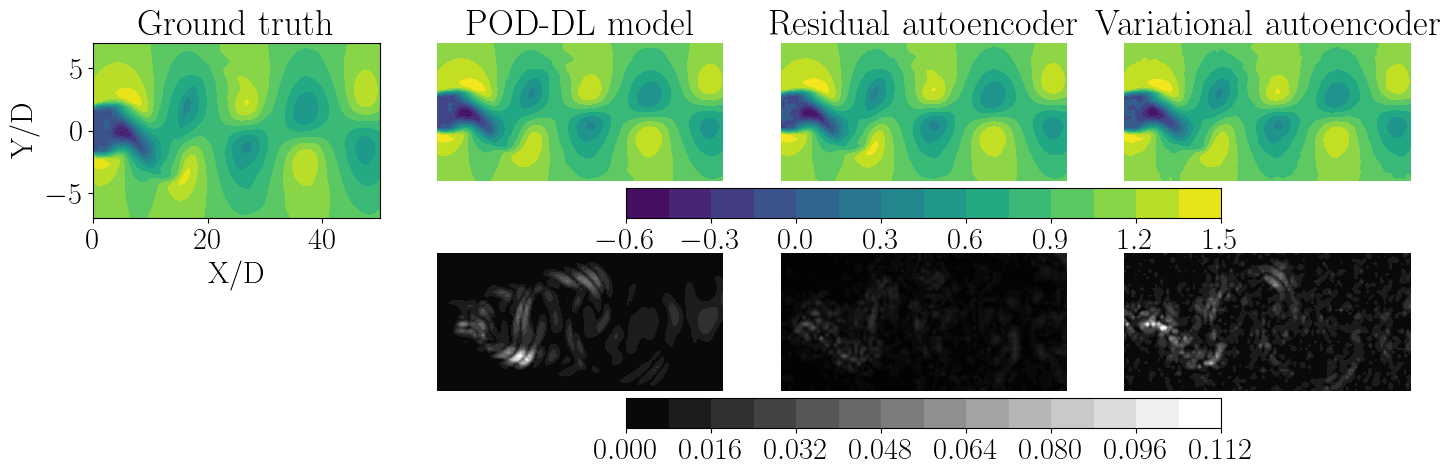}
         \caption{Ground truth snapshot, predictions and absolute error at time instant $400$.}
     \end{subfigure}
     \hfill
     \begin{subfigure}[b]{0.62\textwidth}
         \centering
         \includegraphics[width=\textwidth]{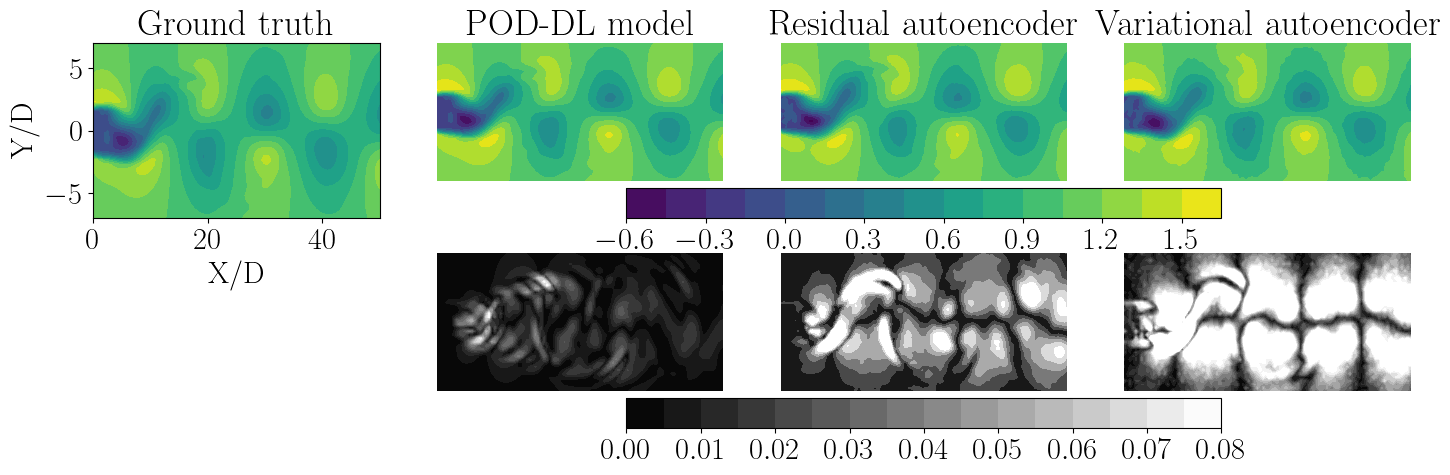}
         \caption{Ground truth snapshot, predictions and absolute error at time instant $460$.}
     \end{subfigure}
     \hfill
     \begin{subfigure}[b]{0.62\textwidth}
         \centering
         \includegraphics[width=\textwidth]{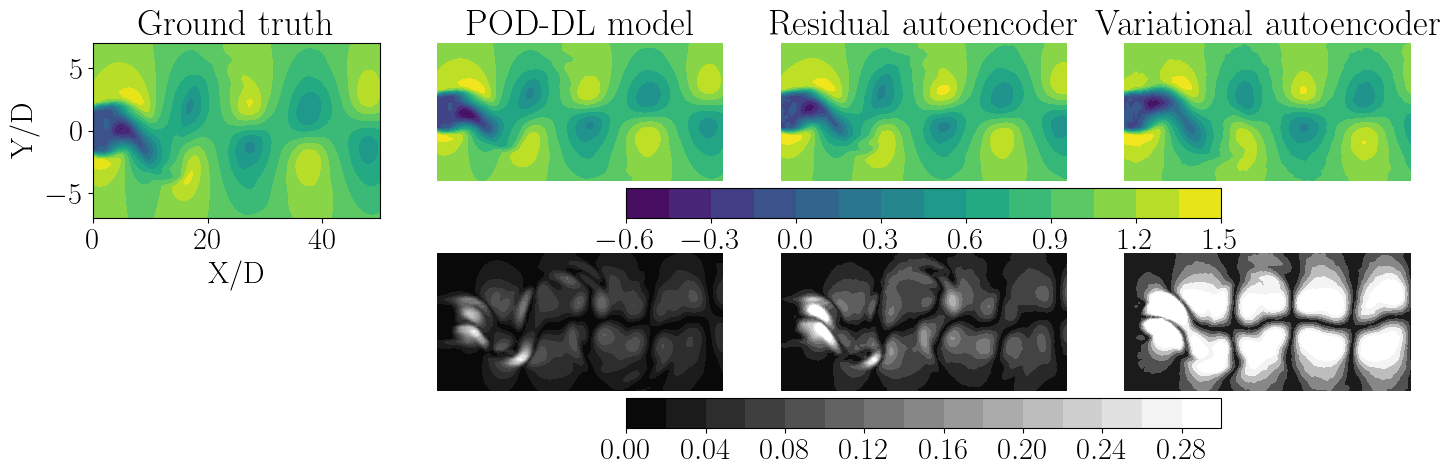}
         \caption{Ground truth snapshot, predictions and absolute error at time instant $599$.}
     \end{subfigure}
        \caption{Snapshots at some representative time instants showing the ground truth streamwise velocity component of the laminar flow, the predictions obtained from the different models and their respective absolute error in gray-scale. The time instants chosen are $400$, $460$ and $599$.}
        \label{fig: synthetic_strw_snapshots_comp}
\end{figure*}

\begin{figure*}
     \centering
     \begin{subfigure}[b]{0.62\textwidth}
         \centering
         \includegraphics[width=\textwidth]{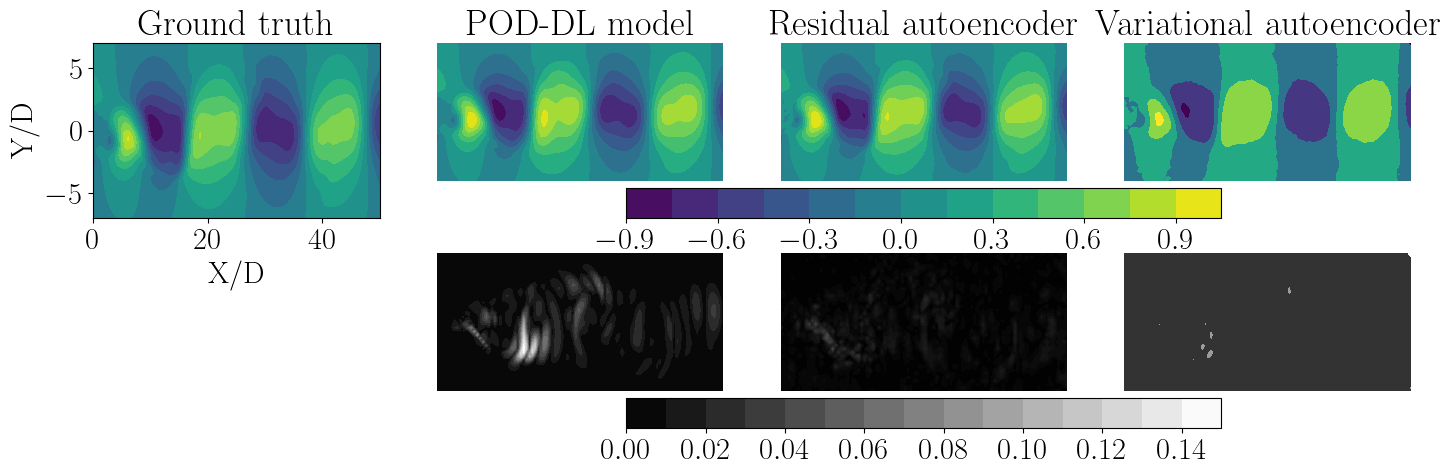}
         \caption{Ground truth snapshot, predictions and absolute error at time instant $400$.}
     \end{subfigure}
     \hfill
     \begin{subfigure}[b]{0.62\textwidth}
         \centering
         \includegraphics[width=\textwidth]{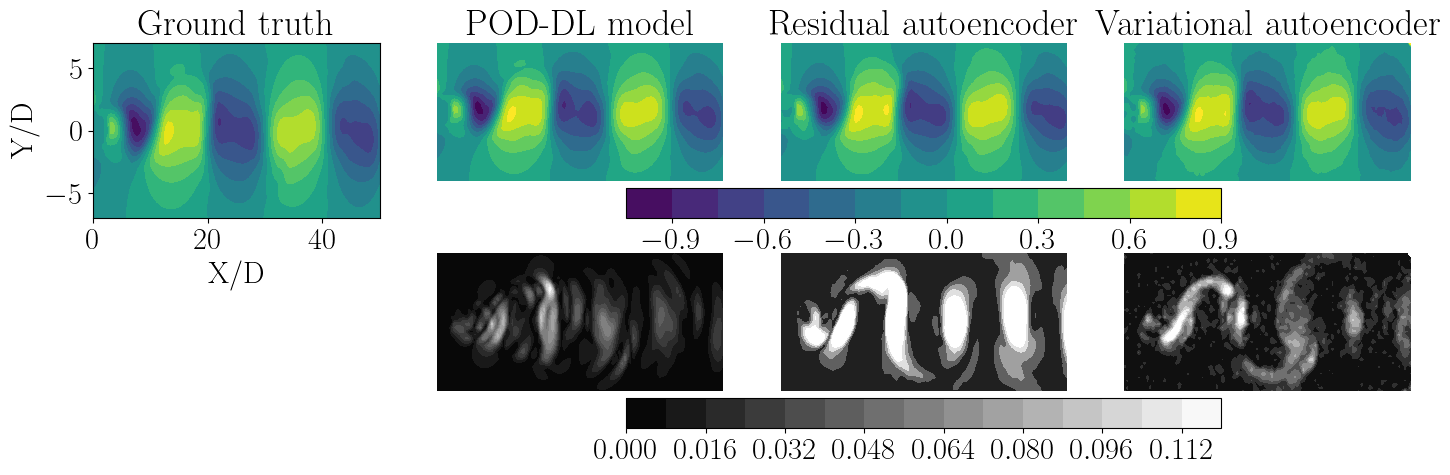}
         \caption{Ground truth snapshot, predictions and absolute error at time instant $460$.}
     \end{subfigure}
     \hfill
     \begin{subfigure}[b]{0.62\textwidth}
         \centering
         \includegraphics[width=\textwidth]{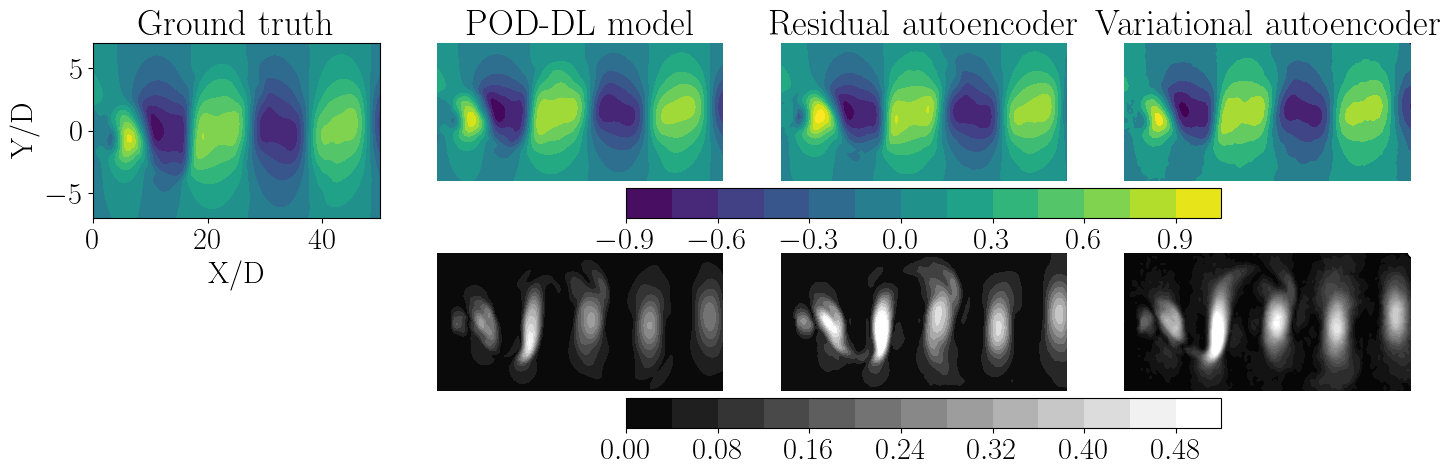}
         \caption{Ground truth snapshot, predictions and absolute error at time instant $599$.}
     \end{subfigure}
        \caption{Snapshots at some representative time instants showing the ground truth wall-normal velocity component of the laminar flow, the predictions obtained from the different models and their respective absolute error in gray-scale. The time instants chosen are $400$, $460$ and $599$.}
        \label{fig: synthetic_normal_snapshots_comp}
\end{figure*}

This stability can also be qualitatively observed in Fig. \ref{fig: synthetic_strw_snapshots_comp} and Fig. \ref{fig: synthetic_normal_snapshots_comp}, where representative snapshots are plotted comparing the ground truth and predictions from the three models. In particular, at time step $599$, predictions from the residual autoencoder and variational autoencoder show noticeable shifts, whereas predictions from the POD-DL model, although still shifted, are less so.

From these observations, we can infer that the model best capturing the flow dynamics is the POD-DL, followed by the residual autoencoder, with the VAE providing the least accurate predictions. This difference in performance maybe due to the complexity of the ELBO loss function \eqref{eq: elbo_new}, which aims to approximate a probability distribution. This loss function is more complex than the Mean Squared Error (MSE) used to train both the POD-DL and residual autoencoder models. Consequently, the VAE may require more training data to achieve comparable performance to the other two models.

\subsection{Three-dimensional wake behind a circular cylinder: turbulent flow} \label{sec: results_experimental}

This section shows and discusses the results obtained from the three models when they are applied to the turbulent flow, which as well represents a flow past a circular cylinder, with the difference that this dataset is two-dimensional in space and the data was obtained by experimental measurements \cite{Mendez_2020_experimental}.

This dataset is derived from an experimental analysis of a three-dimensional flow past a circular cylinder with a diameter of $D = 5$. However, it consists of a set of two-dimensional velocity fields. These experiments were carried out in the L10 low-speed wind tunnel of the von Karman Institute, using time-resolved particle image velocimetry (TR-PIV). The full dataset covers about $4.5$ time units, where the velocity of the free stream $U$ evolves through two steady state conditions, and the transition between them covers approximately $1$ time unit. The full dataset is formed by $13200$ snapshots, where the first $4000$ belong to the first steady state at $\hbox{Re} = 4000$, the next $4000$ belongs to the transitory state and the last $5200$ belong to the second steady state at $\hbox{Re} = 2600$ \cite{Mendez_2020_experimental}. 

\begin{figure}
    \centering
    \includegraphics[width=0.4\textwidth]{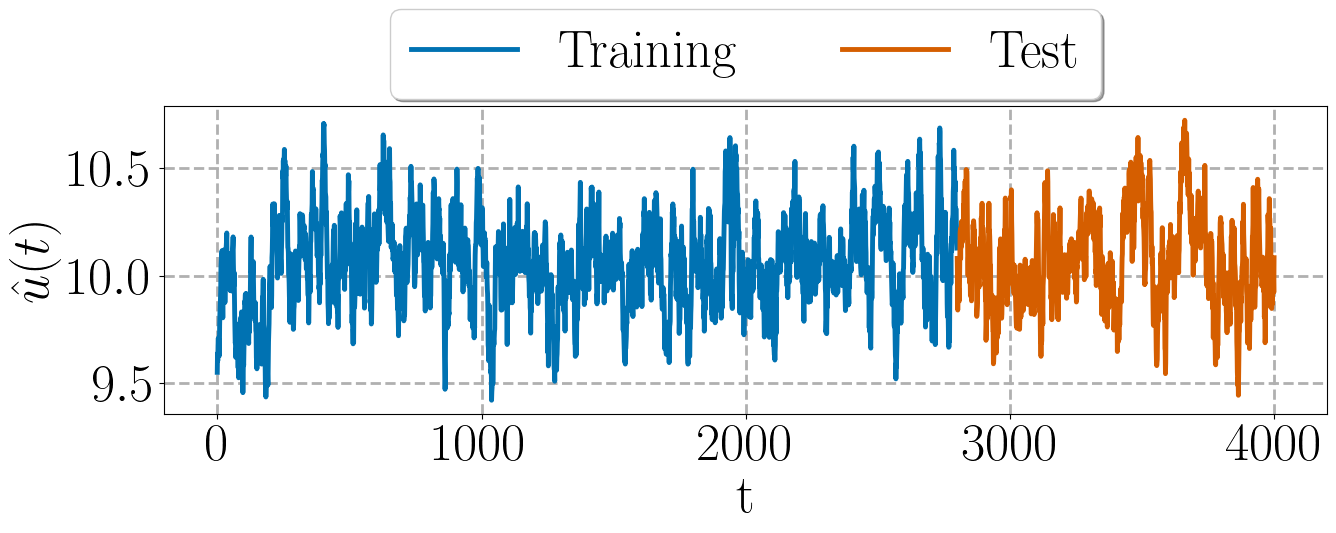}
    \caption{Streamwise velocity $\hat{u}(t)$ evolution in the turbulent flow. Blue and orange lines represent the data used for training and testing the models, respectively.}
    \label{fig: experiment_train_test_split}
\end{figure}

The data used to train and test the models in this work correspond to the first $4000$ representing the steady phase at $\hbox{Re} = 4000$. Where, the first $2800$ snapshots being allocated for training and the remaining $1200$ snapshots reserved for testing. Figure \ref{fig: experiment_train_test_split} illustrates the streamwise velocity evolution for each sample and visually delineates the training and testing data partition.

Examining Fig. \ref{fig: experiment_train_test_split}, it is evident that the flow exhibits significantly greater complexity compared to the laminar flow shown in Fig. \ref{fig: synthetic_train_test_split}. This complexity is attributable to the high Reynolds numbers at which the flow was investigated.

\begin{figure}
     \centering
     \begin{subfigure}[b]{0.22\textwidth}
         \centering
         \includegraphics[width=\textwidth]{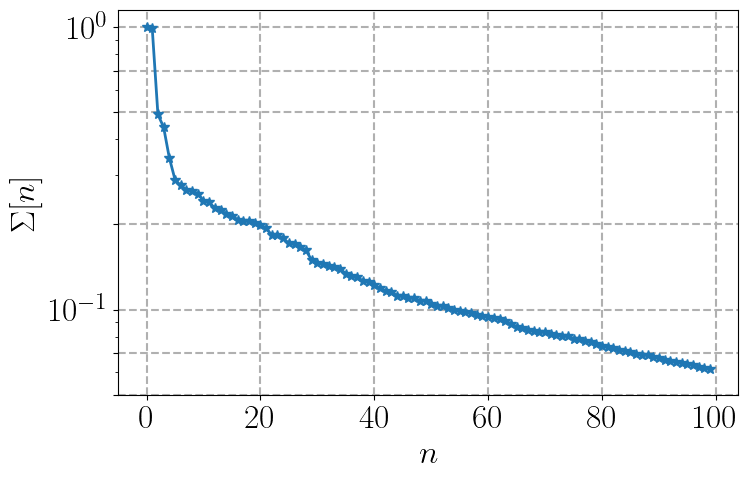}
         \caption{}
     \end{subfigure}
     \hfill
     \begin{subfigure}[b]{0.22\textwidth}
         \centering
         \includegraphics[width=\textwidth]{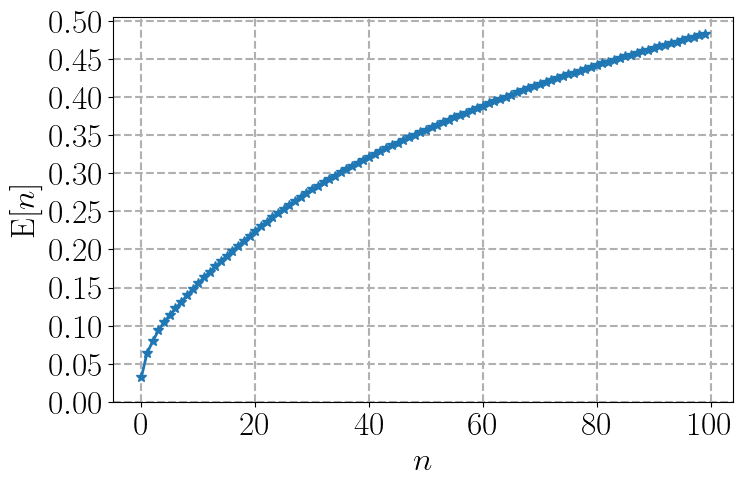}
         \caption{}
     \end{subfigure}
        \caption{(a) Mode decay and (b) cummulative energy for the $100$ most energetic modes in the turbulent flow.}
        \label{fig: experm_singular_values}
\end{figure}

Through extensive trials, it was determined that the optimal approach to deal with this dataset is by simplifying the dynamics via singular value decomposition (SVD). Specifically, we reconstruct the flow by considering only the most energetic modes that encapsulate the principal dynamics. Figure \ref{fig: experm_singular_values} shows the decay of the $100$ most energetic modes and cummulative energy $E$. The cummulative energy is defined as in \ref{eq: cummEnerg}.

The optimal training performance for both the residual and variational autoencoders was achieved when retaining the $20$ most energetic modes. Table \ref{tab: experiment_orig_svd} illustrates the visual differences between the original flow data, derived from experimental measurements, and the flow reconstruction obtained using singular value decomposition (SVD) with these 20 dominant modes. As shown in Fig. \ref{fig: experm_singular_values}, this set of modes captures approximately $23\%$ of the total energy.

The simplified dataset, retaining only the 20 most energetic modes, is utilized to train both the residual and variational autoencoders. In contrast, the POD-DL model, which inherently applies SVD, is trained on the original dataset. However, for consistency with the autoencoders, the POD-DL model is also constrained to the $20$ most energetic modes. In all cases, SVD is applied exclusively to the initial $2800$ samples used for training the models. Similar to the laminar flow, the models are tested on unseen data during training, i.e., in this case the models are tasked to predict from snapshot $2800$.

The predictions from each model are then compared against ground truth data from experimental measurements to assess the ability of these reduced-order models to accurately predict complex flow dynamics.

The rationale behind training the models on a simplified dataset is to enhance the learning efficiency by focusing on the principal dynamics of the flow. This approach is in line with similar methodologies adopted in previous studies, such as the one conducted by Mata {\it et al.} \cite{mata_etal_2023}.

\begin{table}
    \centering
    \caption{Comparison of the original flow from experimental measurements and the reconstruction using SVD, keeping the $20$ most energetic modes to reconstruct the flow field.\label{tab: experiment_orig_svd}}
    \begin{ruledtabular}
    \begin{tabular}{ccccc}
         & \textbf{Original} & \textbf{SVD reconstruction} \\
        \hline
         & \includegraphics[width=4cm]{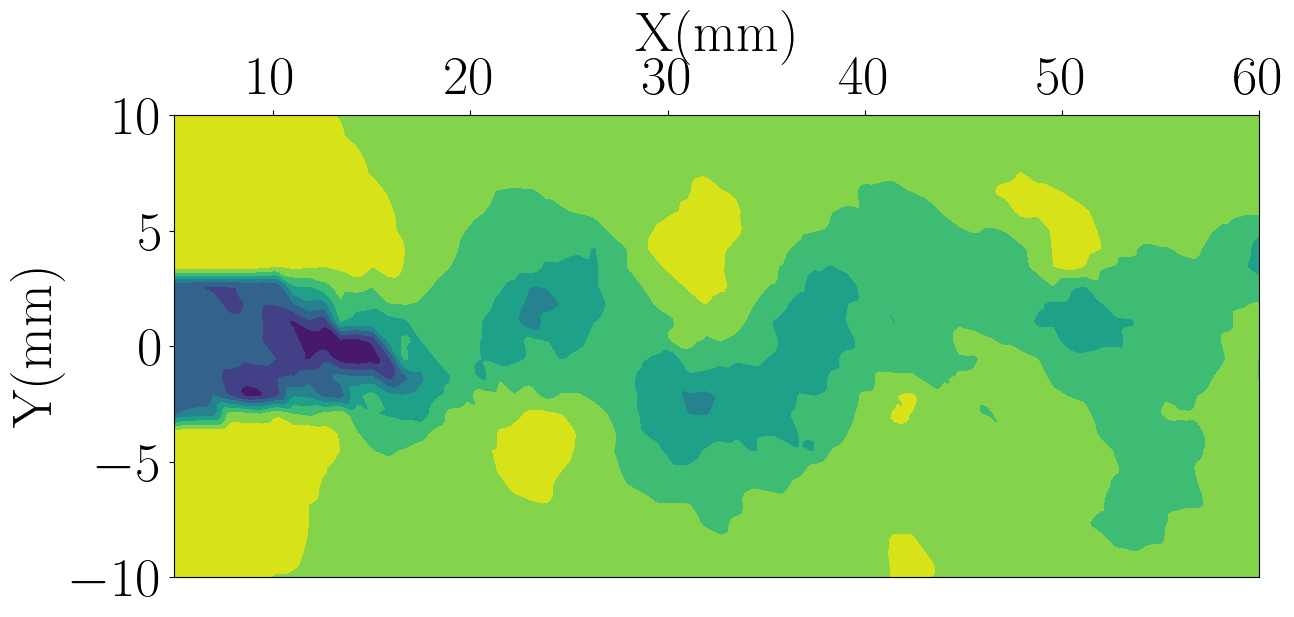} & \includegraphics[width=4cm]{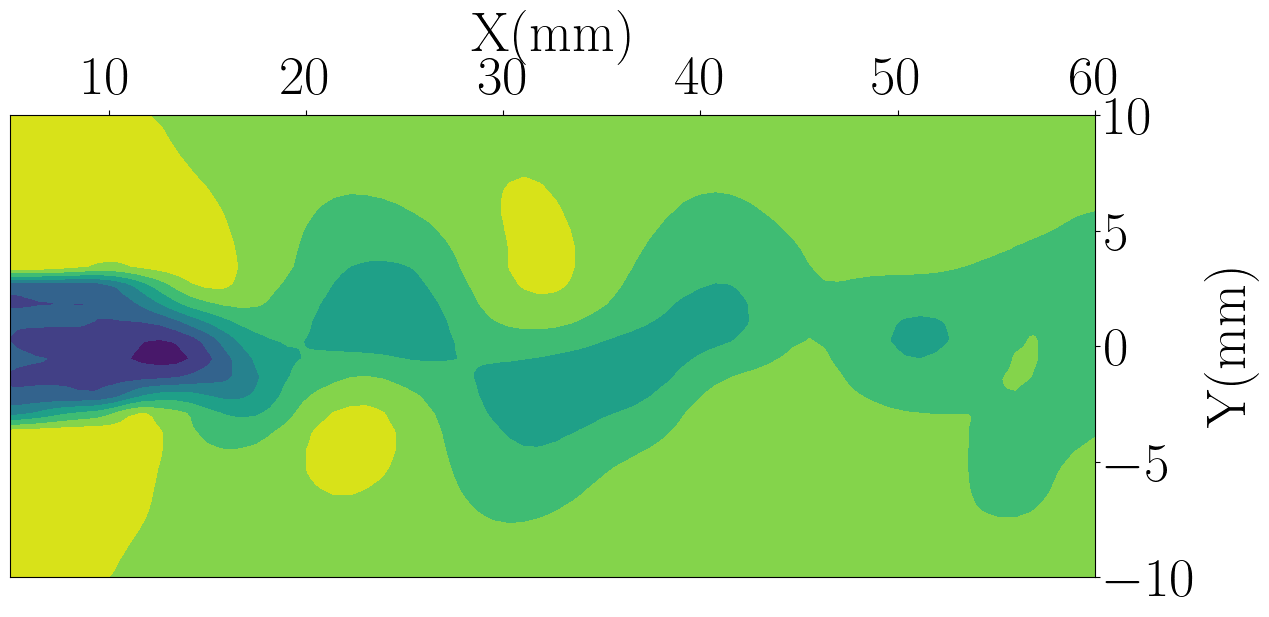} \\
         & \includegraphics[width=4cm]{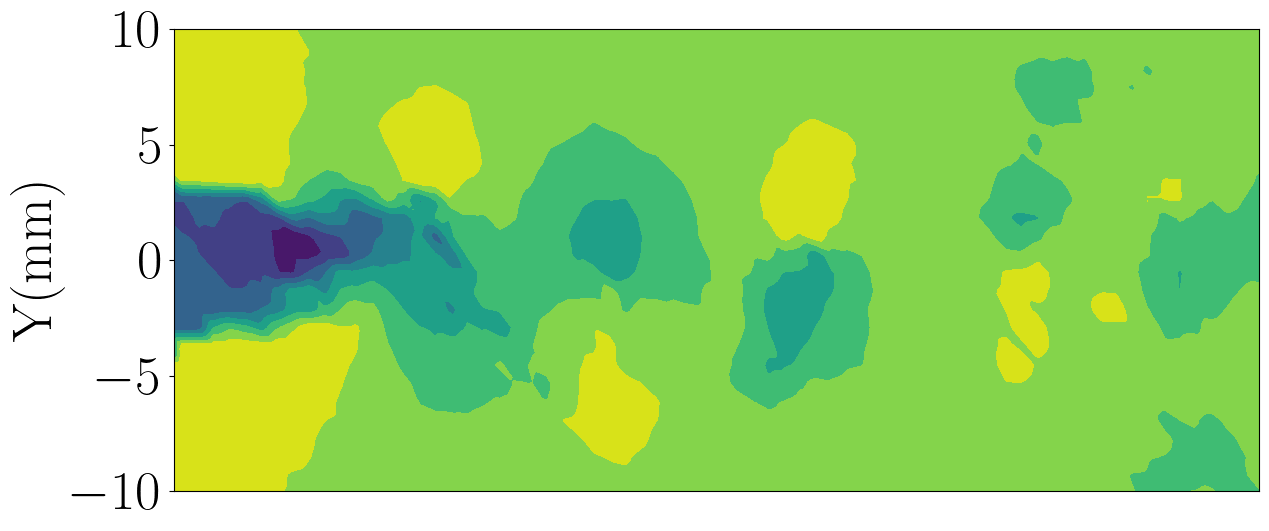} & \includegraphics[width=4cm]{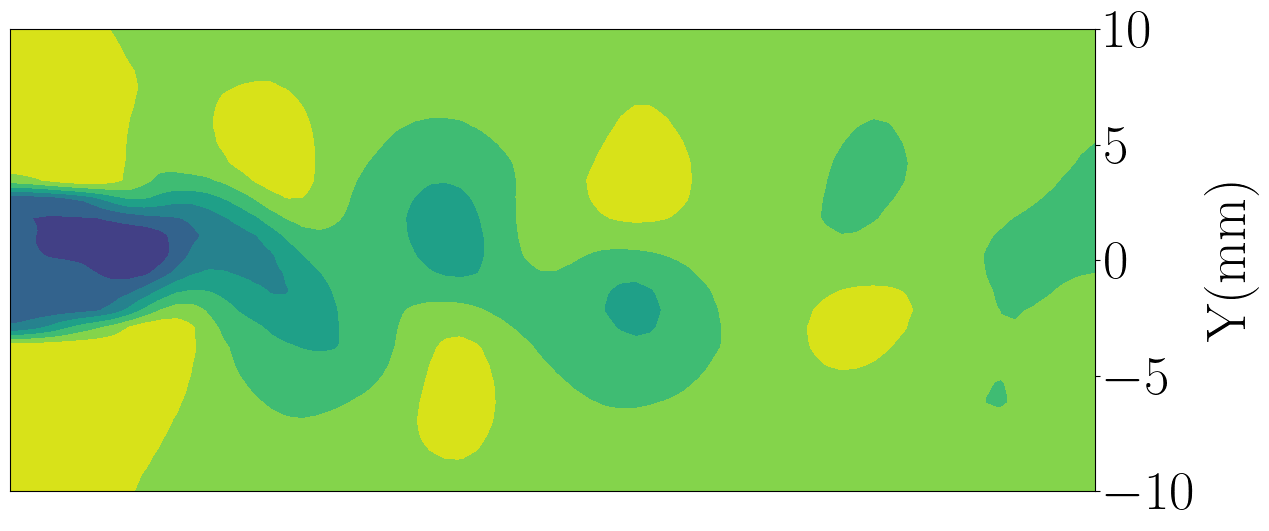}
    \end{tabular}
    \end{ruledtabular}
\end{table}

\begin{figure*}
     \centering
     \begin{subfigure}[b]{0.33\textwidth}
         \centering
         \includegraphics[width=\textwidth]{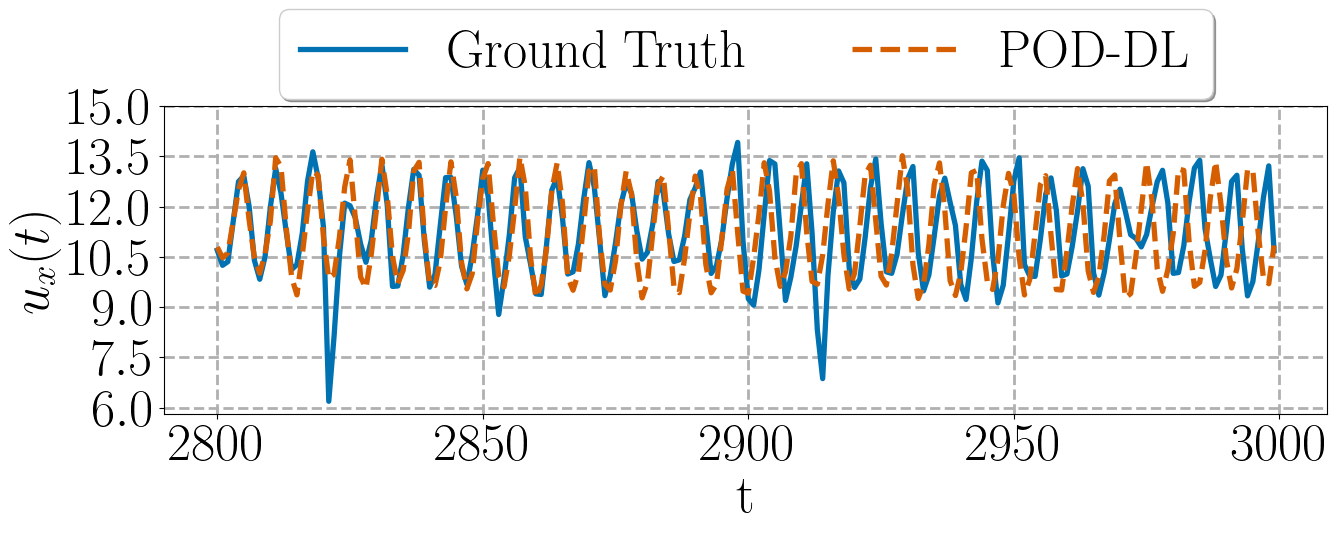}
         \caption{}
     \end{subfigure}
     \hfill
     \begin{subfigure}[b]{0.33\textwidth}
         \centering
         \includegraphics[width=\textwidth]{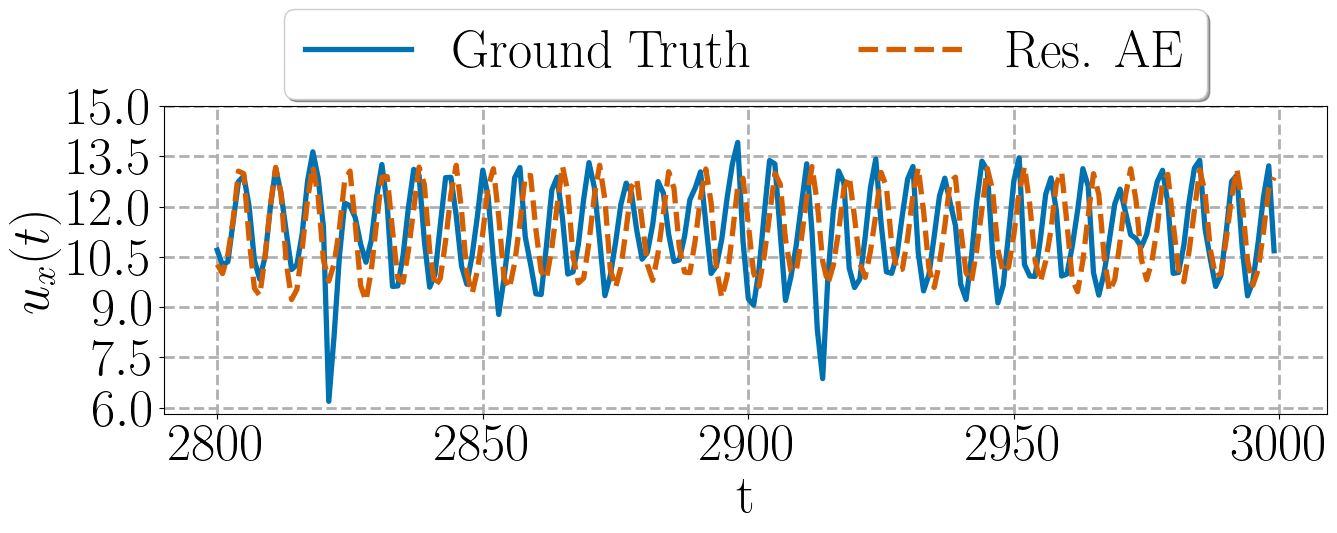}
         \caption{}
     \end{subfigure}
     \hfill
     \begin{subfigure}[b]{0.33\textwidth}
         \centering
         \includegraphics[width=\textwidth]{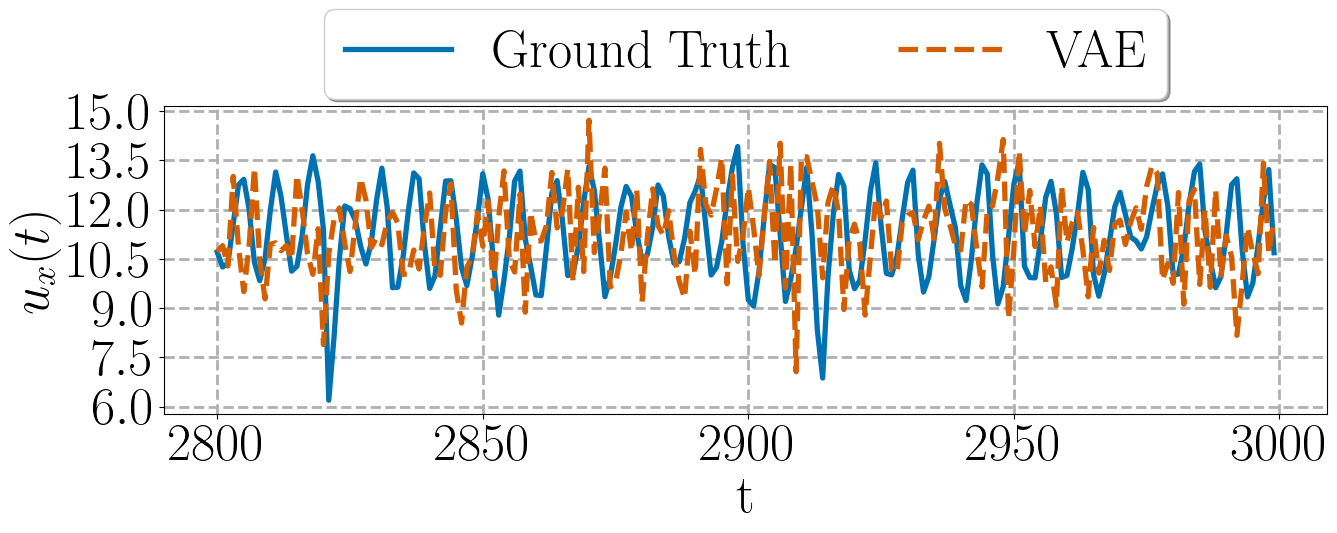}
         \caption{}
     \end{subfigure}
     \hfill
     \begin{subfigure}[b]{0.33\textwidth}
         \centering
         \includegraphics[width=\textwidth]{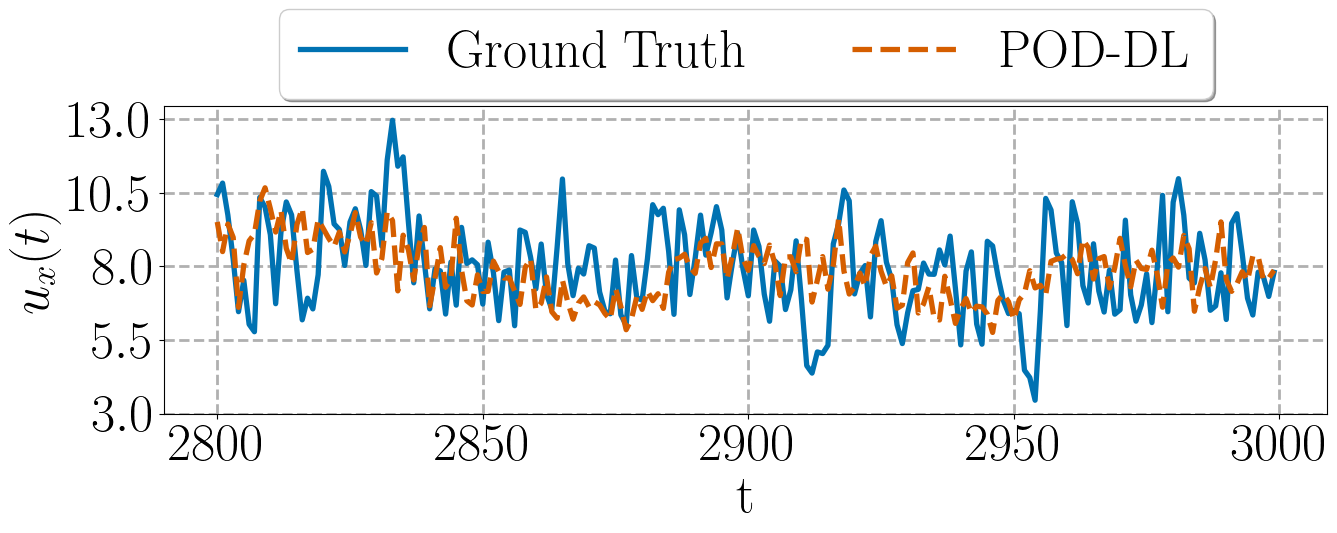}
         \caption{}
     \end{subfigure}
     \hfill
     \begin{subfigure}[b]{0.33\textwidth}
         \centering
         \includegraphics[width=\textwidth]{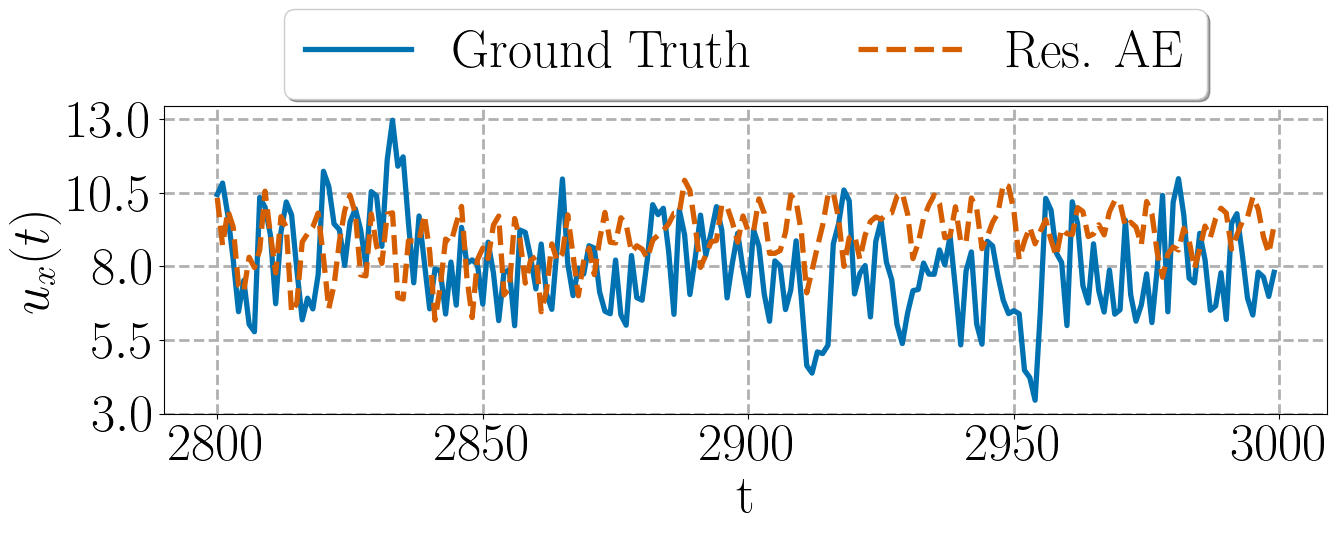}
         \caption{}
     \end{subfigure}
     \hfill
     \begin{subfigure}[b]{0.33\textwidth}
         \centering
         \includegraphics[width=\textwidth]{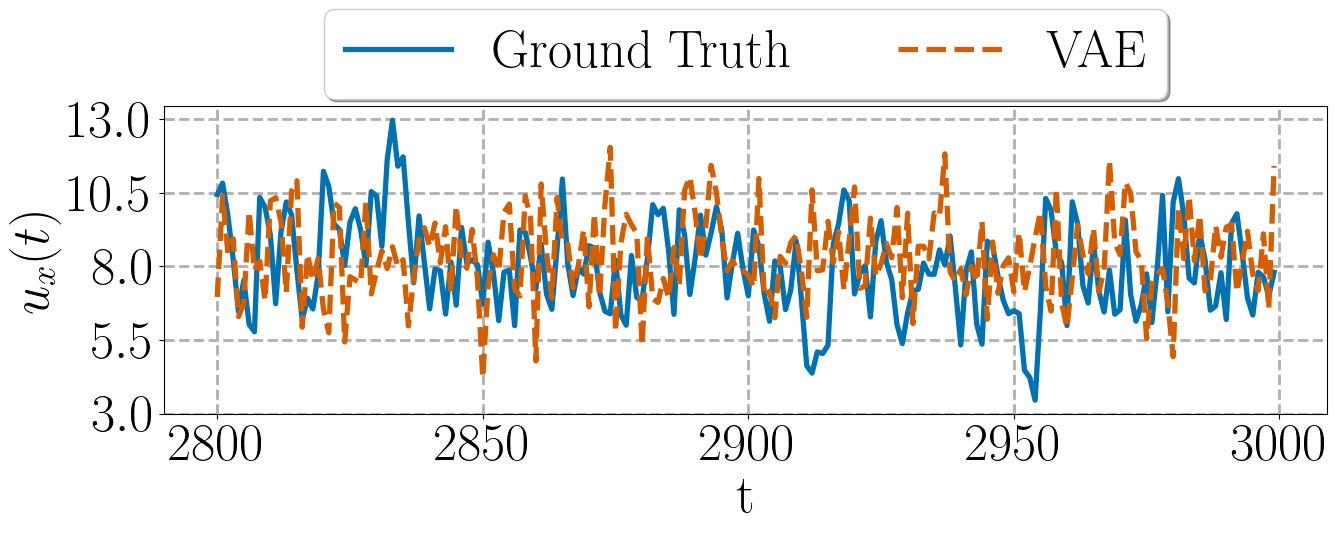}
         \caption{}
     \end{subfigure}
        \caption{Comparison of the ground truth turbulent flow streamwise velocity evolution $u_{x}(t)$, at the spatial coordinate $P_{1}$ shown in Fig. \ref{fig: experm_ref_points}, against the predictions obtained from the (a) POD-DL, (b) residual autoencoder and (c) variational autoencoder models, respectively. Counter part for the spatial coordinate $P_{2}$ in figures (d), (e) and (f). In all figures the ground truth is represented by a solid line and the prediction by a dashed line.}
        \label{fig: experm_strmw_velocity_compare}
\end{figure*}

\begin{figure}
     \centering
     \begin{subfigure}[b]{0.236\textwidth}
         \centering
         \includegraphics[width=\textwidth]{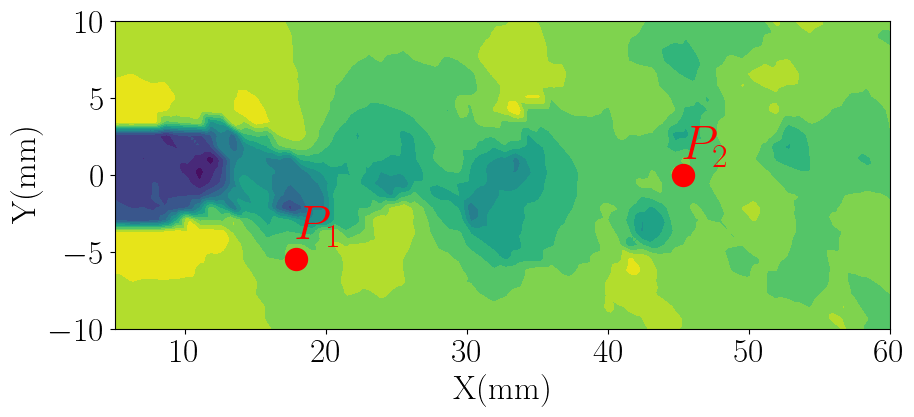}
         \caption{}
     \end{subfigure}
     \hfill
     \begin{subfigure}[b]{0.236\textwidth}
         \centering
         \includegraphics[width=\textwidth]{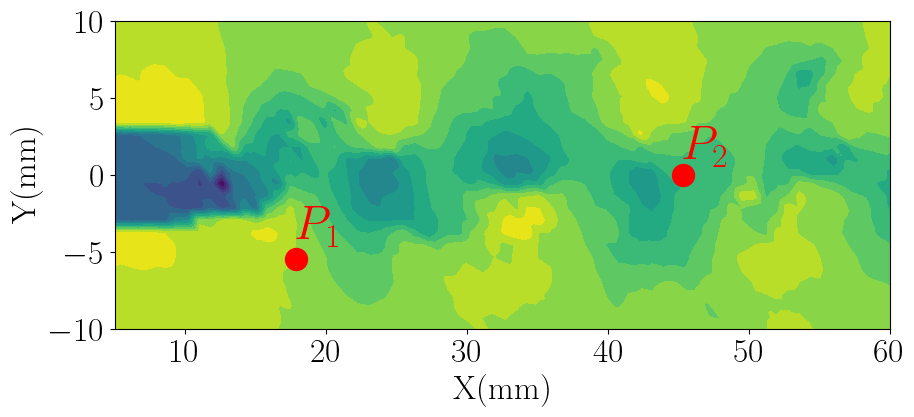}
         \caption{}
     \end{subfigure}
    \caption{Snapshots from the turbulent flow streamwise velocity, at time instants (a) $2801$ and (b) $2850$, representing by red points, $P_{1}$ (left) and $P_{2}$ (right), the spatial coordinates used to compare the streamwise, $u_{x}$ velocity evolution shown in Fig. \ref{fig: experm_strmw_velocity_compare}.}
    \label{fig: experm_ref_points}
\end{figure}

Once the models are trained, they are tasked to iteratively predict $200$ snapshots ahead in time (from snapshots [$2801$ - $3000$]) and compare the statistics of the predicted snapshots with the ground truth snapshots obtained from experimental measurements. Figure \ref{fig: experm_strmw_velocity_compare} presents this comparison, at two representative spatial coordinates shown in Fig. \ref{fig: experm_ref_points}, where left, middle and right columns depict the streamwise velocity predictions from the POD-DL, residual autoencoder, and variational autoencoder (VAE) models, respectively.

The comparison reveals that predictions from the POD-DL model are the only ones that closely follow the trend of the ground truth data. This contrasts with the laminar flow results, where both the residual autoencoder and the POD-DL model provided good predictions. Observing the velocity evolution, it is evident that the predictions from the POD-DL model are significantly more accurate than those generated by the residual autoencoder and the VAE.

This superior performance of the POD-DL model in capturing the essential dynamics and producing accurate predictions highlights its robustness and effectiveness, especially in complex flows.

\begin{figure}
     \centering
     \begin{subfigure}[b]{0.4\textwidth}
         \centering
         \includegraphics[width=\textwidth]{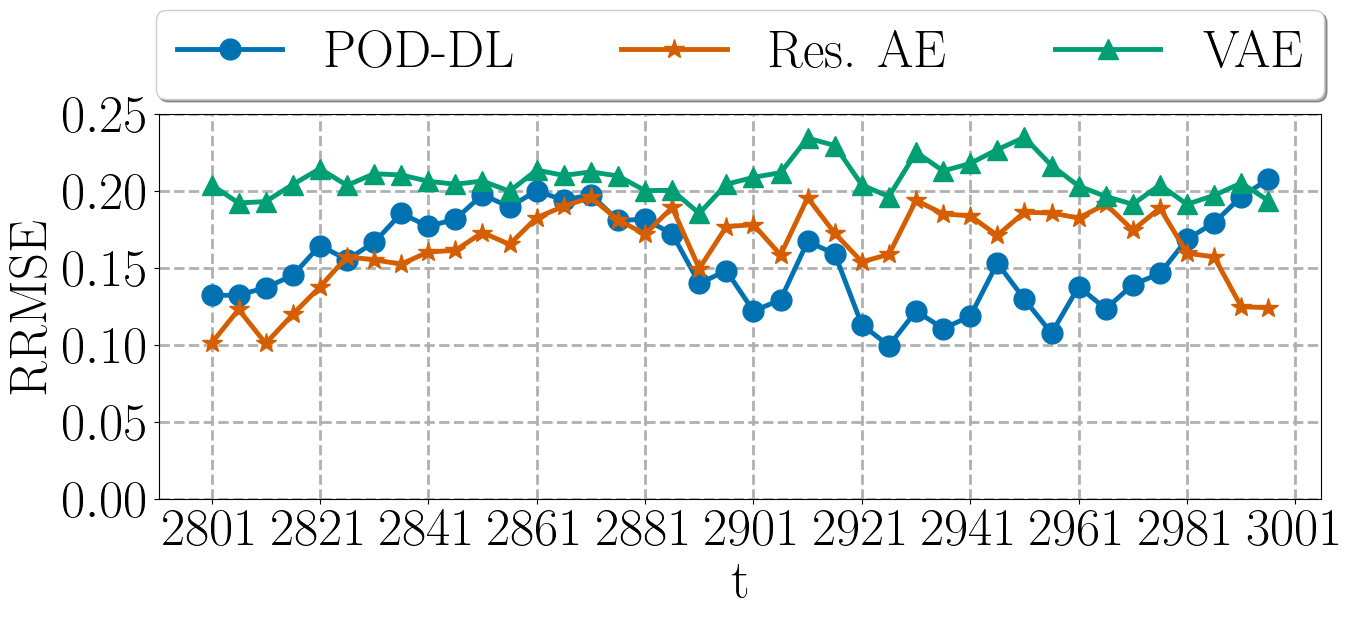}
         \caption{}
     \end{subfigure}
     \hfill
     \begin{subfigure}[b]{0.4\textwidth}
         \centering
         \includegraphics[width=\textwidth]{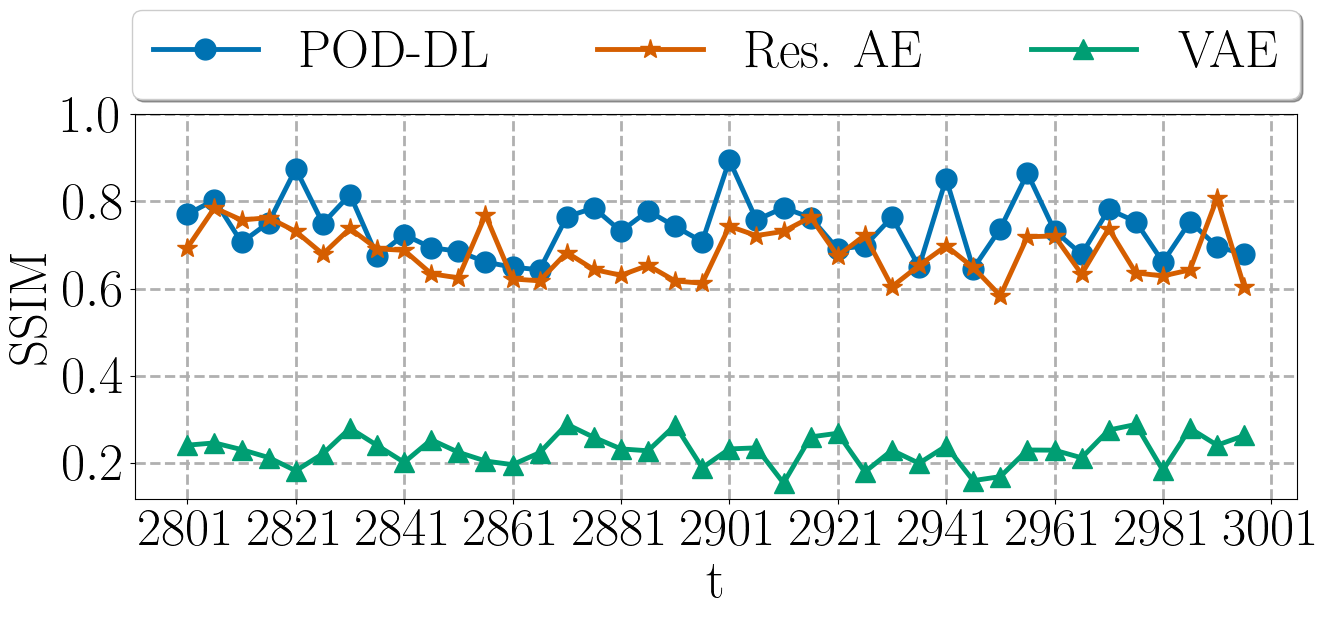}
         \caption{}
     \end{subfigure}
        \caption{Error committed by the models, in each prediction for the streamwise velocity component, measured by the (a) RRMSE and (b) the SSIM.}
        \label{fig: experiment_rrmse_ssmi}
\end{figure}

Similar to the laminar flow, the RRMSE and the SSIM are used to quantify the similarity between the model predictions and the ground truth data. Figure \ref{fig: experiment_rrmse_ssmi} presents the results obtained from these two metrics for the turbulent flow.

The analysis confirms our earlier assumption based on the streamwise velocity comparison: the POD-DL model yields the most accurate predictions, as evidenced by both the RRMSE and SSIM metrics. This is crucial to note because while the RRMSE values in Fig. \ref{fig: experiment_rrmse_ssmi} (a) suggest that all models produce predictions of similar accuracy, the SSIM values in Fig. \ref{fig: experiment_rrmse_ssmi} (b) reveal a stark difference. Specifically, the SSIM shows that the predictions from the VAE are significantly less similar to the ground truth data compared to the other models. This discrepancy arises because the SSIM considers structural information, providing a more nuanced assessment of similarity between two snapshots. This highlights the importance of using complementary metrics like SSIM alongside RRMSE to obtain a comprehensive evaluation of model performance.

\begin{figure*}
     \centering
     \begin{subfigure}[b]{0.62\textwidth}
         \centering
         \includegraphics[width=\textwidth]{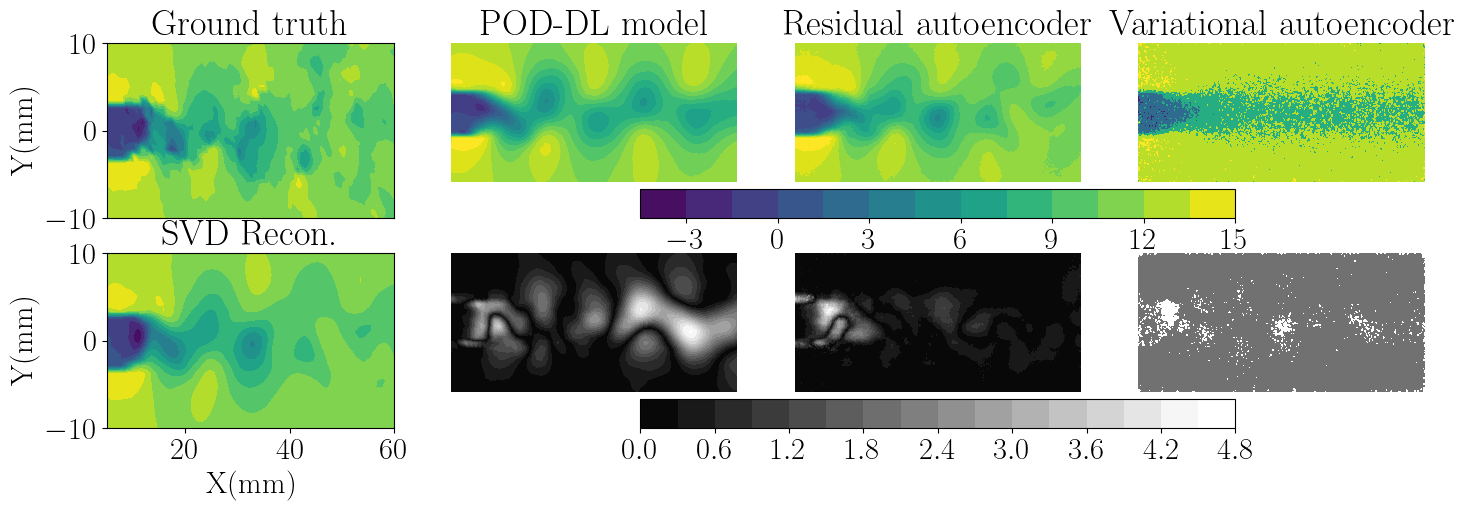}
         \caption{Ground truth snapshot, predictions and absolute error at time instant $2801$.}
     \end{subfigure}
     \hfill
     \begin{subfigure}[b]{0.62\textwidth}
         \centering
         \includegraphics[width=\textwidth]{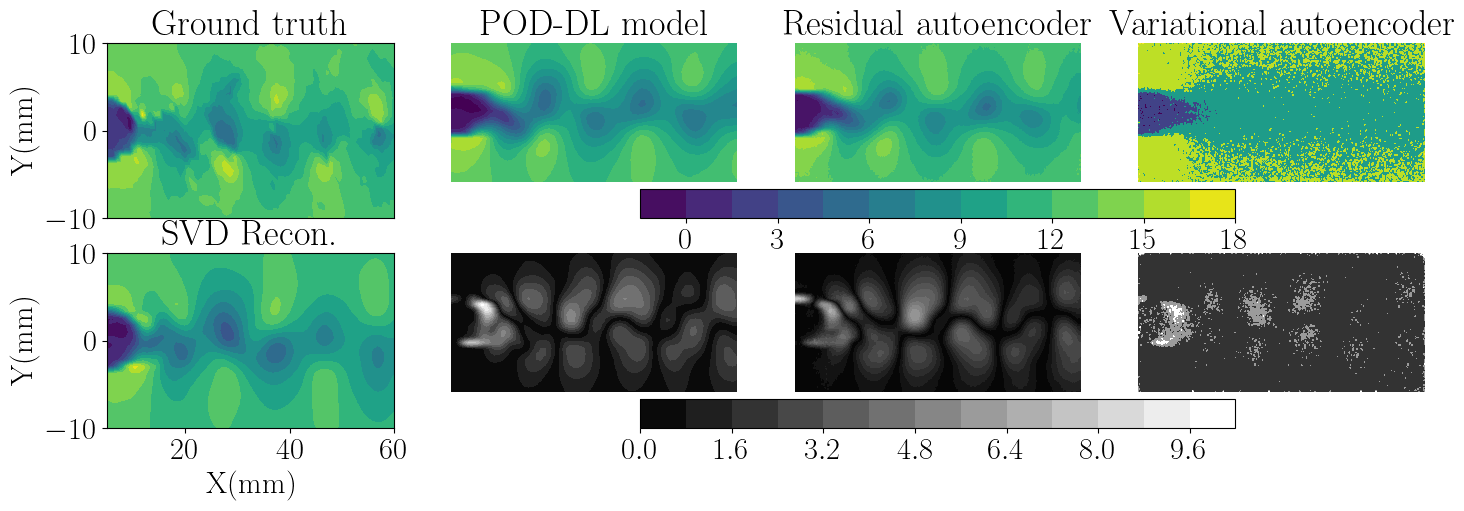}
         \caption{Ground truth snapshot, predictions and absolute error at time instant $2860$.}
     \end{subfigure}
     \hfill
     \begin{subfigure}[b]{0.62\textwidth}
         \centering
         \includegraphics[width=\textwidth]{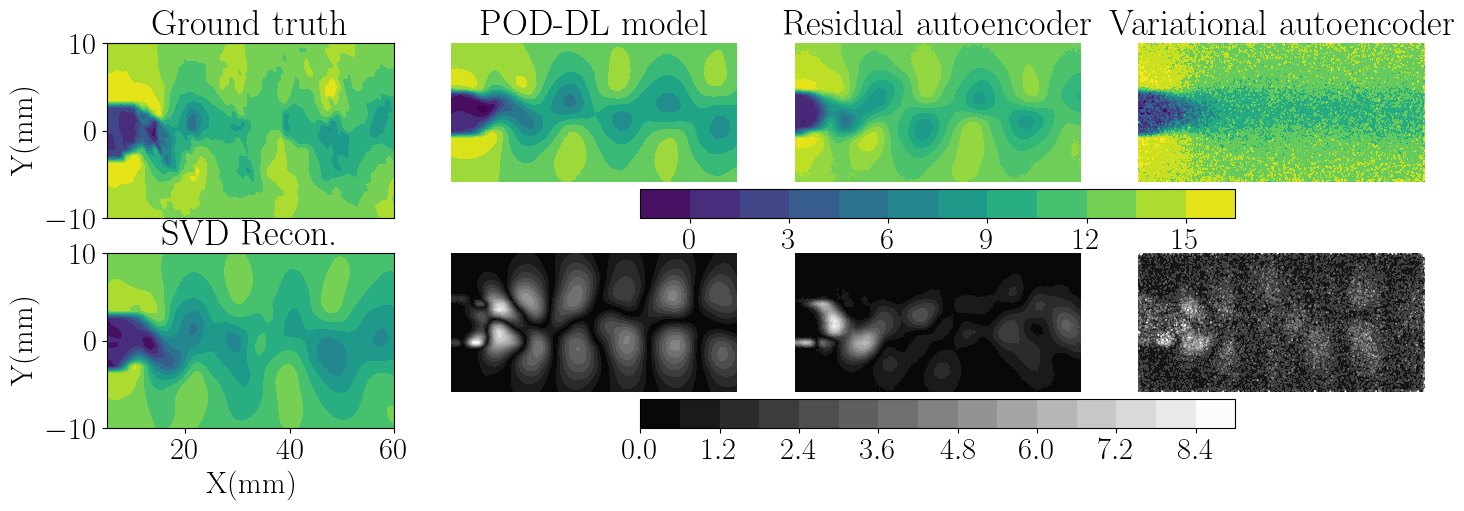}
         \caption{Ground truth snapshot, predictions and absolute error at time instant $3000$.}
     \end{subfigure}
        \caption{Snapshots at some representative time instants showing the ground truth streamwise velocity component of the turbulent flow, the predictions obtained from the different models and their respective absolute error in gray-scale. The time instants chosen are $2801$, $2860$ and $3000$.}
        \label{fig: experiment_strw_snapshots_comp}
\end{figure*}

The pronounced differences in prediction accuracy among the models are further corroborated in Fig. \ref{fig: experiment_strw_snapshots_comp}, where visual comparisons of the streamwise velocity snapshots demonstrate the superior performance of the POD-DL model.

Given the turbulent nature of this flow, predicting the exact state-space velocity flow field is highly challenging due to the presence of small-scale structures and inherent randomness in the dynamics. However, predicting the flow statistics is more feasible.

\begin{figure*}
     \centering
     \begin{subfigure}[b]{0.25\textwidth}
         \centering
         \includegraphics[width=\textwidth]{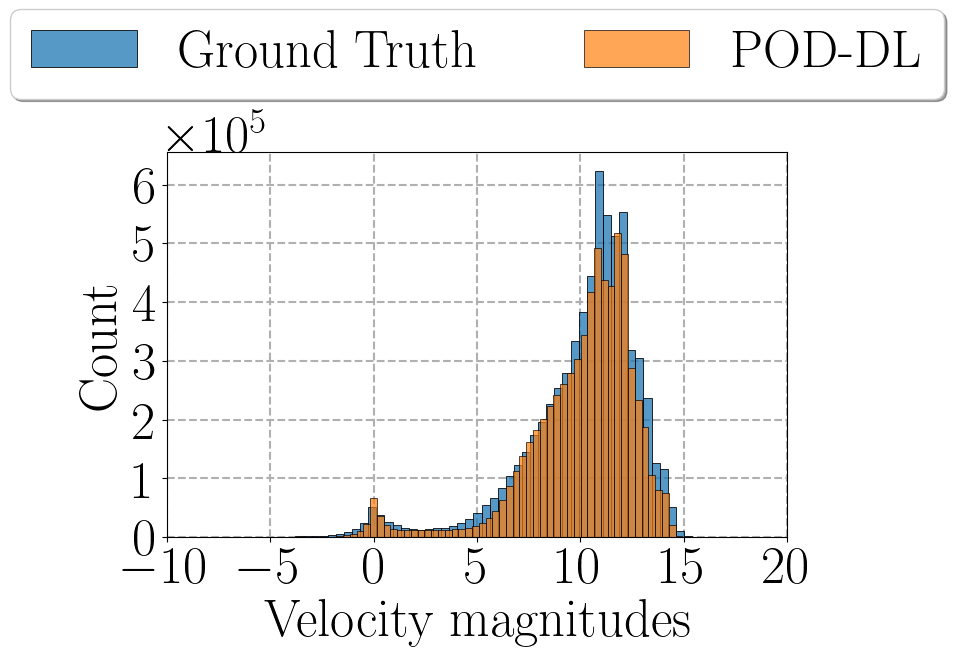}
         \caption{}
     \end{subfigure}
     \hfill
     \begin{subfigure}[b]{0.25\textwidth}
         \centering
         \includegraphics[width=\textwidth]{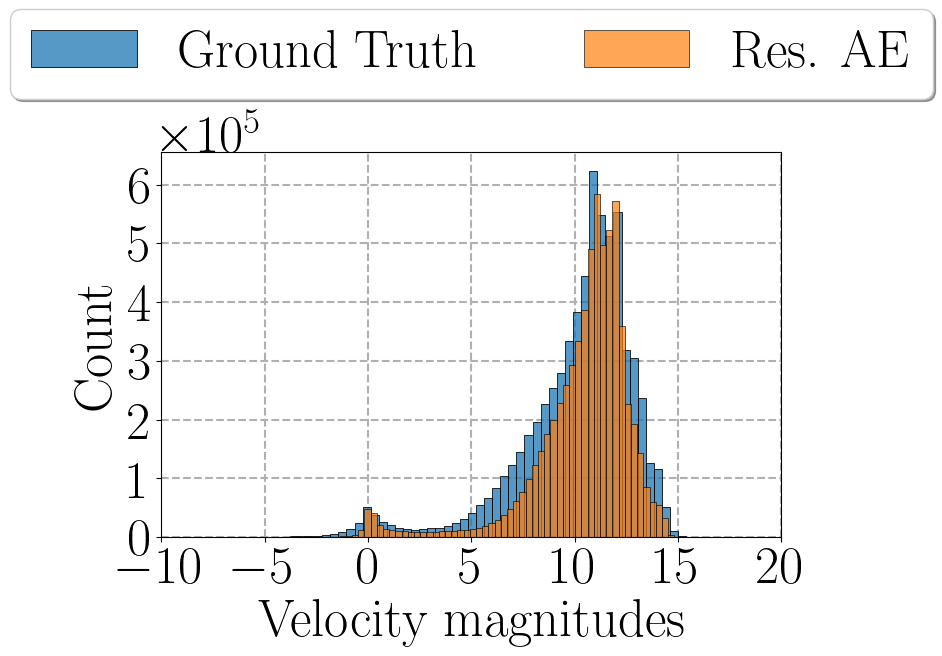}
         \caption{}
     \end{subfigure}
     \hfill
     \begin{subfigure}[b]{0.23\textwidth}
         \centering
         \includegraphics[width=\textwidth]{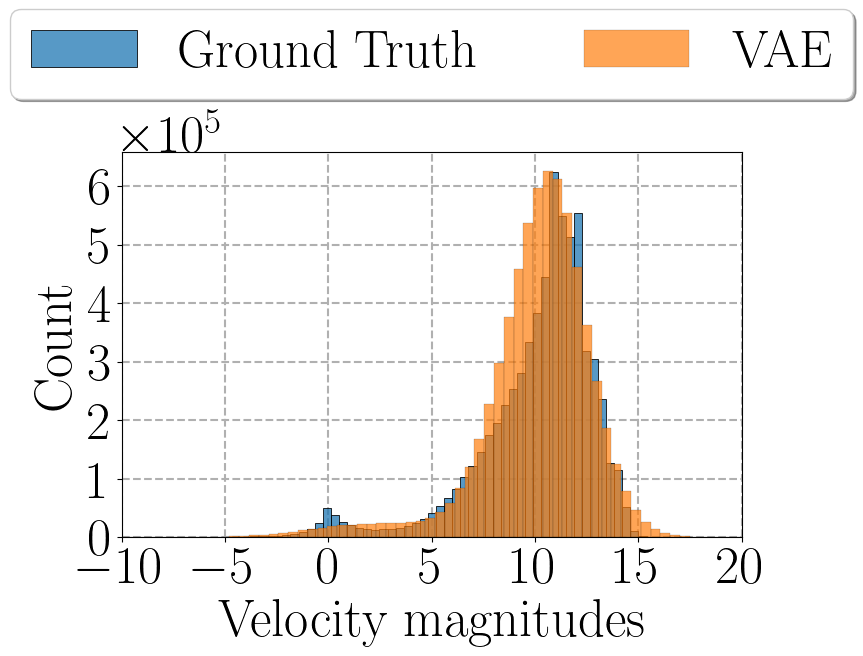}
         \caption{}
     \end{subfigure}
        \caption{Comparison of the velocity histograms between the ground truth and the temporal predictions of the turbulent flow obtained from (a) the POD-DL, (b) the residual autoencoder and (c) the variational autoencoder models, respectively.}
        \label{fig: experiment_histograms}
\end{figure*}

In Fig. \ref{fig: experiment_histograms}, we compare the velocity magnitude histograms of the real snapshots with the predictions from the three models. The results indicate that predictions from the POD-DL model exhibit a distribution that is closer to the ground truth data compared to the other two models, specially the variational autoencoder whose predictions are mostly the mean flow with noise, as shown in Fig. \ref{fig: experiment_strw_snapshots_comp}. This observation aligns with the earlier findings presented in Fig. \ref{fig: experm_strmw_velocity_compare} and Fig. \ref{fig: experiment_rrmse_ssmi}, further validating the superior performance of the POD-DL model in capturing the statistical properties of the turbulent flow.

\section{Discussion} \label{sec: discussion}

In this work, the three forecasting models (POD-DL, residual autoencoder and VAE) are applied to both laminar and turbulent flows to evaluate their generalization capabilities across different flow dynamics. This evaluation considers not only the accuracy of the predictions but also the ease of adapting the models to diverse datasets. By examining performance across varying flow regimes, we aim to assess the robustness of each model in capturing distinct fluid behaviors and its practical applicability to a range of flow scenarios.

From the two test cases studied, modeling a laminar and turbulent wake behind a circular cylinder, we have observed the behavior and stability of the models in iteratively generating temporal predictions of the flow dynamics. Also, we show the generalization capabilities and robustness of the models, which are used in laminar and turbulent flows. The low performance of the VAE model could be explained by the complexity of the ELBO (evidence lower bound) loss function \eqref{eq: elbo_new}, which is designed to approximate a probability distribution. However, other models have been proposed in the field of probabilistic forecasting, such as temporal diffusion models \cite{lin_2023_diffusion_survey} or copula methods for forecasting multivariate time series \cite{PATTON2013899_copulas_multivariate_forecasting, SMITH202381_implicit_couplas_forecasting}. Therefore, we do not rule out that models based on probabilistic forecasting could predict flow dynamics.

Regarding the other two models, both utilize the MSE \eqref{eq: mse} as the loss function. However, their architectures and approaches differ significantly. The POD-DL model combines SVD with a LSTM architecture, whereas the residual autoencoder is entirely based on deep learning, consisting of convolutional networks, transpose convolutional networks and a convolutional LSTM.

Note the POD-DL model does not directly predict the snapshots. Instead, it predicts the POD coefficients through the LSTM network. These predicted modes are subsequently used to reconstruct the snapshots via SVD. This two-step approach leverages the dimensionality reduction and the capability of LSTM to capture temporal dependencies. In contrast, the residual autoencoder predicts the snapshots directly, without the need for any post-processing to reconstruct them.

The main advantage of the POD-DL model lies in its substantial reduction in dimensionality, as it replaces high-dimensional snapshots with a simple vector whose size is determined by the number of retained modes. The SVD decomposes the dataset into three matrices ($\bU$, $\bSigma$, $\bV^{T}$), where $\bSigma$ and $\bV^{T}$ are used for forecasting, and $\bU$ is kept as is to reconstruct the snapshots. This approach inherently assumes that the matrix $\bU$ remains constant, which may not be entirely accurate since $\bU$ could experience slight variations over time.

However, as shown in Fig. \ref{fig: synthetic_strmw_velocity_compare}, Fig. \ref{fig: synthetic_normal_velocity_compare}, Fig. \ref{fig: synthetic_rrmse_ssim}, Fig. \ref{fig: experm_strmw_velocity_compare} and Fig. \ref{fig: experiment_rrmse_ssmi} the POD-DL model generates the most accurate predictions for either the laminar and turbulent flow. This suggests that, although the unaccounted variations in the $U$ matrix might introduce some errors, these errors are not significant enough to overshadow the advantages of the POD-DL model. This ability to capture the main dynamics of the flow, makes it the most effective model among the models tested, despite the potential inaccuracies introduced by assuming a constant $\bU$ matrix.

The poor performance of the models in Fig. \ref{fig: experm_strmw_velocity_compare}, at the spatial coordinate $P_{2}$, may be attributed to the limitations of the hidden layer, which consists of either an LSTM or a ConvLSTM, depending on the model. These architectures may struggle to capture the complex dynamics associated with small-scale flow structures. A more effective approach could involve approximating the statistical behavior of these scales. In this context, temporal diffusion models, known for their capability to approximate complex distributions, could serve as a viable alternative to LSTM or ConvLSTM architectures.

A significant difference between the POD-DL model and the residual autoencoder is that the latter must adjust its architecture based on the size (i.e., resolution) of the incoming snapshots. This issue could be addressed using spatial pyramid pooling \cite{kaiming_etal_2014_pyramid_pooling}, which resizes every input snapshot to a fixed size and aids in identifying multiple scales within the snapshot. Although this technique presents a viable alternative, its implementation adds complexity to the model.

For this reason, we advocate that the POD-DL model possesses the best properties. By reducing the high dimension of the incoming snapshots to vectors, which encapsulate the main dynamics of the flow. The POD-DL model simplifies both the required architecture and the training process. This dimensionality reduction makes it also more intuitive to apply classical multivariate forecasting methods compared to using two-dimensional snapshots.

Moreover, the ability of POD-DL to handle different datasets with minimal modifications to the forecasting model enhances its versatility. This advantage is particularly valuable in practical applications where datasets vary in resolution. By maintaining a consistent and simplified approach to forecasting, the POD-DL model ensures robustness and adaptability across diverse datasets, as shown in Appendix \ref{appen: pretrained_model} where a pre-trained POD-DL is tasked to predict a completely new flow dynamics, with and without retraining the model with the new data.

However, the performance of these data-driven models is intrinsically dependent on the quality and quantity of the training data. Additionally, their predictive accuracy declines as the complexity of the dynamics increases, particularly in turbulent flow regimes. Among the evaluated models, the POD-DL approach exhibits superior performance in managing turbulence compared to the residual autoencoder and VAE. Nevertheless, its effectiveness is highly sensitive to the number of POD modes retained during truncation, where capturing the dynamics of the less energetic modes is particularly challenging due to their chaotic and noisy behavior, often necessitating more advanced models than LSTMs, which in turn may require larger training datasets. This balance between the size of the training dataset and prediction accuracy remains an active area of research.

\section{Conclusions} \label{sec: conclusion}

The main goal of this work is to present the generalization capabilities and robustness of purely deep learning models and hybrid models, which are characterized by the physics of the flow, in fluid dynamics application. We have compared three forecasting models inspired by methodologies commonly proposed in the literature to predict the temporal evolution of flow dynamics: a convolutional autoencoder combined with a convolutional LSTM (ConvLSTM), a VAE combined with a ConvLSTM, and a hybrid model that combines POD with an LSTM (POD-DL). We applied these models to two fluid dynamics datasets, which are characterized by their multidimensional and nonlinear nature.

The high dimensionality of these datasets necessitated the development of methods that utilize the latent dimension. Specifically, we employed models that decompose the high-dimensional dynamics into a reduced-order representation in a bijective manner. This approach allows us to predict the future evolution of the system in the reduced-order space, and then recover the original dimensionality, leveraging the bijective property.

We have tested these models on two datasets representing the velocity field of a flow past a circular cylinder at laminar and turbulent regimes, respectively. Our observations indicate that both the VAE and the residual autoencoder achieved accurate predictions in the laminar flow.

However, the hybrid POD-DL model has demonstrated superior performance, by outperforming the other two models in either the laminar and turbulent flows. This highlights the generalization capabilities of hybrid models that combine modal decomposition with deep learning. The POD-DL model's use of POD to reduce the dimensionality of the incoming snapshots contributes significantly to its success. By applying POD, the model effectively simplifies the architecture of the forecasting component and reduces the amount of training data required.

This combination with the POD, also allows the model to be informed about the physics of the problem, as POD decomposes the spatio-temporal flow dynamics in different POD modes with physical meaning. We believe that this work has demonstrated the potential of combining modal decomposition techniques with deep learning forecasting models, to predict the temporal evolution of flow dynamics. Further research could explore the integration of more advanced deep learning architectures within the POD-DL methodology, substituting the LSTM layer by a temporal diffusion model \cite{Lin_etal_2024_diffusion_forecasting_survey}, which has shown promising results in weather forecasting \cite{Kochkov_etal_2024_diff_model_weather}, as well as the application of these models to a wider range of fluid dynamics problems. Especially those where a larger truncation of POD modes is required to capture the dynamics of interest. Additionally, investigating the use of transfer learning to leverage pre-trained models, from other dynamics, for new flow conditions could be beneficial to a wide variety of applications in both industry and academia.

\begin{acknowledgments}
The authors acknowledge the grants TED2021-129774B-C21 and PLEC2022-009235 funded by MCIN/AEI/10.13039/501100011033 and by the European Union “NextGenerationEU”/PRTR and the grant PID2023-147790OB-I00 funded by MCIU/AEI/10.13039/501100011033/FEDER, UE. The MODELAIR and ENCODING projects have received funding from the European Union’s Horizon Europe research and innovation programme under the Marie Sklodowska-Curie grant agreement No. 101072559 and 101072779, respectively. The results of this publication reflect only the author(s) view and do not necessarily reflect those of the European Union. The European Union can not be held responsible for them.
\end{acknowledgments}

\section*{Data Availability Statement}

The code used in this work is available in \url{https://github.com/RAbadiaH/hybrid_and_purely_dl_models.git}. The data of the turbulent flow that was obtained by Mendez \textit{et al.} \cite{Mendez_2020_experimental} and supports the findings of this study is openly available in \url{https://github.com/mendezVKI/MODULO/tree/master/download_all_data_exercises}.

\nocite{*}
\section*{References}
\bibliography{references_v1}

\appendix

\section{Datasets and preprocessing} \label{sec: datasets_and_preprocessing}
Two public available datasets describing the velocity field of a flow past a circular cylinder are used to test the forecasting models. One of the datasets is a three-dimensional laminar flow at Reynolds number $\hbox{Re} = 280$, generated from a numerical simulation. More information on the characteristics of the simulation can be found in \cite{LeClainche_2018_cilind}. The second dataset is obtained through measurements of a turbulent flow, which describes the wake of a circular cylinder at $\hbox{Re} = 4000$. In this case, the dataset is two-dimensional in space. More details on the experimental setup can be found in \cite{Mendez_2020_experimental}.

Regardless of the dataset, these are represented as spatio-temporal tensors formed by $K$ snapshots $\bv_{k} = \bv(t_{k})$, where $\bv(t_{k})$ is the flow realization at the $k$-th time instant. For convenience, these snapshots are collected in the following snapshot tensor,
\begin{equation}
    \bV = [\bv_{1}, \bv_{2}, \dots, \bv_{k}, \dots, \bv_{K}].
    \label{eq: dataset_tensor_shape}
\end{equation}
Each $\bv(t_{k})$ is a tensor itself, containing both the components and spatial dimensions. In this work, the components dimension represents the velocity components (streamwise and wall-normal). For instance, let us consider the turbulent flow, which is two-dimensional in space and is formed by a single velocity component. 

This dataset represents the velocity field $\bv$ in a Cartesian coordinates systems with spatial dimension $J_{2} \times J_{3}$, as
\begin{eqnarray}
    \bv_{k} = \bv(x_{j_{2}},y_{j_{3}},t_{k}), \quad j_{2} \in \{1,\dots,J_{2}\},\nonumber \\ j_{3} \in \{1,\dots,J_{3}\}, \ k \in \{1,\dots,K\}.
\end{eqnarray}
Therefore, the snapshot tensor $\bV$ can be re-organized in a fourth-order $J_{1} \times J_{2} \times J_{3} \times K$-tensor, formed by $J_{1} = 1$ components.
\begin{equation}
    \bV_{1j_{2}j_{3}k} = \bv^{1}_{k} = \bv^{1}(x_{j_{2}},y_{j_{3}},t_{k}).
\end{equation}
The dataset from the laminar flow, which is three-dimensional in space and is formed by two velocity components, can be represented using the same notation, where now $\bv_{k} = \bv(x_{j_{2}}, y_{j_{3}}, z_{j_{4}}, t_{k})$, and $\bV$ is a $J_{1} \times J_{2} \times J_{3} \times J_{4} \times K$-tensor.
\begin{align}
    \bV_{1j_{2}j_{3}j_{4}k} & = \bv^{1}_{k} = \bv^{1}(x_{j_{2}},y_{j_{3}}, z_{j_{4}},t_{k}), \label{eq: tensor_3d_1}\\
    \bV_{2j_{2}j_{3}j_{4}k} & = \bv^{2}_{k} = \bv^{2}(x_{j_{2}},y_{j_{3}}, z_{j_{4}},t_{k}). \label{eq: tensor_3d_2}
\end{align}
Note $J_{1}$ is the number of components, while $J_{2}$, $J_{3}$ and $J_{4}$ correspond to the spatial dimensions, $x$, $y$ and $z$, respectively. Finally, $K$ represents the temporal dimension.

In the next two sections, we describe a data augmentation process that leverages the three spatial dimensions in the laminar flow, and secondly explain the rolling window technique that it is use to prepare the time series for the forecasting models.

\subsection{Data augmentation}
Note that the laminar flow is three-dimensional in space, making it challenging for a machine learning model to learn both the spatial and temporal correlations, especially when data are limited, as in our case. This challenge has been previously addressed in the literature \cite{forecasting_convAutoEnc_3D_2021}, often at the expense of a large amount of training data. 

Therefore, in this work, a data augmentation technique is applied to this dataset. Specifically, the tensor $\bV$ \eqref{eq: dataset_tensor_shape} is split into $J_{4}$ different sub-tensors, as follows,
\begin{align}
    \bV_{j_{1}j_{2}j_{3}1k} & = \bv(c_{j_{1}},x_{j_{2}},\bv_{j_{3}}, 1,t_{k}), \label{eq: data_aug_1}\\
    \bV_{j_{1}j_{2}j_{3}2k} & = \bv(c_{j_{1}},x_{j_{2}},\bv_{j_{3}}, 2,t_{k}), \\
    \vdots \\
    \bV_{j_{1}j_{2}j_{3}J_{4}k} & = \bv(c_{j_{1}},x_{j_{2}},\bv_{j_{3}}, J_{4},t_{k}). \label{eq: data_aug_end}
\end{align}
As shown in \ref{eq: tensor_3d_1} and \ref{eq: tensor_3d_2}, when the dataset is three-dimensional, the elements of the $z$ dimension in $\bV$ can be considered independently. Allowing to distinguish $J_{4}$ two-dimensional datasets (\ref{eq: data_aug_1} - \ref{eq: data_aug_end}), each one representing a different time-series, instead of a single three-dimensional dataset.

This technique helps to improve the predictions obtained from the forecasting models by increasing the number of available samples by a factor of $J_{4}$. However, this splitting may lead to a loss of information in the $z$ dimension (spanwise) since we no longer consider this dimension as a whole but independently.

\subsection{Rolling window}
\begin{figure}
	\centering
	\includegraphics[height=0.4\columnwidth]{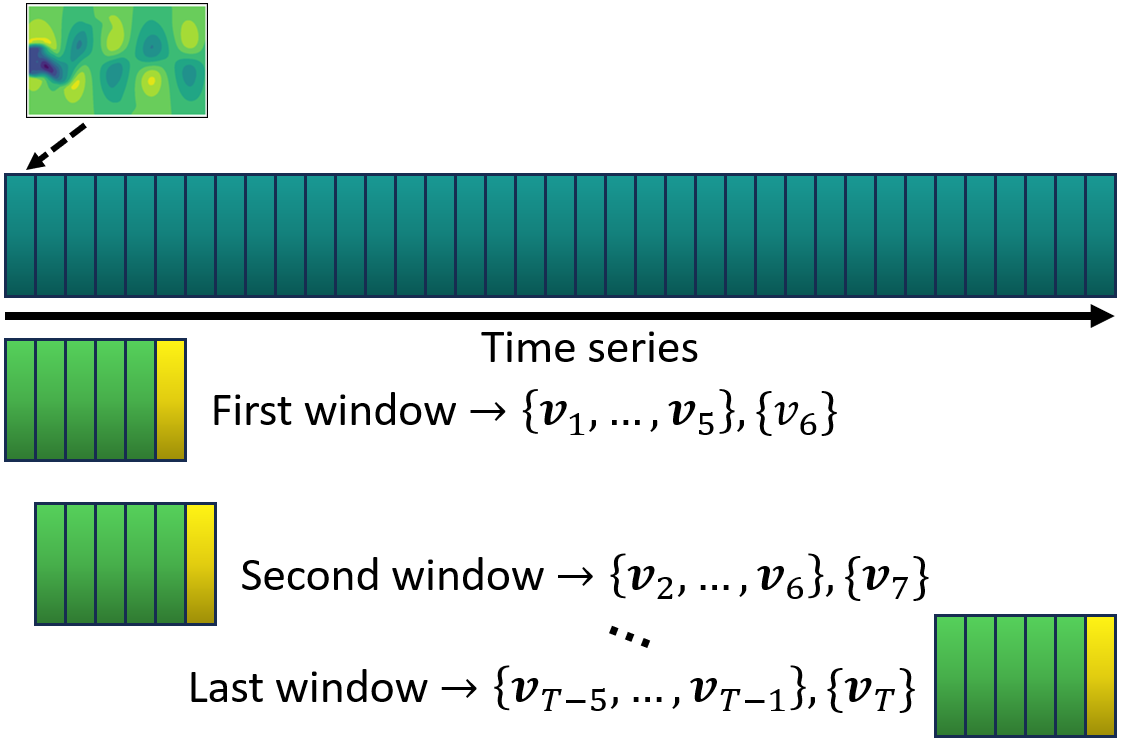}
	\caption{Sliding window method traverses the time series, with a predefined stride, creating windows that contain both the previous samples, which are inputs to the forecasting model, and the time-ahead sample that we want to predict. In this representative case, we take $5$ previous snapshots and ask the model to predict the next one.}
	\label{fig: sliding_window}
\end{figure}

In contrast to the data augmentation technique explained above, which can only be applied to three-dimensional datasets and is used to increase the number of samples available for training, the rolling-window technique is a method used to prepare time series for forecasting and can be applied to any dataset.

In a time series forecasting problem, it must be specify which samples will be used as inputs to the forecasting models (previous sequence) and which ones will be used as targets (time-ahead). The rolling-window technique facilitates this by traversing the time series with a predefined stride, creating windows that contain both the previous samples (inputs) and the time-ahead samples (targets). By applying this technique to each time series separately, we generate a set of windows that can be used to train the forecasting models. Figure \ref{fig: sliding_window} illustrates this process for a single time series.

All these windows are scaled using the min-max scaling, with the global minimum and maximum values of the original time series. This preprocessing step ensures that the data is normalized to the range $[0, 1]$, which helps to improve the training stability and performance of machine learning models, by preventing gradients from diverging or getting stuck. This vanishing or exploding gradient problem is also handle with the architectural choice of the models, for example the residual autoencoder utilizes residual connections explicitly designed to mitigate this problem, and the Variational Autoencoder and POD-DL models, make use of a LSTM layer that was incorporated to maintain long-term dependencies, thereby addressing the vanishing gradient issue inherent in recurrent networks. Additionally, a careful weight initialization were employed across all models to further reduce the risk of gradient instability.

\section{Stability of POD-DL at different numbers of training samples}\label{appen: different_training_size}
This appendix examines the training stability of the POD-DL model when trained with a smaller training set. The dataset chosen for this analysis is the turbulent flow past a cylinder \cite{Mendez_2020_experimental}, due to its inherent complexity.

For this aim, the POD-DL model was independently trained on three distinct training sets, each comprising a different number of samples: $500$, $1000$ and $2800$, where the last one matches the number of samples used to train the residual autoencoder and VAE models, results from this last case can be found in Section \ref{sec: results_experimental}.

\begin{figure}
     \centering
     \begin{subfigure}[b]{0.52\textwidth}
         \centering
         \includegraphics[width=\textwidth]{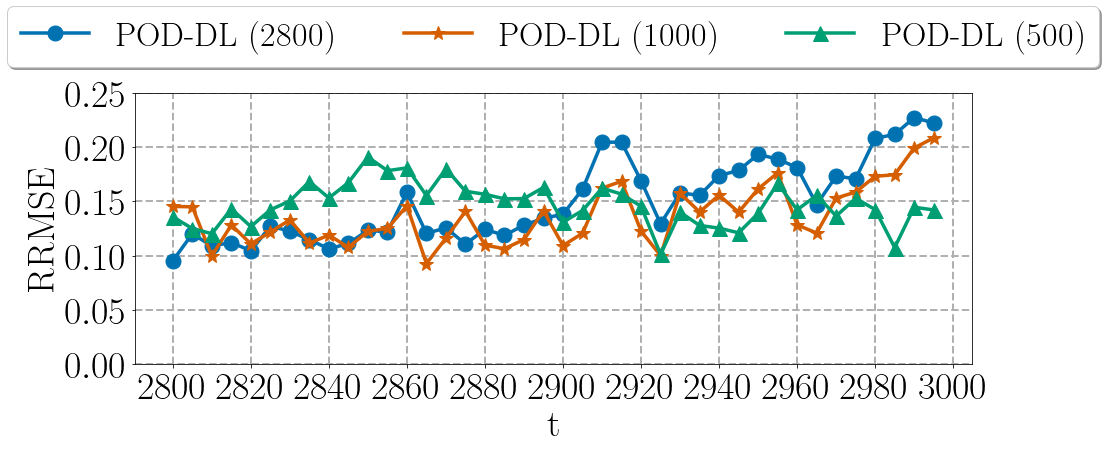}
         \caption{}
     \end{subfigure}
     \hfill
     \begin{subfigure}[b]{0.52\textwidth}
         \centering
         \includegraphics[width=\textwidth]{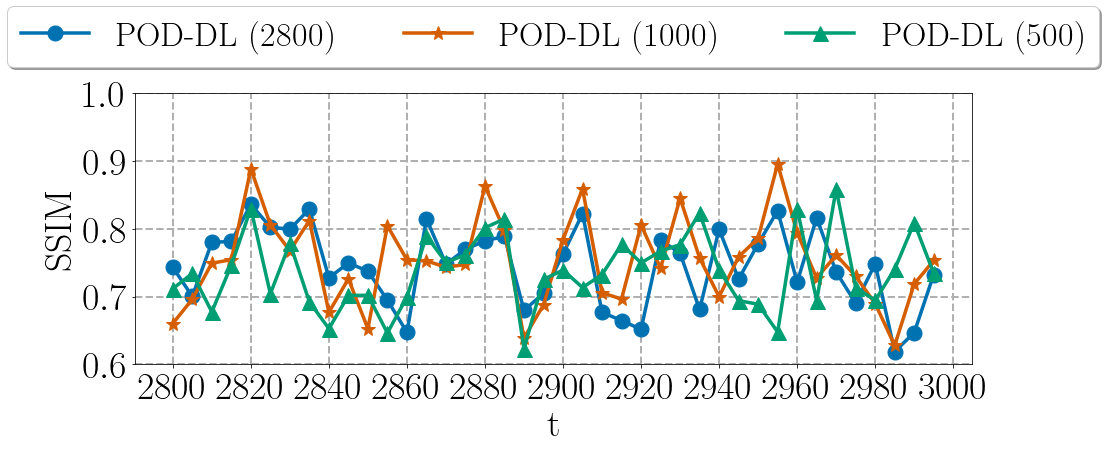}
         \caption{}
     \end{subfigure}
        \caption{Prediction error of the POD-DL model trained with different number of samples (indicated in brackets) for the streamwise velocity component of the turbulent flow past a cylinder. Errors are quantified using (a) the RRMSE and (b) the SSIM defined in \eqref{eq: rrmse} and \eqref{eq: ssim}, respectively.}
        \label{fig: mult_train_rrmse_ssmi}
\end{figure}

Figure \ref{fig: mult_train_rrmse_ssmi} illustrates the prediction error of the POD-DL model over the same $200$ time-steps analyzed in Section \ref{sec: results_experimental}, evaluated across different training sets. The results highlight the model’s training stability, even when fewer training samples are used. In comparison, while the residual autoencoder requires $2800$ samples for training, the POD-DL model maintains relatively consistent performance for smaller training sets, demonstrating robustness to variations in training set size. This robustness arises from the dimensionality reduction achieved by POD, which enables the use of lower-dimensional vectors for training, unlike the high-dimensional snapshots required by the residual autoencoder and VAE models.

\section{Results from a pre-trained model}\label{appen: pretrained_model}

This appendix shows the results from the POD-DL model when is pre-trained on the turbulent flow and task it to forecast the dynamics from a completely different flow. This is done by performing transfer learning, where the weights from a previous training are loaded before forecasting.

Two cases are study in this appendix: (1) forecasting the new dynamics without retraining with snapshots from the new dataset and (2) retraining the POD-DL model with the first $800$ snapshots from the new dataset. This is done to test the generalization capabilities of the POD-DL model to unseen dynamics.

\begin{table}
    \centering
    \caption{Comparison of the original flow from the numerical simulation of the Jet LES and the reconstruction using SVD, retaining the $20$ most energetic modes to reconstruct the flow field.\label{tab: jetLES_orig_svd}}
    \begin{ruledtabular}
    \begin{tabular}{ccccc}
         & \textbf{Original} & \textbf{SVD reconstruction} \\
        \hline
         & \includegraphics[width=4cm]{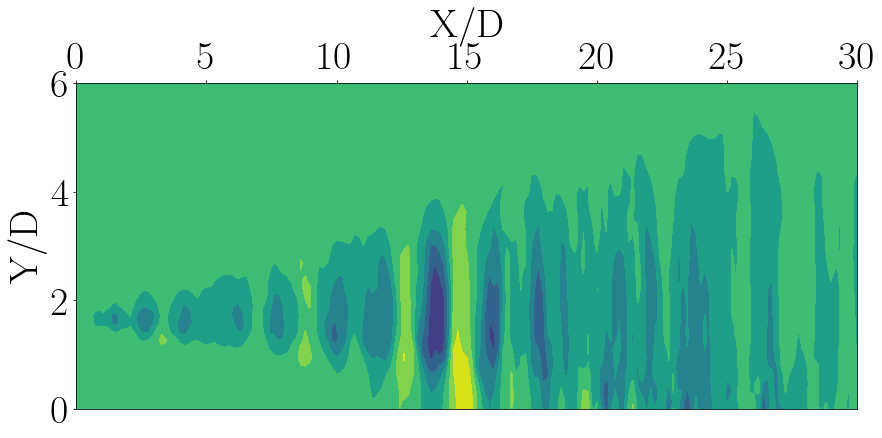} & \includegraphics[width=4cm]{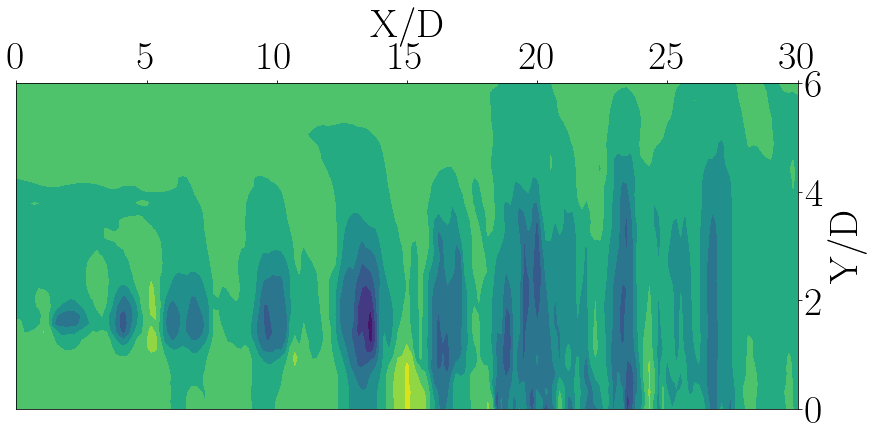} \\
         & \includegraphics[width=4cm]{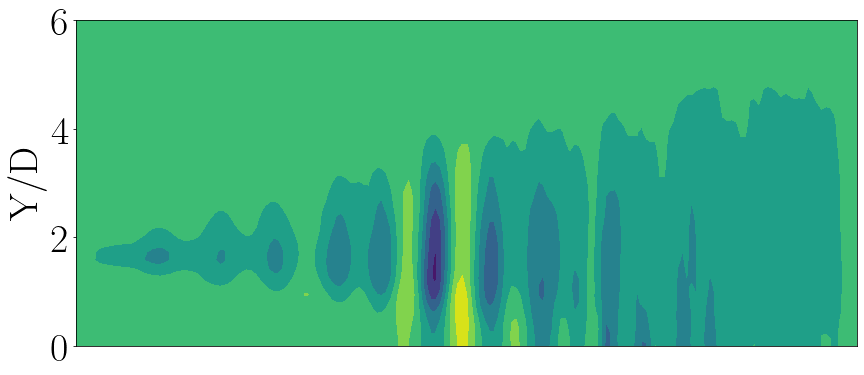} & \includegraphics[width=4cm]{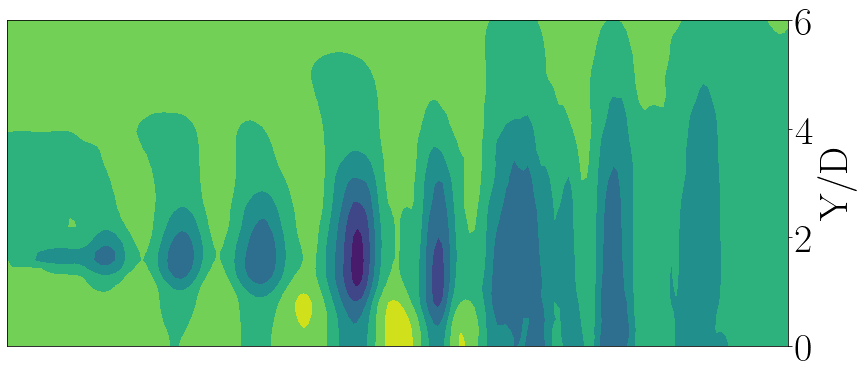}
    \end{tabular}
    \end{ruledtabular}
\end{table}

\begin{figure}
     \centering
     \begin{subfigure}[b]{0.5\textwidth}
         \centering
         \includegraphics[width=\textwidth]{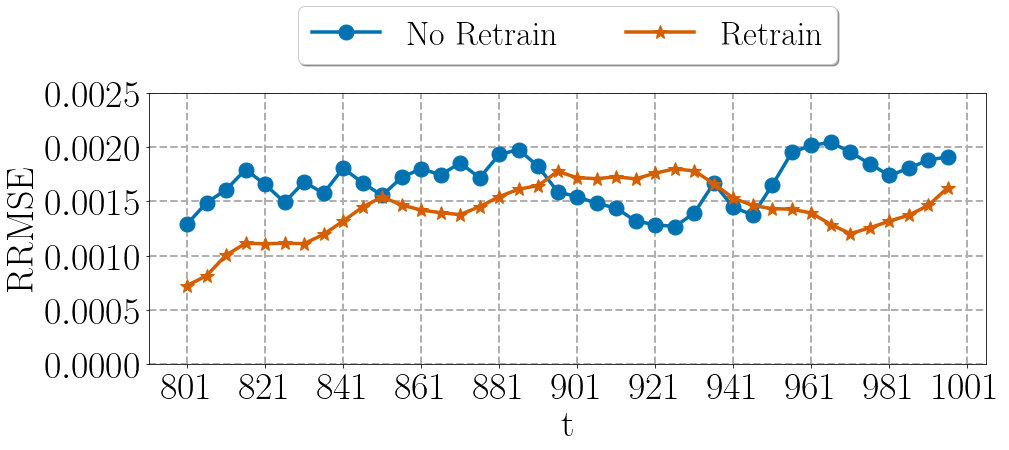}
         \caption{}
     \end{subfigure}
     \hfill
     \begin{subfigure}[b]{0.5\textwidth}
         \centering
         \includegraphics[width=\textwidth]{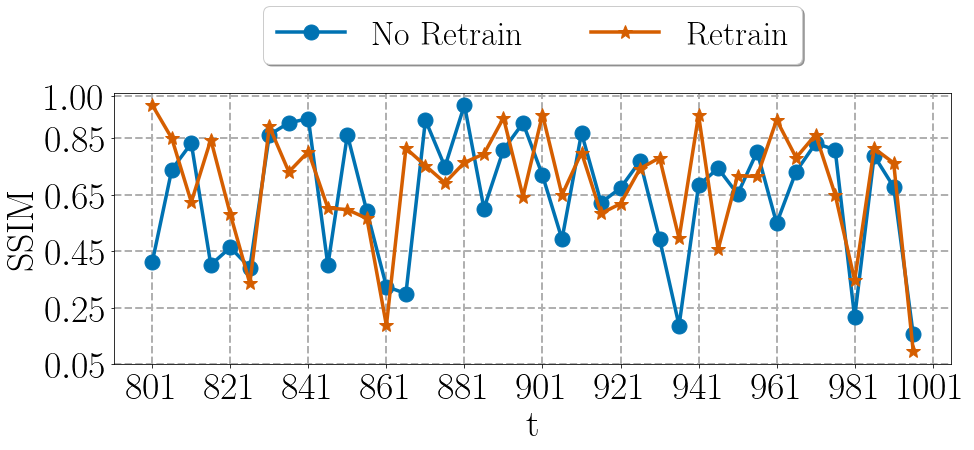}
         \caption{}
     \end{subfigure}
        \caption{Error committed by the POD-DL model with and without retraining, on the streamwise velocity component of the Jet LES. Errors are quantified using (a) the RRMSE and (b) the SSIM defined in \eqref{eq: rrmse} and \eqref{eq: ssim}, respectively.}
        \label{fig: new_data_rrmse_ssmi}
\end{figure}

\begin{figure}
     \centering
     \begin{subfigure}[b]{0.3\textwidth}
         \centering
         \includegraphics[width=\textwidth]{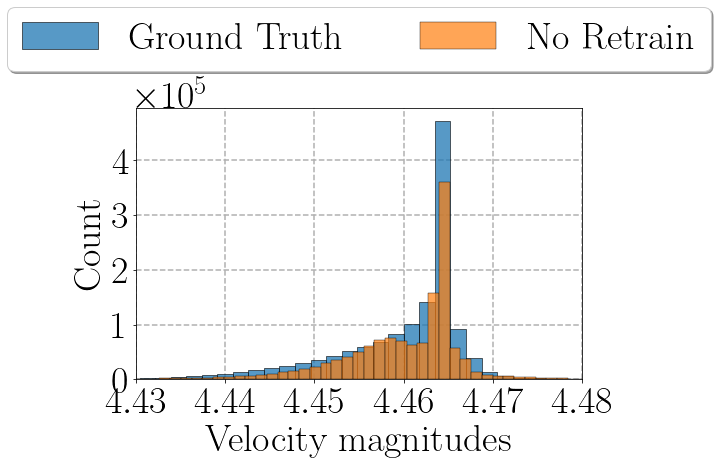}
         \caption{}
     \end{subfigure}
     \hfill
     \begin{subfigure}[b]{0.27\textwidth}
         \centering
         \includegraphics[width=\textwidth]{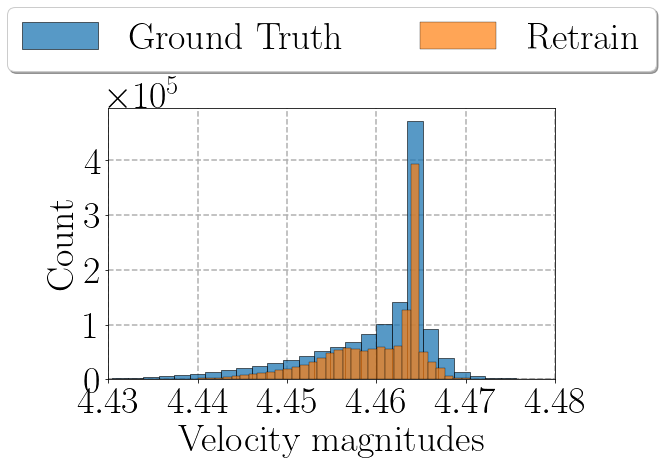}
         \caption{}
     \end{subfigure}
        \caption{Comparison of the velocity histograms between the ground truth and the temporal predictions of the Jet LES obtained from (a) the POD-DL without retraining and (b) the POD-DL with retraining, respectively.}
        \label{fig: new_data_histograms}
\end{figure}

This task was infeasible for the residual autoencoder and the VAE due to their architectures, which is rigidly tied to the dimensions of the snapshots used during the training. Adapting these architectures for snapshots of different dimensions would necessitate significant modifications to the model structure, making it impractical or impossible to reuse the pre-trained weights. This limitation could be addressed using techniques such as Spatial Pyramid Pooling \cite{kaiming_etal_2014_pyramid_pooling}, which standardizes incoming snapshots to a fixed size.

However, the inherent design of the POD-DL model circumvents this issue entirely, where the dimensionality of the snapshots is irrelevant, as the requirement is solely to truncate to the same number of POD modes as during training.

The dataset chosen to carry out these tests comes from a numerical simulation of an isothermal subsonic jet \cite{guillaume_etal_2018_Jet_LES} originating from a round nozzle with an exit diameter of $D = 50$mm. The simulation was performed using the large eddy simulation (LES) flow solver, CharLES, which solves the spatially-filtered compressible Navier-Stokes equations on unstructured grids. The solver employs a finite-volume method combined with third-order Runge-Kutta time integration. The simulation utilized approximately sixteen million control volumes and was conducted for a duration of 2000 acoustic time units ($t c_{\infty} / D$) and at Re $\approx 10^{6}$.

This dataset comprises $5000$ snapshots of the jet's streamwise velocity, sampled at intervals of $\Delta t = 0.02$ acoustic time units. The data is captured on a structured cylindrical output grid that closely approximates the resolution of the underlying LES. The grid extends over a streamwise distance of $30D$ and a radial distance of $6D$, effectively capturing the jet's spatial dynamics.

To utilize the pre-trained POD-DL model for the turbulent flow, the same number of POD modes must be truncated, specifically $20$. Table \ref{tab: jetLES_orig_svd} presents a comparison between the ground truth snapshots obtained from the numerical simulation of the Jet LES and the SVD reconstruction, retaining only the $20$ most energetic modes.

As illustrated in Fig. \ref{fig: new_data_rrmse_ssmi}, there is no significant difference in the results between the cases with and without retraining, suggesting that the prediction accuracy is not substantially impacted by retraining the model using snapshots from the Jet LES dataset. We attribute this to the similarity found in the dynamics of these problems. Although their nature is different, the main dynamics is driven by a periodic solution, where the neural network model has been trained to forecast the flow evolution following this type of dynamics. Further studies should be carried out to test the POD-DL model in quasi-periodic systems, but this remains an open topic for future research.

\begin{figure*}
     \centering
     \begin{subfigure}[b]{0.6\textwidth}
         \centering
         \includegraphics[width=\textwidth]{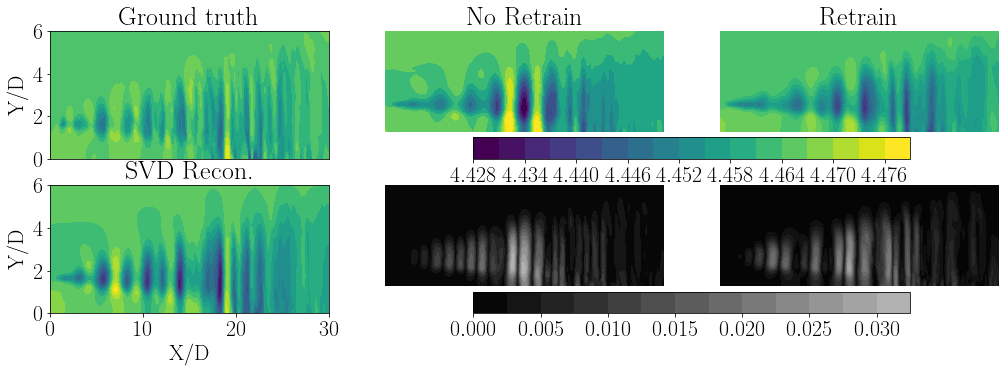}
         \caption{Ground truth snapshot and predictions at time instant $850$.}
     \end{subfigure}
     \hfill
     \begin{subfigure}[b]{0.6\textwidth}
         \centering
         \includegraphics[width=\textwidth]{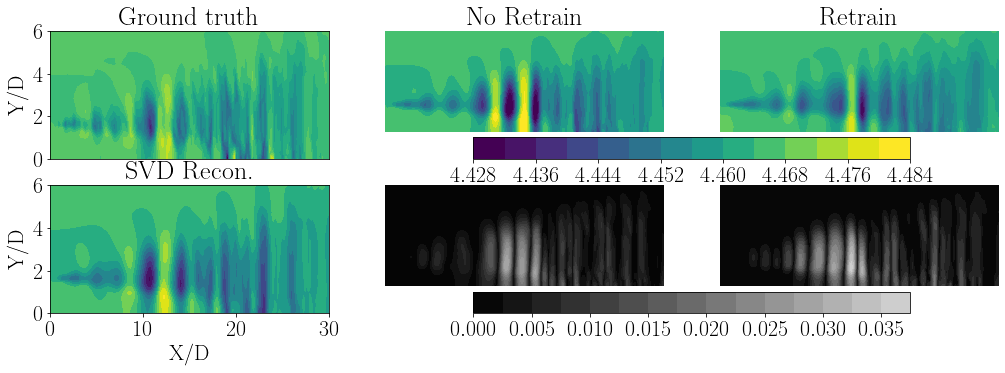}
         \caption{Ground truth snapshot and predictions at time instant $900$.}
     \end{subfigure}
     \hfill
     \begin{subfigure}[b]{0.6\textwidth}
         \centering
         \includegraphics[width=\textwidth]{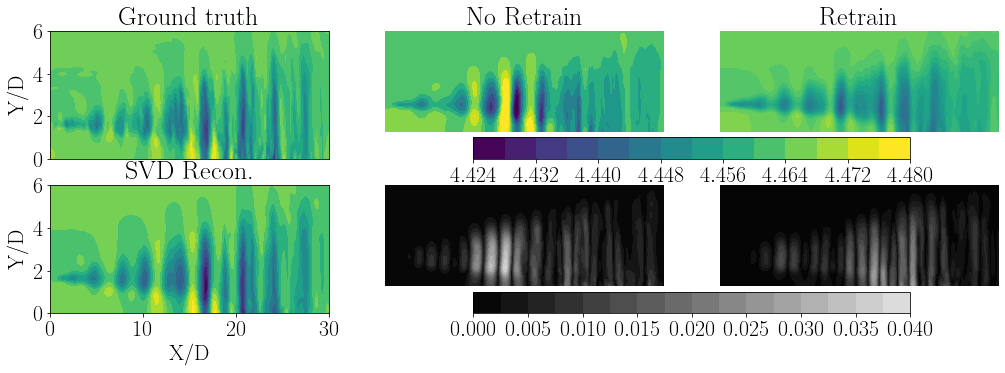}
         \caption{Ground truth snapshot and predictions at time instant $950$.}
     \end{subfigure}
        \caption{Snapshots at some representative time instants showing the ground truth streamwise velocity (left column) and the predictions, from the POD-DL model, for the Jet LES. No retrain (middle column) makes reference to the predictions obtain from case (1) and Retrain (right column) from case (2), described at the beginning of Appendix \ref{appen: pretrained_model}.}
        \label{fig: new_data_strw_snapshots_comp}
\end{figure*}

Figure \ref{fig: new_data_histograms} compares the probability distributions of the ground truth dataset with those predicted by the POD-DL model, highlighting the model's relative accuracy. Notably, the predictions are slightly more accurate when no retraining is applied. This trend is also qualitatively observed in Fig. \ref{fig: new_data_strw_snapshots_comp}, which provides a visual comparison between the ground truth and predicted snapshots.

\begin{figure}
    \centering
    \includegraphics[width=0.5\textwidth]{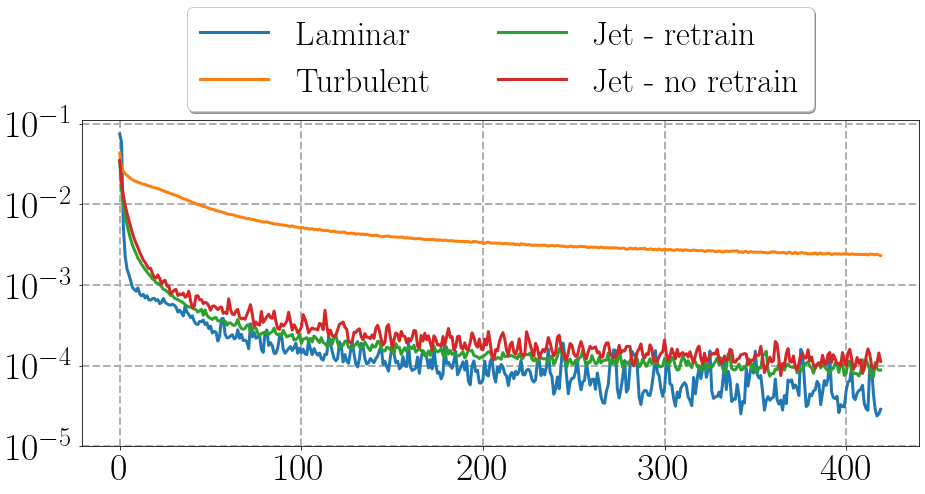}
    \caption{Loss evolution of the POD-DL model during training for different datasets: laminar flow past a cylinder (Laminar), turbulent flow past a cylinder (Turbulent), Jet LES with retraining (Jet - retrain), and Jet LES without retraining (Jet - no retrain).}
    \label{fig: pod_dl_loss}
\end{figure}

Figure \ref{fig: pod_dl_loss} presents the loss evolution of the POD-DL model during training across all dataset cases examined in this work: laminar flow, turbulent flow, and Jet LES with and without retraining. Notably, training on the turbulent flow dataset stagnates at a higher loss value compared to other cases, underscoring the increased complexity of this scenario. Additionally, for the Jet LES case, the training dynamics of the POD-DL model appear largely unaffected by retraining.

\section{Models architecture and residual autoencoder latent space}\label{appen: model_arch_tab}

This appendix show the tables describing the detailed architecture for the models presented in Section \ref{sec: methodology}: POD-DL (Tab. \ref{tab: pod_dl}), residual autoencoder (Tabs. \ref{tab: encoder}, \ref{tab: hidden}, \ref{tab: decoder}) and variational autoencoder (Tabs. \ref{tab: vae_encoder}, \ref{tab: vae_decoder}). Note for the autoencoders the architecture described is conditioned on the incoming dataset, which in this case is the laminar flow \cite{LeClainche_2018_cilind}, since the architecture of these models directly depends on the resolution of the incoming snapshots.

\begin{figure*}
	\includegraphics[trim={0cm 0cm 0cm 0cm},height=2.3\columnwidth]{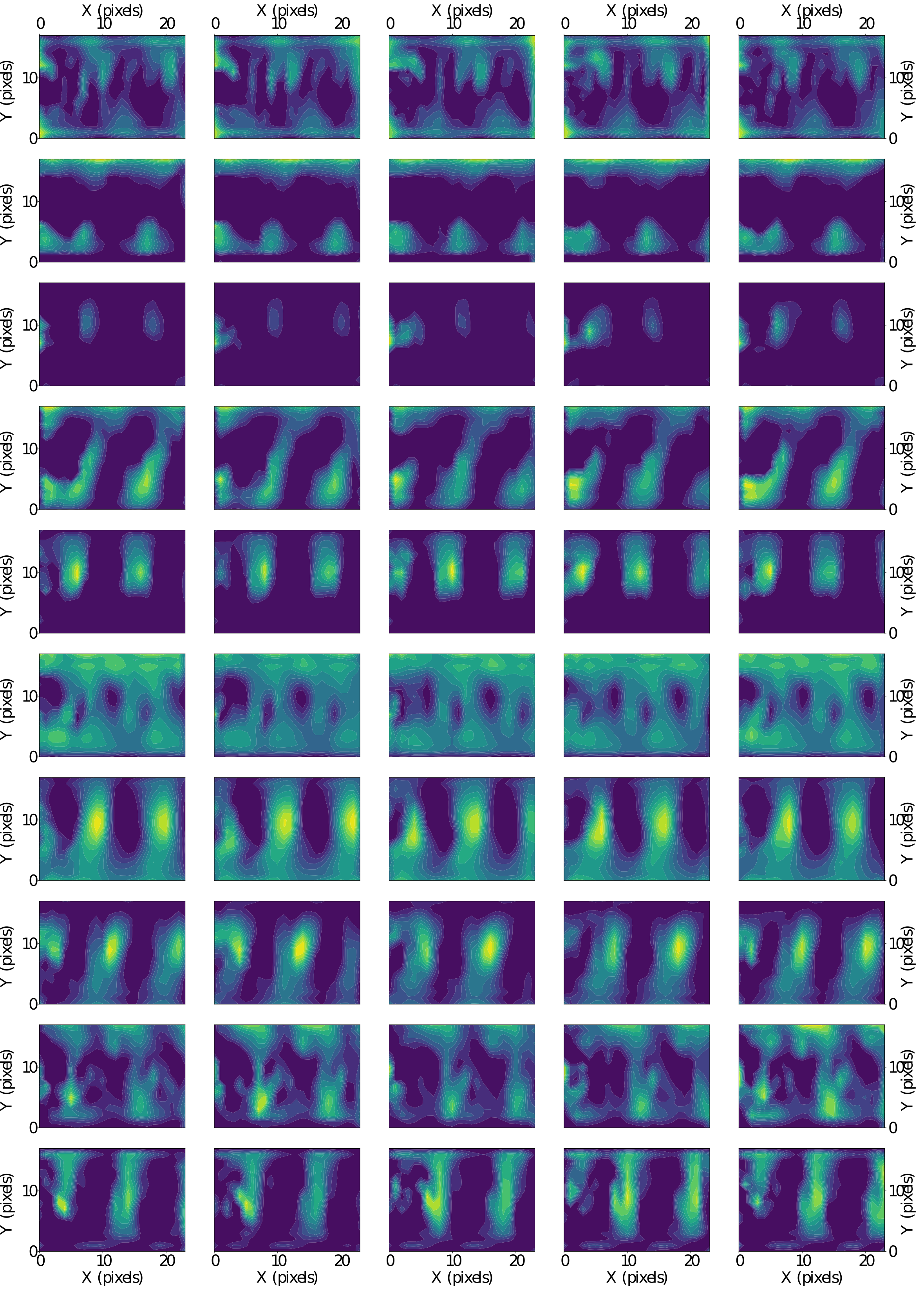}
	\caption{Snapshots of the latent variables obtained from the encoder, in the residual convolutional autoencoder, when the input dataset is the laminar flow \cite{LeClainche_2018_cilind}. Note the encoder takes as entries a sequence of $10$ previous snapshots $(10 \times 100 \times 40 \times 3)$ and returns a tensor with size $(10 \times 24 \times 18 \times 64)$. From top to bottom are represented the channels $1,2,3,4,5,10,20,35,55$ and $64$. From left to right is the temporal variation, forward in time, of these channels along $5$ snapshots. Recall from Section \ref{sec: deterministic_framework} that neither the encoder and decoder modifies the temporal dimension. Unlike the Singular Value Decomposition, which returns the modes sorted by its energy, the encoder does not sort the latent variables. However, several methods have been proposed to sort the latent variables produced by an autoencoder \citep{pod_autoencoders_ConvAE_Fukami_2020_ordered, munoz_etal_2023_sort_latent_variables}, providing information on which latent variable contributes most to the overall dynamics, but this is beyond the scope of this work. Note the dimensions of the latent variables shown in this figure are measured in pixels.}
	\label{fig: enc_out_kth}
\end{figure*}

Figure \ref{fig: enc_out_kth} presents the latent variables generated by the encoder of the residual autoencoder model when applied to the laminar flow. The visualization highlights distinct spatial structures within the flow dynamics, suggesting that the encoder effectively decomposes the flow into its constituent dynamic components.

\begin{table*}
    \caption{Architecture details in the POD-DL model. Unlike the other two models that receive snapshots as inputs, the POD-DL receive vectors representing the retained POD modes. In the case of the laminar flow is $20$, as explained in section \ref{sec: results_synthetic}. The term ($6 \times$) in the second and third layer, mean that $6$ layers with the same configuration are trained in parallel.\label{tab: pod_dl}}
    \begin{ruledtabular}
    \begin{tabular}{ccccc}

        \textbf{Layer} & \textbf{Layer details} & \textbf{\# Neurons} & \textbf{Activation function} & \textbf{Dimension}\\
        \hline
        $0$ & Input & $N$ & - & $10 \times 20$\\
        $1$ & LSTM & $6 * 100$ & ReLU & $1 \times 600$\\
        $2$ & Reshape & - & - & $6 \times 1 \times 100$\\
        $3$ & ($6 \times$) FC & ($6 \times$) $80$ & ReLU & $6 \times 1 \times 80$ \\
        $4$ & ($6 \times$) FC & ($6 \times$) $20$ & Linear & $6 \times 1 \times 20$ \\
    \end{tabular}
    \end{ruledtabular}
\end{table*}

\begin{table*}
    \caption{Encoder architecture in the Residual Autoencoder, when the input are snapshots coming from the laminar flow. Note, this architecture may vary depending on the size of the input snapshots (i.e., its resolution). By varying we mean adding or removing Indentity Blocks (Identity B.) and Convolutional Blocks (Conv B.).\label{tab: encoder}}
    \begin{ruledtabular}
    \begin{tabular}{cccccccc}
        \textbf{\# Layer} & \textbf{Layer details} & \textbf{Kernel size} & \textbf{Stride} & \textbf{Padding} & \textbf{Activation} & \textbf{\# Params.} & \textbf{Dimension}\\
        \hline
        $0$& Input & - & - & - & - & - & $10 \times 100 \times 40 \times 2$\\
        $1$& Conv 2D & $3 \times 3$& $2 \times 1$& Valid & ReLU & $448$ & $10 \times 49 \times 38 \times 16$\\
        $1$& LayerNorm. & - & - & - & - & $32$ & $10 \times 49 \times 38 \times 16$\\

        $2$& Conv 2D & $3 \times 3$& $1 \times 1$& Valid & ReLU & $4640$ & $10 \times 47 \times 36 \times 32$\\
        $2$& LayerNorm. & - & - & - & - & $64$ & $10 \times 47 \times 36 \times 32$\\

        $3$& Identity B. & $3 \times 3$& $1 \times 1$& Same & ReLU & $18560$ & $10 \times 47 \times 36 \times 32$\\
        $4$& Identity B. & $3 \times 3$& $1 \times 1$& Same & ReLU & $18560$ & $10 \times 47 \times 36 \times 32$\\
        $5$& Identity B. & $3 \times 3$& $1 \times 1$& Same & ReLU & $18560$ & $10 \times 47 \times 36 \times 32$\\

        $6$& Conv B. & $3 \times 3$& $2 \times 2$& Valid & ReLU & $57600$ & $10 \times 24 \times 18 \times 64$\\

        $7$& Identity B. & $3 \times 3$& $1 \times 1$& Same & ReLU & $73984$ & $10 \times 24 \times 18 \times 64$\\
        $8$& Identity B. & $3 \times 3$& $1 \times 1$& Same & ReLU & $73984$ & $10 \times 24 \times 18 \times 64$\\
        $9$& Identity B. & $3 \times 3$& $1 \times 1$& Same & ReLU & $73984$ & $10 \times 24 \times 18 \times 64$\\
    \end{tabular}
    \end{ruledtabular}
\end{table*}

\begin{table*}
    \caption{Same as Table \ref{tab: encoder} for the hidden layer in the Residual Autoencoder.\label{tab: hidden}}
    \begin{ruledtabular}
    \begin{tabular}{cccccccc}
        \textbf{\# Layer} & \textbf{Layer details} & \textbf{Kernel size} & \textbf{Stride} & \textbf{Padding} & \textbf{Activation} & \textbf{\# Params.} & \textbf{Dimension}\\
        \hline
        $0$& Input & - & - & - & - & - & $10 \times 24 \times 18 \times 64$\\
        
        $1$& ConvLSTM & $3 \times 3$ & $1 \times 1$& Same & Tanh/Sigmoid & $884736$ & $1 \times 24 \times 18 \times 128$\\
        $1$& LayerNorm. & - & - & - & - & $256$ & $1 \times 24 \times 18 \times 128$\\
    \end{tabular}
    \end{ruledtabular}
\end{table*}

\begin{table*}
    \caption{Same as Table \ref{tab: encoder} for the decoder in the Residual Autoencoder. As in the encoder, the number of Transpose Convolutional layers depends on the size of the input snapshots.\label{tab: decoder}}
    \begin{ruledtabular}
    \begin{tabular}{cccccccc}
        \textbf{\# Layer} & \textbf{Layer details} & \textbf{Kernel size} & \textbf{Stride} & \textbf{Padding} & \textbf{Activation} & \textbf{\# Params.} & \textbf{Dimension}\\
        \hline
        $0$& Input & - & - & - & - & - & $1 \times 24 \times 18 \times 128$\\

        $1$& Conv 2D T. & $3 \times 3$ & $2 \times 2$& Valid & ReLU & $73728$ & $1 \times 49 \times 37 \times 64$\\
        $1$& LayerNorm. & - & - & - & - & $128$ & $1 \times 49 \times 37 \times 64$\\

        $2$& Conv 2D T. & $4 \times 4$ & $2 \times 1$ & Valid & ReLU & $32768$ & $1 \times 100 \times 40 \times 32$\\
        $2$& LayerNorm. & - & - & - & - & $64$ & $1 \times 100 \times 40 \times 32$\\

        $3$& Conv 2D & $1 \times 1$& $1 \times 1$& Valid & Linear & $96$ & $1 \times 100 \times 40 \times 2$\\
    \end{tabular}
    \end{ruledtabular}
\end{table*}

\begin{table*}
    \caption{Encoder architecture in the variational autoencoder, when the input are snapshots coming from the laminar flow. Note, this architecture may vary depending on the size of the input snapshots (i.e., its resolution). By varying we mean adding or removing convolutional layers (Conv 2D).\label{tab: vae_encoder}}
    \begin{ruledtabular}
    \begin{tabular}{cccccccc}
        \textbf{\# Layer} & \textbf{Layer details} & \textbf{Kernel size} & \textbf{Stride} & \textbf{Padding} & \textbf{Activation} & \textbf{\# Params.} & \textbf{Dimension}\\
        \hline
        $0$& Input & - & - & - & - & - & $10 \times 100 \times 40 \times 1$\\
        
        $1$& Conv 2D & $3 \times 3$& $2 \times 1$& Valid & ReLU & $400$ & $10 \times 94 \times 34 \times 8$\\
        $1$& LayerNorm. & - & - & - & - & $16$ & $10 \times 94 \times 34 \times 8$\\

        $2$& Conv 2D & $3 \times 3$& $1 \times 1$& Valid & ReLU & $1168$ & $10 \times 46 \times 16 \times 16$\\
        $2$& LayerNorm. & - & - & - & - & $32$ & $10 \times 46 \times 16 \times 16$\\

        $3$& Conv 2D & $3 \times 3$& $1 \times 1$& Valid & ReLU & $4640$ & $10 \times 22 \times 14 \times 32$\\
        $3$& LayerNorm. & - & - & - & - & $64$ & $10 \times 22 \times 14 \times 32$\\

        $4$& ConvLSTM & $3 \times 3$& $2 \times 1$& Valid & ReLU & $221440$ & $1 \times 11 \times 14 \times 64$\\

        $5$& Flatten & - & - & - & - & - & $1 \times 9856$\\

        $6$& Dense & - & - & - & - & $3449950$ & $1 \times 350$\\

        $7$& Posterior dist. & - & - & - & - & - & $1 \times 25$\\
    \end{tabular}
    \end{ruledtabular}
\end{table*}

\begin{table*}
    \caption{Same as Table \ref{tab: vae_encoder} for the decoder in the variational autoencoder. As in the encoder, the number of transpose convolutional layers depends on the size of the input snapshots.\label{tab: vae_decoder}}
    \begin{ruledtabular}
    \begin{tabular}{cccccccc}
        \textbf{\# Layer} & \textbf{Layer details} & \textbf{Kernel size} & \textbf{Stride} & \textbf{Padding} & \textbf{Activation} & \textbf{\# Params.} & \textbf{Dimension}\\
        \hline
        $0$& Input & - & - & - & - & - & $1 \times 25$\\

        $1$& Dense & - & - & - & - & $256256$ & $1 \times 9856$\\

        $2$& Reshape & - & - & - & - & - & $1 \times 11 \times 14 \times 64$\\

        $3$& Conv 2D T. & $3 \times 3$ & $2 \times 2$& Valid & ReLU & $73728$ & $1 \times 23 \times 16 \times 64$\\
        $3$& LayerNorm. & - & - & - & - & $128$ & $1 \times 23 \times 16 \times 64$\\

        $4$& Conv 2D T. & $4 \times 4$ & $2 \times 1$ & Valid & ReLU & $32768$ & $1 \times 47 \times 18 \times 32$\\
        $4$& LayerNorm. & - & - & - & - & $64$ & $1 \times 47 \times 18 \times 32$\\

        $5$& Conv 2D T. & $3 \times 3$ & $2 \times 2$& Valid & ReLU & $73728$ & $1 \times 96 \times 37 \times 16$\\
        $5$& LayerNorm. & - & - & - & - & $128$ & $1 \times 96 \times 37 \times 16$\\

        $6$& Conv 2D T. & $3 \times 3$ & $2 \times 2$& Valid & ReLU & $73728$ & $1 \times 100 \times 40 \times 8$\\
        $6$& LayerNorm. & - & - & - & - & $128$ & $1 \times 100 \times 40 \times 8$\\

        $7$& Likelihood dist. & - & - & - & - & $353565$ & $1 \times 100 \times 40 \times 1$\\ 
    \end{tabular}
    \end{ruledtabular}
\end{table*}

\end{document}